\documentclass[%
 aip,
 amsmath,amssymb,
 reprint,%
]{revtex4-1}
\newcommand{\comment}[1]{}
\usepackage{graphicx}
\usepackage{dcolumn}
\usepackage{bm}
\setcitestyle{numbers,square}
\usepackage[utf8]{inputenc}
\usepackage[T1]{fontenc}
\usepackage{mathptmx}
\usepackage{xcolor}
\begin{document}
\setcitestyle{numbers,square}

\preprint{AIP/123-QED}

\title[Rheology of Wormlike Micellar Gels]{Rheology of Wormlike Micellar Gels Formed By Long-Chained Zwitterionic Surfactants}

\author{Ronak Gupta}
 \affiliation{ 
Department of Mechanical Engineering, University of British Columbia, 6250 Applied Science Ln, Vancouver, BC, V6T1Z4, Canada
}%

\author{Rodrigo Mitishita}%
 \affiliation{ 
Department of Mechanical Engineering, University of British Columbia, 6250 Applied Science Ln, Vancouver, BC, V6T1Z4, Canada
}%

\author{Ian A. Frigaard}
\affiliation{ 
Department of Mechanical Engineering, University of British Columbia, 6250 Applied Science Ln, Vancouver, BC, V6T1Z4, Canada
}%
\affiliation{ 
 Department of Mathematics, University of British Columbia, 1984 Mathematics Road, Vancouver, BC, V6T 1Z2, Canada
}%

\author{Gwynn J. Elfring}
 \homepage{Email address for correspondence : gelfring@mech.ubc.ca}
\affiliation{ 
Department of Mechanical Engineering, University of British Columbia, 6250 Applied Science Ln, Vancouver, BC, V6T1Z4, Canada
}%

\date{\today}

\begin{abstract}
Long-chained surfactant solutions have found widespread use in the oil and gas industry due to a host of attractive properties. In this paper, we characterize one such commercially used viscoelastic surfactant that forms a wormlike micellar gel at room temperature and a viscoelastic solution at higher temperatures. We probe both states by conducting linear and nonlinear rheological tests and analyze their behaviour under the framework of micellar rheology. Our study outlines departure from behaviour exhibited by more conventional micellar systems and uncovers interesting dynamics like shear-induced fracture and possible shear-banding in these materials. In doing so we provide a detailed understanding of a novel class of wormlike micellar solutions.
\end{abstract}

\maketitle

\section{\label{intro}Introduction}
\noindent
The amphiphilic nature of surfactant molecules leads to unfavourable contact between its hydrophopic tail section and the solvent, resulting in a wide variety of spontaneous self assembled structures. The morphology of these assemblies is dictated by energetic and entropic considerations and is a complex interplay of geometry and physiochemical interactions which can be controlled by a gamut of parameters like temperature, surfactant concentration, solvent pH and added molecules (salts, co-surfactants etc). Increasing surfactant concentration favours the formation of linear worm-like chains which can entangle to form a dynamic network of wormlike micelles (WLM). \textcolor{black}{Further increasing surfactant concentration can lead to the formation of a network of branched micelles}. For this paper, our system of investigation comprises of aqueous solutions of entangled WLM's.

The study of micellar solutions as complex fluids has been an active field of research for a long time owing to interesting flow dynamics \cite{fardin2014flows} and rheology \cite{cates2006rheology,rehage1991viscoelastic}. While, at a static coarse-grained level, entangled WLM solutions share similarities with entangled polymer suspensions, there are two crucial differences. The contour length of worms in WLM solutions is not fixed and has a broad distribution that depends on external factors like temperature \cite{cates1990statics}. Also, unlike polymers, surfactant worms are held together by non-covalent bonds. These weak bonds allow frequent monomer exchange between worms - a breaking-recombination process that leads to the formation of a \textit{transient} entangled network and lend wormlike micellar solutions the monicker of living polymers. The dynamic nature of micellar networks allows the complex fluid to have an additional mechanism for stress relaxation. Much like a system of entangled polymers, entangled worms can relax stress by \textit{reptation} - diffusing out of surrounding constraints. But, while the worms reptate out of their confining tube, they can break and recombine multiple times. In the \textit{fast breaking} limit, the solution is a viscoelastic liquid with a single relaxation time \cite{cates1987reptation}. This property makes WLM solutions a fertile system for experimental and theoretical investigation as they can also be used as a proxy for viscoelastic fluids with Maxwellian rheology. Additionally, WLM solutions re-form after shearing unlike polymer solutions that degrade permanently when exposed to strong shearing flows. WLM solutions thus prove to be ideal systems to probe nonlinear flow and rheological behaviour \cite{rehage1991viscoelastic,sood1999linear,cates2006rheology}. Entangled WLM solutions, often in the semi dilute regime, display a host of instabilities in different canonical fluid flow scenarios \cite{rothstein2008strong,fardin2012instabilities,zhao2014microfluidic}. Understanding the genesis of these novel dynamical phenomena and tying them to microstructural details is a major challenge in complex fluid dynamics and rheology.  \cite{berret2006rheology,fardin2014flows}. 

Aside from being interesting from a fundamental perspective, WLM solutions are crucial in a number of industrial applications \cite{yang2002viscoelastic}. Properties that make them attractive rheology modifiers include high shear rheology, viscoelasticity and self-healing ability. A combination of these properties are often exploited in biomedical, pharmaceutical and personal health applications. But perhaps the most active use of WLM solutions has been in the oil-gas industry \cite{ezrahi2006properties}. WLM solutions can suspend particles due to their high viscosity, but do not incur energy penalties while pumping owing to viscoelasticity enabled drag reduction capabilities. Drag reduction in viscoelastic fluids is a research problem of immense interest in physics and engineering \cite{virk1975drag}. While, extensively investigated in polymeric fluid systems \cite{white2008mechanics}, drag reduction by surfactant additives has also gained attention \cite{li2012turbulent}. Here too, there is a strong demand for a fundamental understanding of the rheology WLM solutions.

In the last two decades, a specific type of surfactant based fluid has stoked the interest of rheologists, primarily motivated by hydraulic fracturing \cite{sullivan2007oilfield} and gravel packing operations \cite{kefi2004expanding} in oil and gas production. In the former, a carrier fluid drives the suspended proppant particles to fill fractures in bedrock. In the latter, gravel carried by a suspending fluid is used to 'pack'  behind wellbore screens and prevent sand contamination. Both processes utilize the above mentioned beneficial properties of WLM solutions, namely excellent solid suspension ability and drag-reduction, which they can retain unlike polymer suspensions. Further, WLM's do not require the use of complex gel breakers and easily break down into spherical micelles on contact with oil, providing little hindrance for oil production \cite{maitland2000oil,boek2002molecular}. 

The surfactant of interest here differs from standard surfactant systems in one key aspect. It is made of a backbone of 22 carbon atoms, unlike previously investigated systems that have tails of 12,16 or 18 carbon atoms and is thus referred to as \textit{long-chained}. The first such WLM solution studied was EHAC and it showed some rheological oddities \cite{raghavan2001highly}. The WLM solution behaved like an elastic gel at room temperature, a marked difference from conventional WLM rheology which typically displays Maxwellian viscoelasticity. Studies on long-chained surfactant systems have since then been pioneered by Raghavan \cite{raghavan2001highly,kumar2007wormlike}, Feng and co-workers \cite{chu2010wormlike,chu2011thermo,han2011wormlike}. In \cite{kumar2007wormlike}, a detailed rheological study of a long-chained (22 C atoms) zwitterionic surfactant system called EDAB unveiled gel-like rheology and a yield stress at room temperature. Solutions of EDAB reverted to classical Maxwellian rheology at higher temperatures i.e WLM solutions of EDAB showed a elastic gel-viscoelastic sol transition with temperature.
The gel-like nature of these solutions is antithetical to traditional WLM solutions and is thought to arise from purely topological interactions  \cite{raghavan2012conundrum,raghavan2017wormlike}.

Because of the wide-ranging usage of wormlike micellar solutions, there is a consistent push for the design of smart micelles that can respond actively/passively to various stimuli \cite{chu2013smart}. In this context, surfactant gels are increasingly being considered as a viable replacement for polymer based hydrogels in biomedical applications like tissue engineering \cite{lee2001hydrogels}. Further, researchers have proposed \cite{raghavan2012conundrum,raghavan2009distinct} that gel formation in surfactant systems shares similarities with some molecular self-assembled systems like entangled F-Actin \cite{mackintosh1997actin}. Infact, strain stiffening, a signature of many biopolymer networks was also observed in a surfactant based organo-gel \cite{tung2008self}. Certain dendritic surfactant systems \cite{bhattacharya2011surfactants} have utility in the fabrication of nanomaterials. Recently, one such system was shown to exhibit structural and rheological properties of gels formed by entangled wormlike micelles \cite{xie2017unique}. Thus, wormlike micellar gels are not only an interesting soft matter system by themselves, but can prove to be a fertile experimental playground to understand behaviour of many analogous systems too. A prime goal of this study is to properly investigate the linear and nonlinear rheology of wormlike micellar gels formed by long chained zwitterionic surfactants and place this unique surfactant system, first within the narrow context of wormlike micellar solutions and then in the broader context of gels and amorphous soft matter.

The surfactant system we focus on is a viscoelastic surfactant solution (VES) provided by Schlumberger Oilfield Services. The solution contains a zwitterionic surfactant similar to previously studied systems \cite{kumar2007wormlike}. Among surfactants, zwitterionic systems are popular in commercial applications as they have environmentally friendly qualities \cite{christov2004synergistic} like high biodegradability \cite{zhou2018surface} and minimal toxicity and have yet received only scant interest from academia. The specific VES system was used in \cite{goyal2017} in which the authors showed that the VES displayed gel-like behaviour. While \cite{goyal2017} primarily focused on shear-rheology in the context of flow-loop studies, here we carry out a more extensive analysis of VES's rheological behaviour by using oscillatory and shear-rheology probes. We characterize the microstructure visually and then indirectly via rheology based analysis and then move on to discussing dynamics under an imposed shear-rate.  
\section{Materials and Methods}
\label{experimental}
\subsection{Surfactant and Preparation}
\noindent Our model system is a drag-reducing gravel packing surfactant product called J590 that is supplied to our laboratory by Schlumberger in liquid form \cite{goyal2017}. VES solutions are usually a combination of a zwitterionic surfactant and a polar solvent - an alcohol for example. 
Our VES solutions contain a mixture of \textit{erucic amidopropyl dimethyl betaine} and propan-2-ol. The surfactant in the VES solution has a CAS number whose associated structure is similar to the zwitterionic surfarctants used in previous studies \cite{kumar2007wormlike,mccoy2016structural}. In \cite{kumar2007wormlike}, the surfactant is called EDAB and \cite{mccoy2016structural} refers to it as EAPB. We refrain from calling our surfactant solution either of those two names, and refer to it as VES throughout the paper. However, we expect that our results should be closely related to studies reported on EDAB/EAPB. Further, because the exact percentage of surfactant in the VES is unknown in the commercially supplied product, we only mention concentration (by \% weight or volume) of the provided VES solution which is proportional to the \textit{true} surfactant concentration. We use the VES as supplied. Inorganic salts, if present, have little effect on the rheological behaviour of zwitterionic surfactants \cite{kumar2007wormlike} and small amounts of alcohol do not greatly affect EDAB's rheological properties \cite{beaumont2013}. To prepare the micellar solution, we use tap water as solvent in order to simulate typical working conditions of the fluid in practice more closely and allow compliance with flow studies being conducted currently in our group. A single concentration of VES was mixed with \textcolor{black}{deionized (DI)} water and tested at two temperatures and we found no appreciable change in relevant quantities. Concentrations between 1 and 10 \% of surfactant was mixed vigorously with an IKA control 60 mixer at 2000 rpm for 20 minutes, with a high shear rate impeller. After mixing, the solution ends up being foamy due to the high mixing speeds. To de-gas the fluid, we can centrifuge at 3000 rpm for 10 minutes and then store the solution in a thermal bath for $\sim$1hr at 80$^{\circ}$C. A well mixed solution appears transparent on visual inspection.
 
\subsection{Methods}
 \label{met}
\noindent Rheological studies were performed with a Malvern Kinexus stress-controlled rheometer in a concentric cylinder geometry with a cup of 37.00 \textit{mm} diameter and height of 66.00 \textit{mm}, and bob with 33.65 \textit{mm} diameter, 37.50 \textit{mm} height and cone angle of 15$^{\circ}$. The inner surface of the cup and the outer surface of the bob were roughened to prevent wall slip. Prior to beginning experiments, the fluid was left to stabilize at a desired temperature for $\sim$20 minutes. A solvent trap was used to prevent evaporation of the sample. To erase memory of loading and ensure a well defined initial state, we applied a pre-shear at 100 \textit{s}$^{-1}$ for 6 minutes, followed resting time of 30 minutes at 0 $Pa$ before every experiment. This protocol is critical as the VES solution experiences a growth of elastic modulus ($G'$) with time, indicating structure buildup after structure destruction due to shearing (as experienced in the loading process). 

\subsection{Cryo-TEM}
\noindent
Cryo-TEM was performed in the Life Sciences Institute in the University of British Columbia. Samples of 4.5 \% and 10\% for cryo-TEM were prepared in a low humidity room to minimise contamination. Before vitrification, excess sample on the grid is removed by blotting, which shears the micellar structure. Keeping this in mind, samples were rested for $\sim$ 5 minutes after blotting, to allow microstructure to relax after blotting. We cannot guarantee that the chosen resting time between blotting and imaging is "sufficient" to let the microstructure fully relax. However, the waiting time in combination with other factors involved in this technique provided good micrographs. We acknowledge that some images showed a high degree of preferred alignment which is most likely a shear induced artifact. However, we expect that resting time should not affect the general morphology of micelles. The surfactant sample is then quenched very quickly in ethane at -180$^{\circ}C$. Visualization of the micellar structure in the vitrified samples was performed with a 300kV FEI Titan Krios electron microscope, equipped with a Falcon III camera.

\section{Results}
\label{results}
\subsection{Visual Observations}
\begin{figure}[h!]
\centering
\begin{tabular}{cc}
{\includegraphics[width=3.3cm]{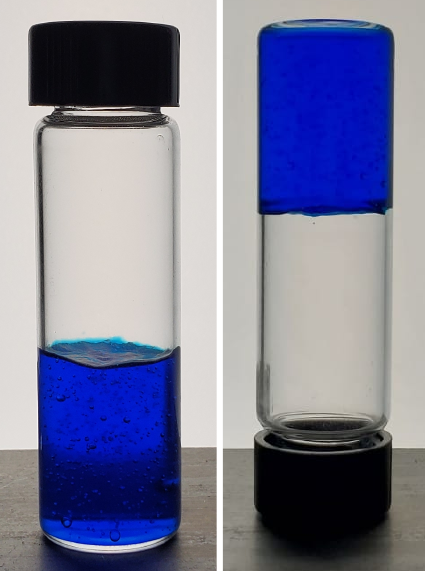}}
{\includegraphics[width=3.1cm]{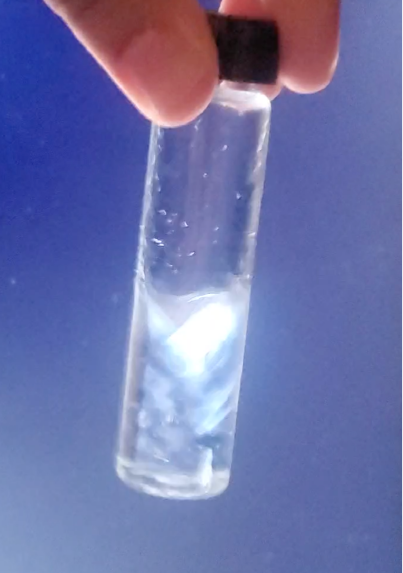}}
\end{tabular}
\caption{(a): Upturned vial of 10\%\textit VES holds its own weight. (b): Slightly tilted vial of 3\% VES solution showing strong birefringence. \textcolor{black}{Images taken at room temperature.}}
\label{fig:fig1}
\end{figure}

\begin{figure}[h!]
\centering
\begin{tabular}{cc}
{\includegraphics[width=4.1cm]{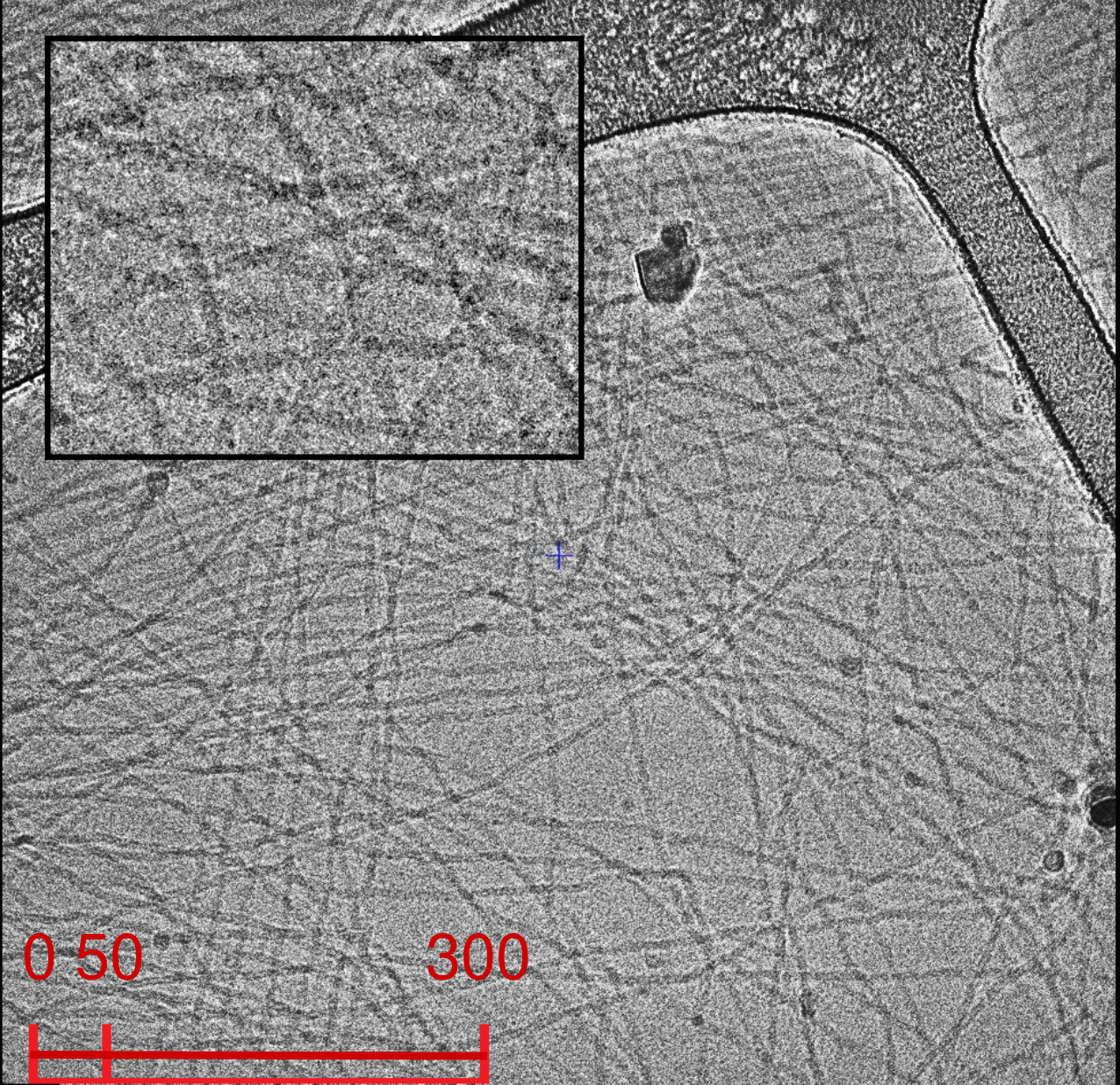}}
{\includegraphics[width=4.1cm]{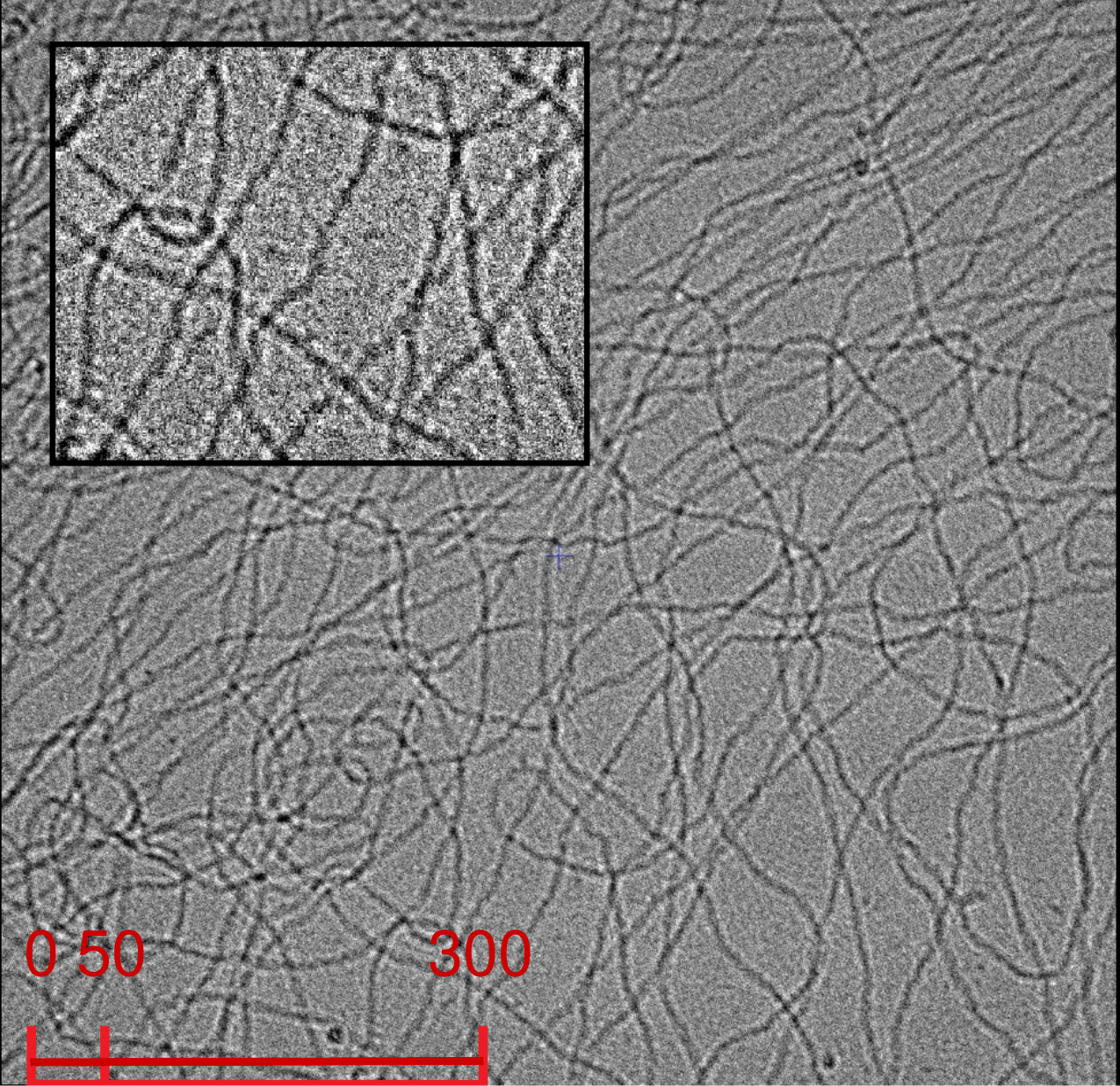}}
\end{tabular}
\caption{Cryo-TEM images of VES samples (a) 4.5\% (b) 10\%. Scale bar shown in red and numbers are length in $nm$. Inset shows a zoomed portion of the main image.}
\label{fig:fig2}
\end{figure}

\noindent After preparation, the VES solution appears gel-like at room temperature. Bubbles trapped remain so indefinitely and a vial of 10 \% solution holds it own weight when upturned for many days (Fig~\ref{fig:fig1}(a)). Vials of VES sample show strong birefringence when tilted and viewed in the presence of plane polarized light and a polarizer as seen in Fig~\ref{fig:fig1}(b), implying the presence of long structures that align when oriented, presumably micelles. The VES contains a zwitterionic surfactant in which closely situated opposite charges on the headgroup almost cancel each other leading to minimal headgroup repulsion. Such surfactants are therefore expected to form cylindrical wormlike micelles without the addition of salts \cite{kumar2007wormlike}. That wormlike micelles are indeed present in our VES solutions is confirmed on looking at Cryo-TEM images of our samples. 
Cryo-TEM is an ideal technique for visually probing complex fluids owing to the method's ability to directly image material microstructure in a natural state \cite{gonzalez2005cryo}. We image VES samples of two different concentrations (4.5,10\%) and show representative images in Fig~\ref{fig:fig2}. Cryo-TEM images have low contrast owing to the vitreous ice layer. However, our images are clear enough to confirm that VES samples indeed contain long wormlike/threadlike micelles. It is noteworthy that there are no visible cross-links or branching. The latter manifests as junctions in individual micelle threads that have the same contrast as the original micelle. Instead, we see a network of wormlike micelles in an overlapping - \textit{entangled} mesh. Points of entanglement can be identified as they appear darker than the rest of the micelle \cite{gonzalez2005cryo,kesselman2017direct}. Quantifying total and persistence lengths from such images is a challenging task that requires analysis of images from a variety of tilt angles \cite{clausen1992viscoelastic}. However, we see from Fig~\ref{fig:fig1}(a,b) that micelles are long and of O(1$\mu m$). For the 4.5\% sample,  the worms look rigid and persistent, more so than the micelles in the 10\% sample. It is possible that the visible rigidity in the 4.5\% sample is due to the straightening of micelles during blotting, whereas the visual difference between the two samples could be a result of difference in their relaxation times. Finally, we note that the images of our VES samples look very similar to those reported for EDAB in \cite{raghavan2012conundrum,raghavan2017wormlike}. Essentially, our Cryo-TEM images re-affirm that at room temperature, the VES samples contain long and entangled wormlike micelles.   
\subsection{Behaviour in Small Amplitude Oscillatory Sweep (SAOS) tests}
\begin{figure}[!h]
	\begin{center}
		  \includegraphics*[width=85mm]{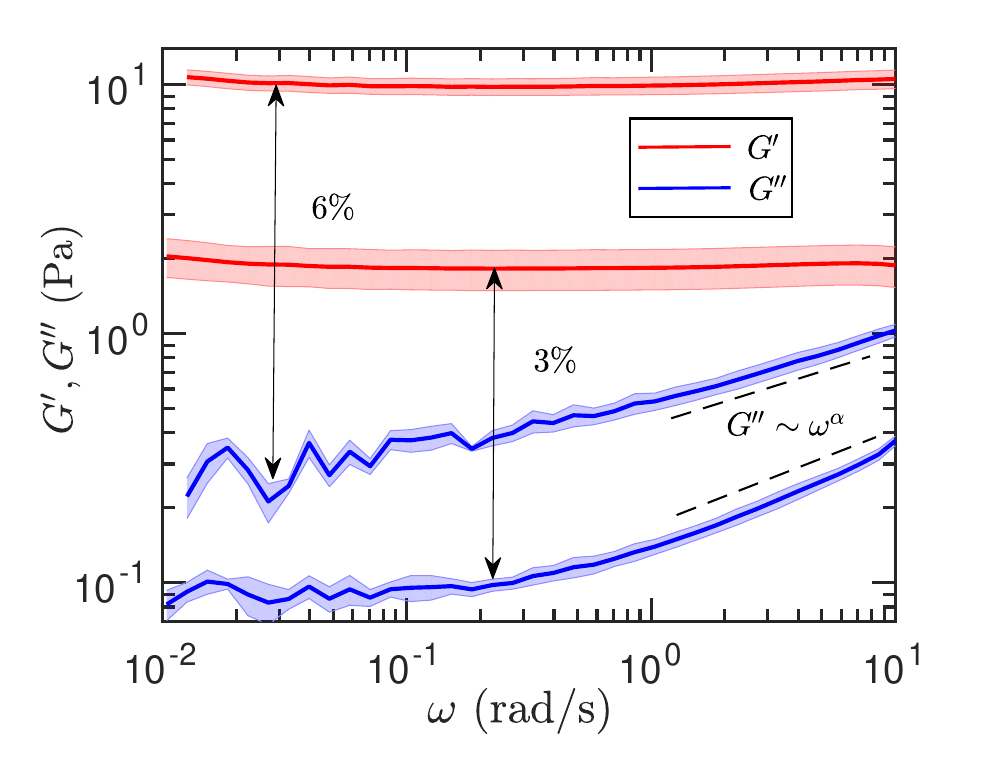}
	\end{center}
	\vspace*{-5mm}
	\caption{Frequency sweeps in the linear viscoelastic region for a 3\% and 6\% VES solutions at \textcolor{black}{25C}. Data is collected over $\sim$ 40 points in the $\omega$ range and plotted as a continuous line by linearly interpolating between discrete points. Shaded area represents standard deviation calculated by averaging data for three independent loadings and plotted the same way the curves for moduli are. }
	\label{saos1}
\end{figure}

To gain better insight into specifics of the rheological behaviour displayed by VES samples we first perform basic SAOS tests.\comment{These are tests in which material is linearly perturbed by oscillations of varying frequency at a fixed imposed strain amplitude.} Typically, entangled wormlike micelles display a classical Maxwellian viscoelastic behaviour in this test. The storage ($G'$) and loss ($G''$) moduli crossover at a critical frequency $\omega_c$ which demarcates the behaviour of the solution as elastic ($\omega>\omega_c$) and viscous ($\omega<\omega_c$) and endows it with a single dominant relaxation time $\tau_r = {\omega_c}^{-1}$. The VES samples at 25C show a marked deviation from this expected behaviour. As seen in Fig.~\ref{saos1}, the storage modulus is independent of $\omega$ for almost 3 decades, with no crossover in the two material moduli for the range of frequencies probed. $G'/G''>10$, for the range of frequencies probed. Thus, VES samples behave like elastic gels or solids with no structural relaxation at least on experimental timescales probed. This is contrary to conventional entangled wormlike micellar solutions, which may have a tendency to appear gel-like but show clear frequency dependence and crossover for $G',G''$.

\begin{figure}[!h]
	\begin{center}
		\includegraphics*[width=85mm]{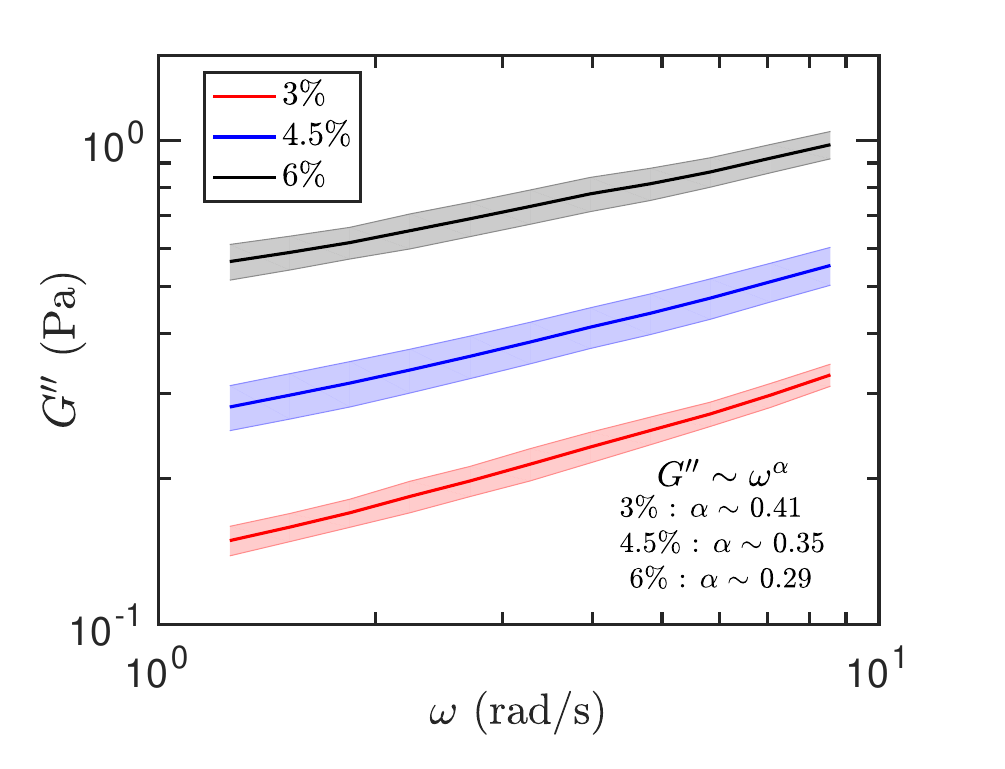}
	\end{center}
	\vspace*{-5mm}
	\caption{SAOS data for three concentrations at \textcolor{black}{25C} for $1<\omega<10$ plotted the same way as data in Fig.~\ref{saos1}. }
	\label{saoshighfreq}
\end{figure}

The gel-like rheology displayed by VES at room temperature is a unique feature shared by long-chained surfactant solutions, previously established for other surfactant systems like EHAC \cite{raghavan2001highly} and EDAB \cite{kumar2007wormlike}. In this, entangled wormlike micellar solutions formed by such surfactants share similarities with other molecular gel systems like F-Actin which can form gels despite the lack of chemical or physical cross-links \cite{raghavan2017wormlike}. Cross-links or jamming of constituents in classical gel systems constrain relaxation, leading to infinite relaxation times. However, as observed in our Cryo-TEM images, VES samples do not show cross-linking and only contain worms with topological constraints akin to entanglements. In \cite{raghavan2017wormlike}, it was proposed that such physical constraints can lead to very long reptation times and in combination with the unfavourable breaking process in worms, explain gel-like rheology in long-chained surfactant systems like the one we employ in this paper. We also notice that, unlike the frequency independence of $G'$, $G''$ seems to display two regimes. For $\omega<1s^{-1}$, $G''$ doesn't vary with $\omega$, whereas for $\omega>1s^{-1}$, $G''\sim\omega^\alpha$; the exponent $\alpha$ varies with concentration. This kind of power law dependence of loss modulus on frequency is reminiscent of other soft matter systems. For instance, $G''\sim\omega^{0.5}$ was observed in an emulsion system and interpreted as a dissipative contribution to the loss modulus activated at higher frequencies due to in-plane slip of material regions \cite{liu1996anomalous}. A similar trend exists for microgel systems in which $G''\sim\omega^{\alpha}$, $\alpha\in(0.3-0.5)$ for a range of concentrations \cite{divoux2011stress,conley2019relationship,migliozzi2020investigation}. Although, these power laws are established for a large range of frequencies starting typically from $\omega>1s^{-1}$, we were unable to get data for $\omega>10 s^{-1}$ owing to errors induced by instrument inertia. However, our system shows a consistent power law for $1s^{-1}<\omega<10s^{-1}$ for independent measurements and different concentrations as seen in \textcolor{black}{Fig.~\ref{saoshighfreq}}. It is possible that the mechanism responsible for frequency dependence of $G''$ in microgels and emulsions is active in our wormlike micellar gels too, although the microstructure underlying above mentioned emulsions and microgels bear a stark contrast to entangled wormlike micelles in VES. 

\begin{figure}[!h]
	\begin{center}
		\includegraphics*[width=65mm]{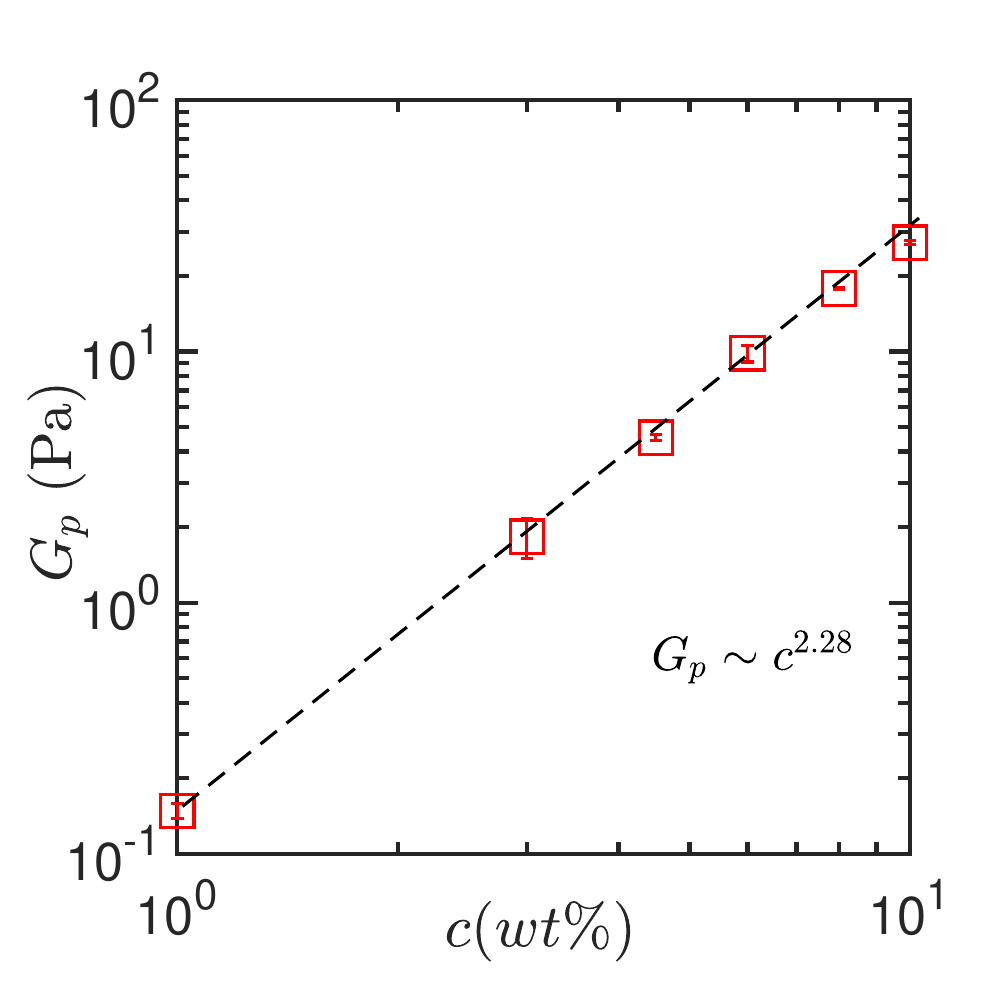}
	\end{center}
	\vspace*{-5mm}
	\caption{$G'$ averaged over \textcolor{black}{$\omega\in(0.02\pi-0.2\pi)$ rad/s} and reported as $G_p$ vs surfactant concentration $c$ at \textcolor{black}{25C}. Black line denotes power law fit (straight line in log-log scale) }
	\label{conc60}
\end{figure}

Finally, we show the concentration dependence of the plateau modulus, $G_p$ in \textcolor{black}{Fig.~\ref{conc60}}. $G_p$ is calculated by averaging the elastic modulus over a frequency range of \textcolor{black}{$\omega\in(0.02\pi-0.2\pi)$ rad/s} over which the moduli is independent of $\omega$. Fitting to a power law results in $G_p\sim c^\beta$, $\beta$ = 2.28. It is not straightforward to glean information from the value of $\beta$, but we note that it lies in the range of values reported for conventional micellar systems for which the plateau modulus, $G_p$ typically scales with concentration as $G_p\sim c^\beta$, $\beta\in(1.8-2.4)$ \cite{berret2006rheology}. 

For some systems exhibiting gel-like rheology the largeness of the exponent can be related to the relative flexibility of constituent fibres. For instance, collagen and fibrin hydrogels were shown to have exponents of 2.48 and 1.86 respectively; concurrently fibrin hydrogels had flexible chains compared to the stiffer elements forming collagen networks \cite{lee2019effect}. We note that physical networks of sterically hindered fibres like our VES systems have been shown to have an exponent of $\sim$ 2.2 : F-Actin made up of semi-flexible fibres with bending rigidity was predicted to show this exponent \cite{mackintosh1995} and recently the same concentration dependence was found in nanofibrillar networks \cite{poling2020yielding}. In \cite{poling2020yielding}, this exponent increased for higher concentrations and the regime in which $G'\sim c^{2.2}$ was primarily a `fluid' regime for which $\tan{\delta}=G''/G'>1$. For the VES however, this scaling was observed for a decade in concentration and $\tan{\delta}<1$ for all frequencies probed.

\begin{figure}[!h]
	\begin{center}
		\includegraphics*[width=85mm]{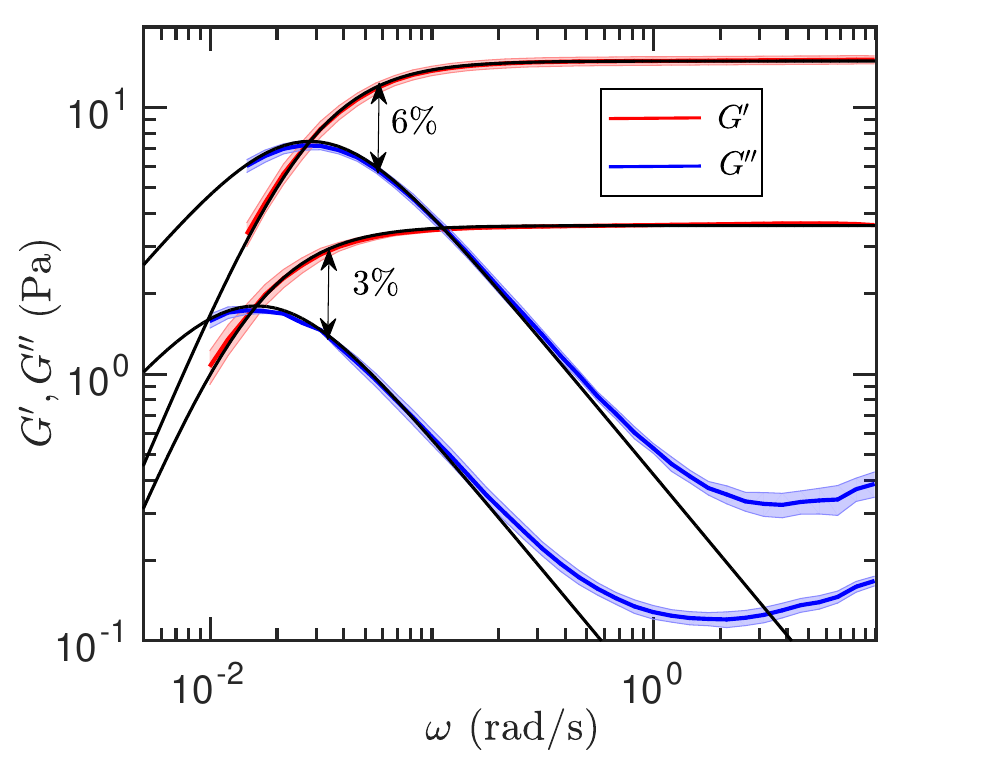}
	\end{center}
	\vspace*{-5mm}
	\caption{Frequency sweeps in the linear viscoelastic region for a 3\%\textit{v} and 6\%\textit{v} VES solutions at \textcolor{black}{60C}. Data plotted the same way as in Fig.~\ref{saos1}. Black  solid lines denote fits to Maxwell's linear viscoelastic model.}
	\label{saos3}
\end{figure}

We now move on to SAOS tests carried out at 60C. Representative plots for two concentrations are shown in Fig.~\ref{saos3} and we clearly see that VES rheology at this temperature is different from the gel-like rheology displayed in SAOS tests at 25C. Both, $G',G''$ are functions of $\omega$ and the dependence is well captured by the Maxwell-fluid model in which viscoelasticity is idealized by a series connection of spring (viscous) and dashpot (elastic) elements resulting in a viscoelastic fluid with a single relaxation time. The functional dependence of the moduli on frequency predicted by this model are :
\begin{equation}
\label{max1}
G'(\omega) = \frac{G_p \omega^2 \tau_r^2}{1 + \omega^2 \tau_r^2}
\end{equation}
\begin{equation}
\label{max2}
G''(\omega) = \frac{G_p \omega \tau_r}{1 + \textcolor{black}{\omega^2 \tau_r^2}}
\end{equation}
Here, $\tau_r$ is the single relaxation time which is the inverse of the frequency ($\omega_c$) at which $G'$ and $G''$ crossover, signalling a regime shift in the behaviour of the solution. $G_p$ is the value of plateau in the elastic modulus attained at higher frequencies. Both moduli show terminal behaviour at lower frequencies. These features are typical of entangled wormlike micellar solutions in the linear viscoelastic regime \cite{rehage1991viscoelastic}. 
The adherence to viscoelasticity defined by a single relaxation time is also a marker of the micelles in solution being in the fast breaking limit in which their dynamic breaking time, $\tau_{br}$ is orders of magnitudes smaller than the reptation time, $\tau_{rep}$ for micelles in an entangled network. Here, we note that the VES moduli at 25C did not show terminal behaviour at low frequencies. In particular, the absence of linear dependence of $G''$ with $\omega$, at the lower range of frequencies probed, points to very slow relaxation process more typical of soft glassy matter than wormlike micellar solutions. 

\begin{figure}[!h]
	\begin{center}
		\includegraphics*[width=65mm]{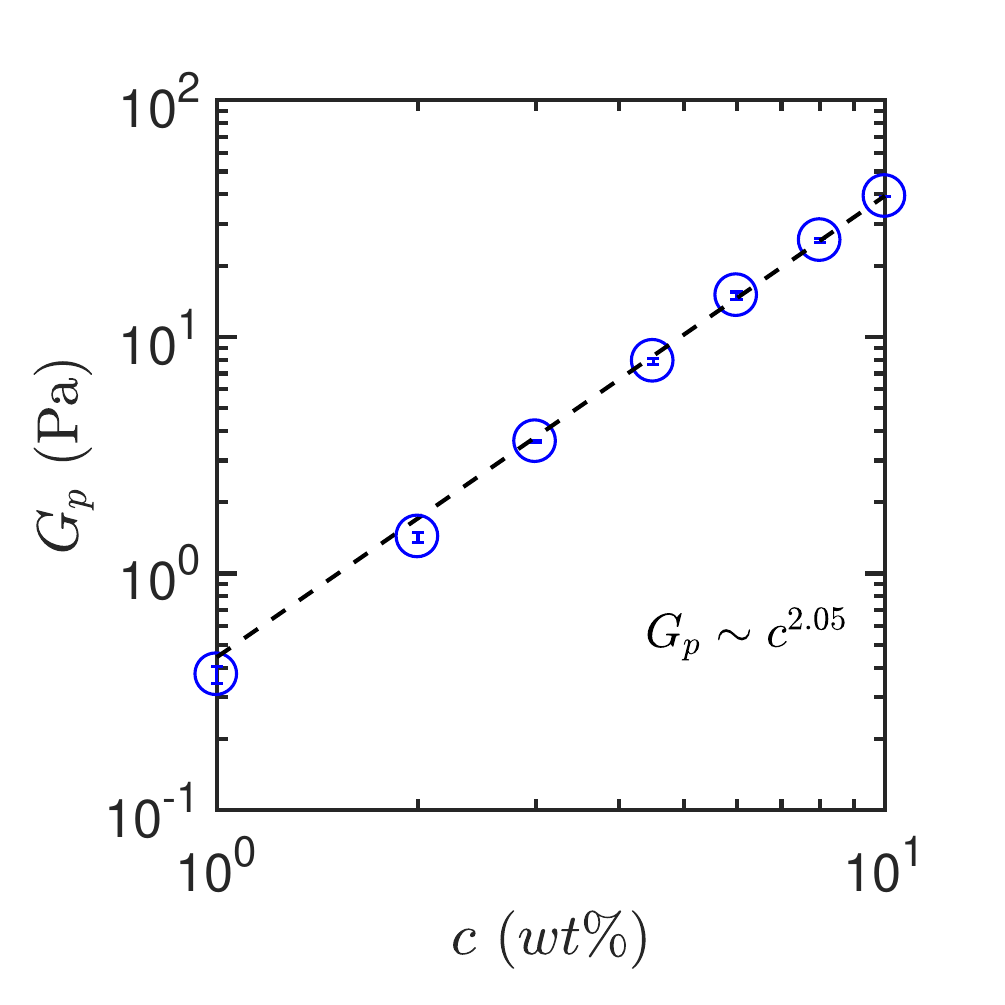}
	\end{center}
	\vspace*{-5mm}
	\caption{$G_p$ vs surfactant concentration $c$ at \textcolor{black}{60C} obtained from Maxwell fit. The black line denotes a power law fit}
	\label{saos2}
\end{figure}

A notable feature of SAOS behaviour of VES at 60C is the deviation from predictions of the Maxwell model at higher frequencies which occur due to the dynamic nature of micelles that can break and recombine. While collective network dynamics dominate in the plateau modulus regime, higher frequencies trigger intra-filament modes like bending and Rouse modes which are not accounted for in the Maxwell model \cite{doi1988theory}. Activation of such relaxation modes leads to an upturn of $G',G''$ at high values of $\omega$, an effect which cannot be fully captured by bulk rheometry owing to instrument inertia limitations and can be captured by microrheology probes \cite{buchanan2005high}.\comment{Deviations from Maxwell model are better visualized in a Cole-Cole plot seen in the inset of Fig.~\ref{saos3}, where we can clearly see deviations from a semi-circle expected from a Maxwell fluid.} We fit SAOS data at 60C to the Maxwell model (Eq.\ref{max1},\ref{max2}) and extract values of $G_p$ and $\tau_r$ for different surfactant concentrations, $c$. The data also allows us to quantify $\tau_{br}$ which is the breaking and reformation time of micelles in equilibrium. While $\tau_{br}$ can be calculated via different techniques \cite{granek1992stress}, we use the inverse of frequency ($\omega_{min}$) at which $G''$ attains a minimum as an estimate of $\tau_{br}$, $\tau_{br}=1/\omega_{min}$. $G_p$ follows a power law dependence on concentration, $G_p\sim c^{2.05}$ as seen in \textcolor{black}{Fig.~\ref{saos2}}, which we note is a weaker dependence than observed at 25C. The reason for this is unclear at this point.

\begin{figure}[!h]
	\begin{center}
		\includegraphics*[width=85mm]{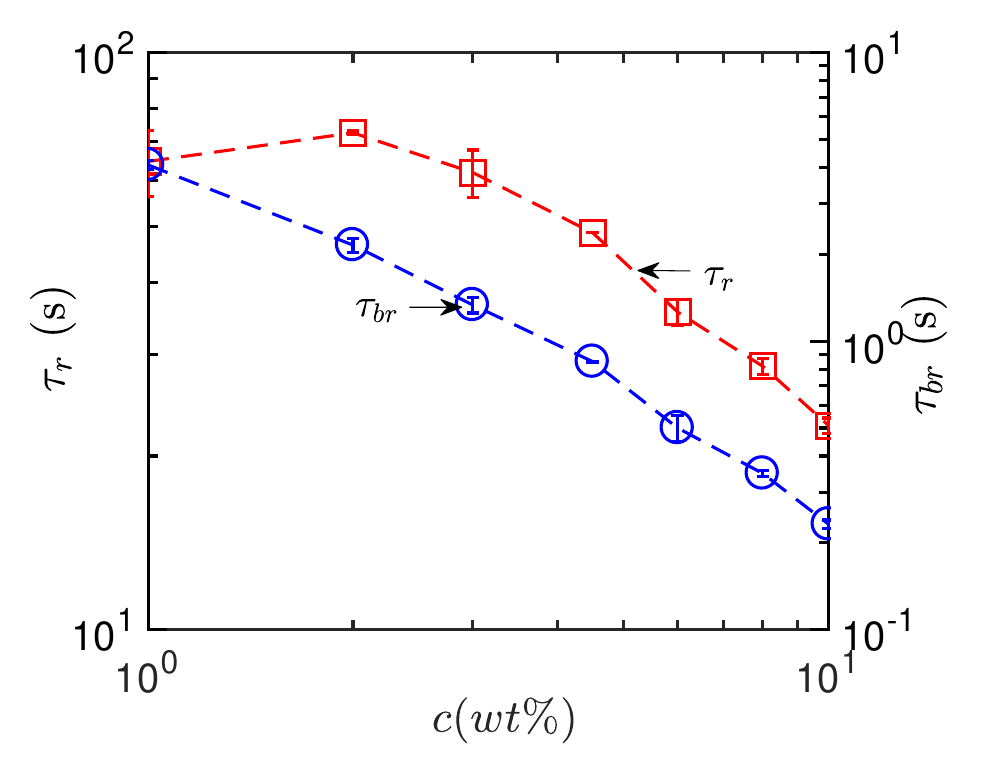}
	\end{center}
	\vspace*{-5mm}
	\caption{Breaking time ($\tau_{br}$) and terminal relaxation time ($\tau_r$) in seconds vs concentration $c[\%]$ at 60C}
	\label{times}
\end{figure}

Before we comment on the variation of $\tau_r,\tau_{br}$ with $c$ in our system, it is instructive to re-visit what theory predicts for these dependencies in wormlike micellar solutions. The two processes that control stress relaxation in these systems vary differently with micellar lengths and hence with $c$. If $\tau_{br}$ is taken to be the rate of unimolecular scission occurring randomly at any point along the length of a micellar chain with reaction rate $k$, $\tau_{br} \sim 1/k\bar{L}$ where $\bar{L}$ is the average length of a micellar chain \cite{cates1990statics}. A mean-field prediction for reptation time, $\tau_{rep}$ in unbreakable polymers is encapsulated in the relation $\tau_{rep} \sim c^2{\bar{L}}^3$ and the mean-field prediction for average chain length dependence on surfactant concentration is $\bar{L}\sim c^{1/2}$ \cite{cates1988dynamics}. For $\tau_{br}\ll\tau_{rep}$ the single relaxation time, $\tau_r$ is given by $\tau_{r}=\sqrt{\tau_{rep}\tau_{br}}$ \cite{cates1987reptation}. Combining the above mentioned equations gives $\tau_r\sim c^{3/2}$ and $\tau_{br}\sim c^{-1/2}$. In Fig.\ref{times} we plot $\tau_{br},\tau_{r}$ for different concentrations at 60C and see that while $\tau_{br}$ expectedly decreases with increasing $c$, it is a much sharper decrease than the $1/2$ power law exponent predicted by scaling theory. Force fitting the data in Fig.~\ref{times} gives $\tau_{br}\sim c^{-1.25}$. For $\tau_r$, the trend seen in experiments is opposite to that predicted by scaling theory - $\tau_r$ decreases with $c$ with a possible maximum between $1-2\%$. 

The scaling theory is developed for neutral micelles, and it is thus not exceptional that many surfactant systems which might be strongly influenced by added counterions, insufficient screening or structural transitions, do not follow the expected scaling \cite{kern1994dynamic,soltero1996rheology,candau2001linear}. However, zwitterionic surfactants are self-screening and are thus expected to be close approximations of neutral micelles in solution. Hence, such a large deviation from theory is surprising. Studies on zwitterionic systems are relatively scarce. \textcolor{black}{While the concentration scaling of shear viscosity has received attention \cite{chu2010wormlike,zhang2013single,wang2017wormlike}}, to the best of our knowledge, there aren't any reports of concentration dependence of \textcolor{black}{$\tau_r,\tau_{br}$} for long-chained zwitterionic surfactants. However, we extracted data from Figure 3,4 in  \cite{kumar2007wormlike} for EDAB and found that $\tau_r$ is a decreasing function of $c$ at 60C - a trend in agreement with what is observed for the VES (although the dependence in  \cite{kumar2007wormlike} is much weaker than in Fig.~\ref{times}).

This discrepancy in $\tau_r$ vs $c$ data has been ascribed to the presence of micellar branching in systems where the trend followed by $\tau_r$ is mirrored by shear viscosity $\eta = G_p\tau_r$ \cite{candau2001linear}. Here, $\eta$ is an increasing function of $c$ throughout the investigated range (not shown) and the strength of the variation is close to that deduced from data in \cite{kumar2007wormlike}. While we haven't directly imaged VES at 60C, it is important to note that no sign of branching was observed in Cryo-TEM images at room temperature. Finally, the dependence of plateau modulus, $G_p$ on $c$ in branched systems is usually weaker than what is observed for VES \cite{chu2010wormlike}. A possible reason for the anomalous variation of $\tau_r$ with $c$ in our VES system could be a deviation in $\bar L$'s dependence on $c$ from the $\bar{L}\sim c^{0.5}$ growth law predicted by theory. Both $\tau_{br}$ and $\tau_{r}$ are functions of $\bar L$ and strongly influenced by how micellar length depends on surfactant concentration. This deviation could be induced owing to subtle dependence on end-cap energy $E_c$ on $c$ \cite{kern1994dynamic}.  $E_c$ sets the favourability of creating two end-caps from none and hence critically decides the length of cylindrical wormlike micelles in solution. We do not investigate this situation further. However, our results call for more experimental and theoretical work on self-assembly in zwitterionic micelles and related systems \cite{kumar2007wormlike,sarmiento2010microrheology}. \textcolor{black}{We conclude this discussion on scaling of relevant parameters with surfactant concentration by noting that while the reported scalings for $\tau_{r}$ and $\tau_{br}$ differ from that predicted by mean-field theory, the plateau modulus scaling is close to the mean-field theory prediction \cite{khatory1993linear} and values reported for some smaller chain surfactants (for example in \cite{kern1992rheological,berret1993linear}). We also note that the reported scaling for $\tau_{br}$ is close to the newly uncovered value of -1.1 reported in \cite{tan2021determining} by conducting mesoscopic simulations with the inclusion of fast Rouse modes}. 

\comment{We also note that at 60C, $\eta$ for 4.5\% VES is $~10^6$ that of water that that temperature. This requires very large micellar lengths. Further, we see that the reptation time extracted from $\tau_r = \sqrt{\tau_{rep}\tau_{br}} \sim 2600s$. The micellar length for this $\tau_{rep}$ calculated from formulae provided in \cite{tan2021determining} for reptation of cylindrical rods, is of the same order of magnitude as the $\bar L$ we calculate here.}

We can extract more information about the microstructure of VES from SAOS tests at 60C. The pore size or mesh spacing $\xi$ is related to the plateau modulus as \textcolor{black}{$G_p = 9.75\frac{k_bT}{\xi^3}$ \cite{zou2014mesoscopic,tan2021determining}}. Plugging in the values of $G_p$ for $c=4.5\%$ we get \textcolor{black}{$\xi\approx179$ nm}. The value of $\xi$ is typically independent of temperature and we expect that these mesh gaps characterize the entangled micellar network throughout this high temperature range. To get an idea of other length scales that describe the entangled wormlike microstructure, we require an estimate of persistence length $l_p$. Persistence length is a difficult quantity to measure and and is performed either by sophisticated scattering techniques \cite{mccoy2016structural} or high-frequency rheology \cite{willenbacher2007broad}. To the best of our knowledge, the only $l_p$ measurement for a zwitterionic long-chained surfactant system was carried out in \cite{mccoy2016structural}. For a $2\%$ EAPB system, the value reported was $\approx$150 nm, a value much higher than $l_p$ for conventional micellar systems which are influenced by salt content and lies between 30-60 nm. Assuming $l_p$ is unaffected by temperature, we use this value as an estimate for $l_p$ for $4.5\%$ VES at 60C. For $\xi$= 83.6 nm and $\xi = l_p^{2/5}l_e^{3/5}$ \cite{doi1988theory} we get \textcolor{black}{$l_e\approx$ 200  nm} : $l_e$ is the length of micellar sections between points of entanglements. To get $\bar{L}$ we use the formula - \textcolor{black}{$G_p/G''_{min}\approx 0.317 (\bar{L}/l_e)^{0.82}$ \cite{tan2021determining}}. \comment{The inverse of this quantity is also called the entanglement number $Z$, because it denotes the average number of points of entanglement across the length of a chain. }The values of $G''_{min}/G_p$ at 60C decrease with $c$ and for 4.5\% VES solution is $\approx$ 0.024 which gives approximately \textcolor{black}{188} entanglements points per chain length and a value of \textcolor{black}{$\bar{L}\approx$ 33$\mu$m}. While, this value of $\bar L$ seems exceedingly large for micelles at 60C, we note that it lies close to the range reported for micelles formed by cationic long chained surfactants \cite{raghavan2001highly} and calculated for zwitterionic long-chained surfactants - EDAB \cite{kumar2007wormlike}, provided the updated formulae from \cite{tan2021determining} are employed. Long wormlike micelles imply very long $\tau_{rep}$, which satisfies one of the conditions that allow for the gel-like rheology at room temperatures, the other being large values of $\tau_{br}$.

\begin{figure}[!h]
	\begin{center}
		\includegraphics*[width=73mm]{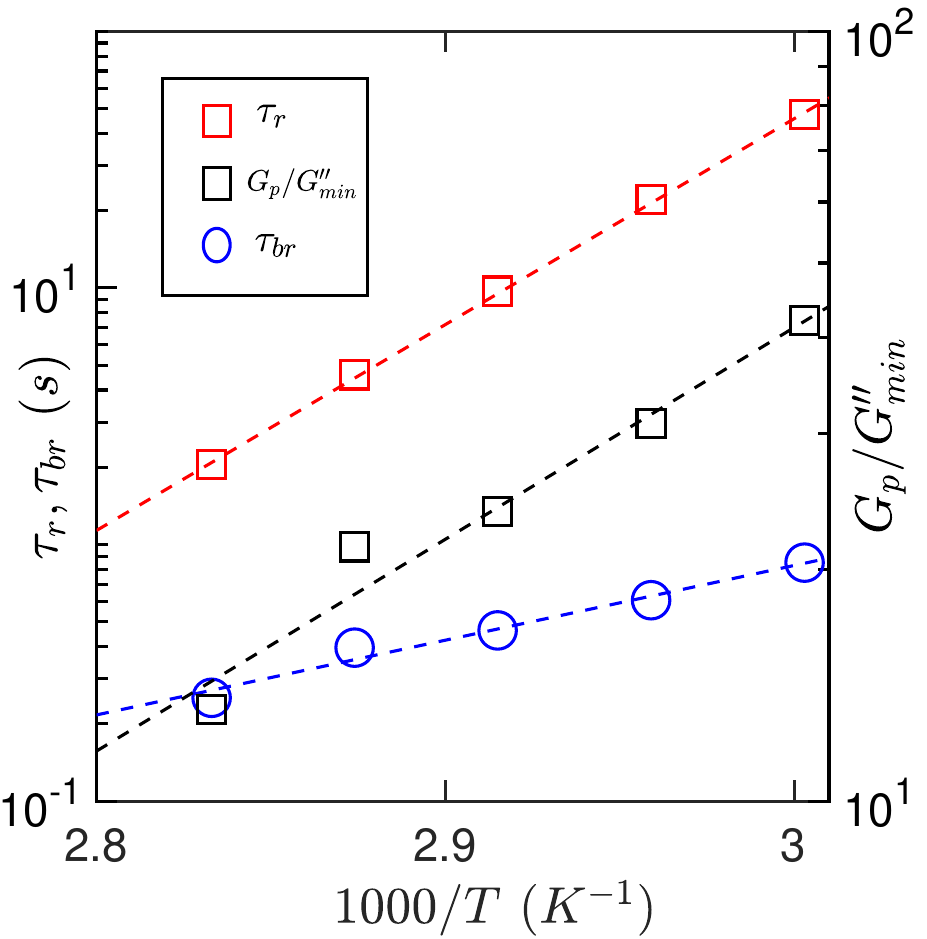}
	\end{center}
	\vspace*{-5mm}
	\caption{\textcolor{black}{$\tau_{br}$, $\tau_r$ and $G_p/G''_{min}$ vs 1000/T (K) plotted in semilog scale} for 4.5\% VES. All dashed lines are linear fits.}
	\label{arren}
\end{figure}
\begin{figure}[!h]
	\begin{center}
		\includegraphics*[width=85mm]{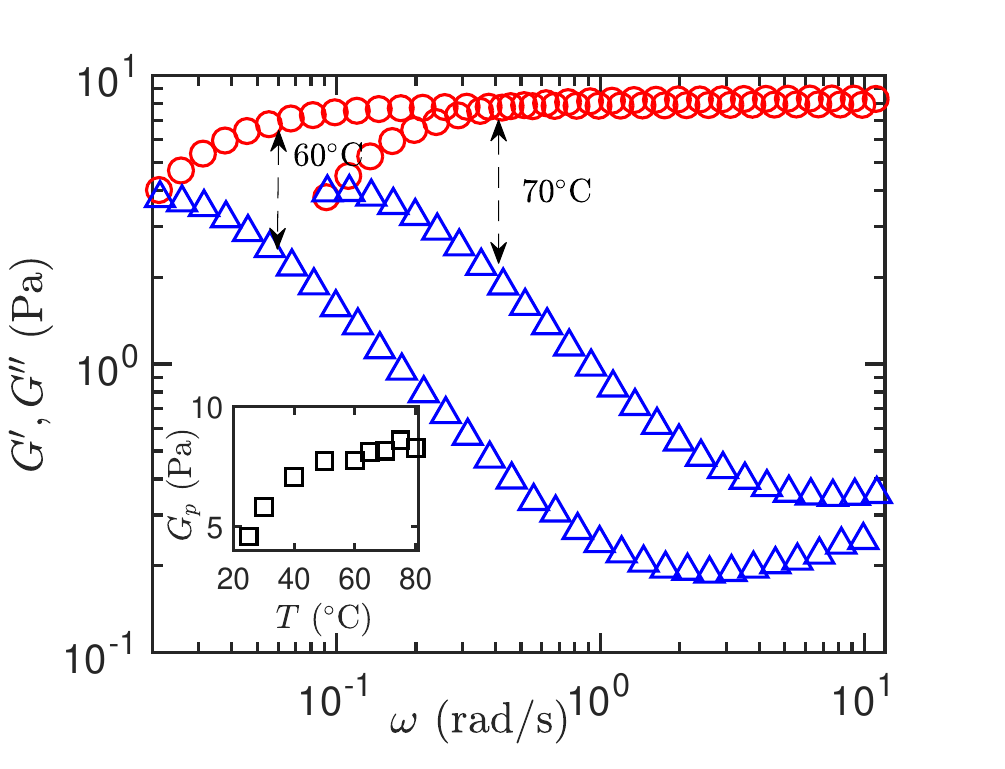}
		\end{center}
	\vspace*{-5mm}
	\caption{Frequency sweeps in the linear viscoelastic regime for 4.5\% VES at two temperatures. Red symbols: $G'$, Blue symbols: $G''$. Inset: Plateau modulus $G_p$ vs $T$($^{\circ}$C) for VES solution.}
	\label{high_temp}
\end{figure}

 
We carry out SAOS tests at temperatures between 60 and 80C for 4.5\% VES. Slight changes in $G_p$ can be accounted for by plateau moduli's linear dependence on $T$. This validates our assumption of temperature independence of $\xi,l_e,l_p$ at these temperatures. We also find that $\tau_r,\tau_{br},G_p/G''_{min}$ can be fit by a straight line in a semi-log plot, indicating an Arrenhius type variation for these quantities as seen in Fig.~\ref{arren}. From the Arrenhius variation $\tau_r\sim e^{\frac{E_r}{k_b T}}$, we obtain the flow activation energy, $E_r$. \comment{which is basically the amount of energy needed to move a single chain in its surrounding environment.}Our data gives $E_r \approx 61k_bT$ or 152 kJ/mol, close to the value obtained for EDAB in \cite{kumar2007wormlike,wang2017wormlike}. The decrease of relaxation time with increasing $T$ is mediated via (1) - drastic reduction in reptation time due to exponentially decreasing mean micellar length and (2) - decreasing breaking time with $T$ as seen in Fig.\ref{arren}. As $G''_{min}/G_p\approx l_e/\bar{L}$ and $l_e$ is constant in this range of $T$, a reduction in $\bar L$ implies an increase in the value $G''_{min}$; a decrease in $\tau_{br}$ with increasing $T$ also pushes the minimum in $G''$ to higher frequencies as seen in Fig.~\ref{high_temp}.
The reductions in both $\bar L$ and $\tau_{br}$ are mediated by activation energies : $\bar L\sim e^{\frac{E_{sciss}}{2k_b T}}$ and $\tau_{br}\sim e^{\frac{E_{b}}{k_b T}}$ , where $E_{b}$ is the activation energy required for breaking a micelle and $E_{sciss}$ is the scission energy. For nonionic micelles, $E_{sciss}\sim E_c$ so we expect that for zwitterionic surfactants, $E_{sciss}$ is very close to the end-cap energy. The Arrenhius fits for $G_p/G''_{min}$ and $\tau_{br}$ give values \textcolor{black}{$E_c \approx 51.2 k_bT$} and $E_b\approx 22 k_bT$. To the best of our knowledge, these are the first values reported for long-chained zwitterionic surfactant systems. They compare favourably to those extracted from Fig.3 in \cite{kumar2007wormlike} for EDAB : \textcolor{black}{$E_c\approx 53.6 k_bT$ and $E_b\approx 28 k_b T$}. However, they are much lower that those in \cite{raghavan2001highly}, where the authors reported (note that authors in \cite{raghavan2001highly} have used $G_p/G''_{min}\sim(\bar{L}/l_e)$ instead of $G_p/G''_{min}\sim(\bar{L}/l_e)^{0.82}$) an unusually high value ($65 k_bT$) of $E_{sciss}$ for long-chained cationic surfactant - EHAC. This was interpreted in the context of the larger lengths of micelles formed by long-chained surfactants, and later aiding the hyopthesis for long breaking times of C22-tailed surfactants \cite{kumar2007wormlike}. We expected the VES and EDAB, both characterized by long carbon chains, to have activation energies similar to that reported in \cite{raghavan2001highly}. Clearly, this is not the case. How do we rationalize this discrepancy in scission energies? 

The formula $\bar L\sim e^{\frac{E_{sciss}}{2k_b T}}$ from Cates theory does not account for entropic contributions to $E_{sciss}$ \cite{cates1990statics} and it was suggested that the values of activation energy calculated from Arrenhius fits only yield the \textit{enthalpy} of scission, $H_{sciss}$, rewriting $E_{sciss}=H_{sciss} - TS_{sciss}$, where $S_{sciss}$ is the entropy of scission \cite{couillet2004growth,jiang2018enthalpy}. The system employed in \cite{raghavan2001highly} consists of a low ionic strength surfactant in conjunction with Salycitate counterions from the salt used. This is likely to lead to the very high values of $H_{sciss}$ and a positive $S_{sciss}$ can then lead to lower values of $E_{sciss}$ \cite{couillet2004growth}, closer to the ones we report here for the VES and EDAB in \cite{kumar2007wormlike}. We note that zwitterionic surfactants should mimic well screened systems and as such, values calculated here and from \cite{kumar2007wormlike} should closely relate to those reported in \cite{couillet2004growth} where the authors employed a highly screened surfactant system. This is not the case and it indicates that it is likely that $H_{sciss}\not\approx E_{sciss}$ for VES and EDAB, unlike the system employed in \cite{couillet2004growth}. It remains to be established if positive entropic contributions can further reduce enthalpies to even lower values.\comment{ This is indeed the case and such lower values of $H_{sciss}$ point towards diminishing effects of the entropic term and thus $H_{sciss}\sim E_{sciss}$ in such systems. Our results seem to indicate that long breaking times need not go hand in hand with high values of end-cap energies as suggested in \cite{kumar2007wormlike}} To better answer the hypothesis of very large breaking times, pin-point its origin, and connection to scission energies, measurements of $\tau_{br}$ at lower temperatures along with energetic and entropic calculations of micellar formation and breaking process are needed. These remain out of the scope of the present study.

Finally, we see that the value of $G_p$ changes with $T$ in an interesting fashion as seen in the inset of Fig.~\ref{high_temp}. From 25-60$^{\circ}$C there is a steeper increase followed by small increments post 60$^{\circ}$C which, as mentioned previously, can be accounted for by the $G_p$'s linear dependence of $T$ in that regime. The steeper increase is hypothesized to be caused by a temperature driven increase in effective surfactant concentration which results from solubilization of molecules at higher $T$ \cite{raghavan2001highly,wang2017wormlike}. Thus the \textit{gel-sol} transition in these systems in driven by an increase in temperature and plateau modulus. \textcolor{black}{Two dominating factors govern this transition induced by temperature - (1): Reduction in micellar mean length that leads to drastic reduction in reptation times, (2): Reduction of breaking times. Both these factors push the WLM system to the fast breaking limit  where they show a Maxwellian viscoelasticity \cite{raghavan2012conundrum,raghavan2017wormlike}.} Before we move on to discuss the shear rheology of wormlike micellar gels, we briefly discuss results from \textcolor{black}{strain}-amplitude sweeps.

\subsection{Behaviour in Strain Amplitude Sweep }
\begin{figure}[!h]
	\begin{center}
		\includegraphics*[width=85mm]{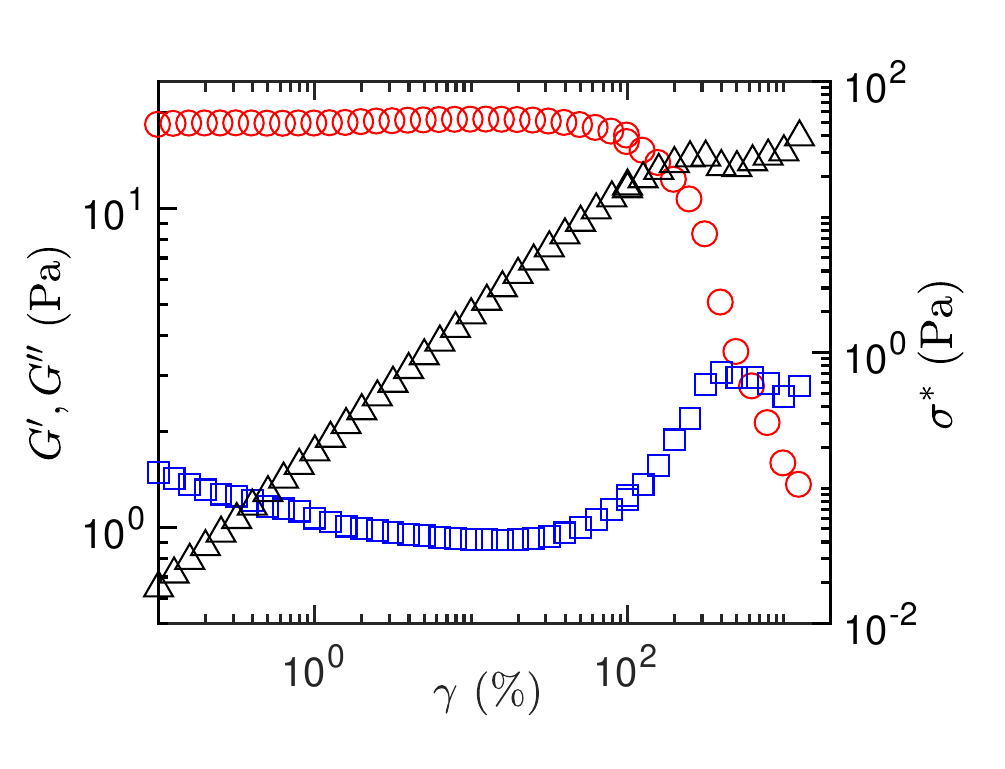}
	\end{center}
	\vspace*{-7mm}
	\caption{Strain controlled amplitude sweep for 8\% VES at \textcolor{black}{2$\pi$ rad/s} frequency at \textcolor{black}{25C}. Red circles: $G'$, Blue squares: $G''$, Black triangle: $\sigma^*$}
	\label{laos}
\end{figure}

The issue of strain-stiffening has been a subject of intense discussion, mostly in the context of biopolymer solutions and gels that are generally characterized by permanent networks comprising of semi-flexible fibres \cite{storm2005nonlinear}. In a strain-amplitude sweep test, strain-stiffening manifests as an increase in storage modulus ($G'$) with increasing strain amplitude ($\gamma\%$). While it is not clear what the exact mechanism behind strain-stiffening is, it is generally ascribed to the rigidity of the network filaments. We want to investigate if micellar gels formed by the VES show strain-stiffening. This line of inquiry is inspired by findings in \cite{tung2008strain} where it was found that even \textit{transient} networks of entangled reverse micelles can also show this phenomenon, and by \cite{tung2008self} in which surfactant based organogels showed strain stiffening. These organogel's linear rheology showed frequency independent moduli and their plateau modulus depended on concentration as - $G_p\sim c^{2.1}$, both features similar to that shown by our wormlike micellar gels. It was noted that these organogels also owe their gel-like characteristics to an entangled network of filaments, much like aqueous solutions of F-Actin \cite{tung2008self}. Again, this is similar to the genesis of gel-like rheology hypothesized for solutions of EDAB \cite{raghavan2012conundrum} and the VES employed in this study. However, as seen in Fig.\ref{laos}, wormlike micellar gels formed by VES do not display strain-stiffening. We note that beyond the linear regime i.e $\gamma\%\gtrsim$ 50, it becomes extremely difficult to obtain steady alternating states, presumably because of the elasticity of the gels. Data in Fig.\ref{laos} suggests that VES strain-softens or thins, which is typical for polymeric networks. 

In such materials, $G'$ reduces with increasing $\gamma\%$, because large deformations weaken networks that lend rigidity. Micellar gels are marked by topological interactions and its possible that large strains weaken \textit{entanglements}, leading to the observed strain-softening response. To understand the strain-stiffening response of transient network of reverse micelles, in \cite{tung2008strain} the authors proposed that an oscillatory shear of large amplitude can add loose micelles to the network, which can then contribute in increasing stiffness. Such a shear can also remove micelles via disentanglement from the network, an effect that becomes weaker with decreasing flexibility of micelles. A balance of these two opposing effects dictates if entangled micellar networks stiffen or soften under increasing deformation amplitude. It is likely that our wormlike micellar gels do not maintain this balance as is also evident from the lack of shear-thickening in imposed shear-rate tests. Note, shear-thickening was an accompanying response observed in strain-stiffening reverse micelle networks. If this mechanism for strain-stiffening proposed by \cite{tung2008strain} holds, it should, in principle, be possible to engineer micellar gels with strain-stiffening properties by precisely tuning polydispersity and persistence length.

\begin{figure}[!h]
	\begin{center}
		\includegraphics*[width=85mm]{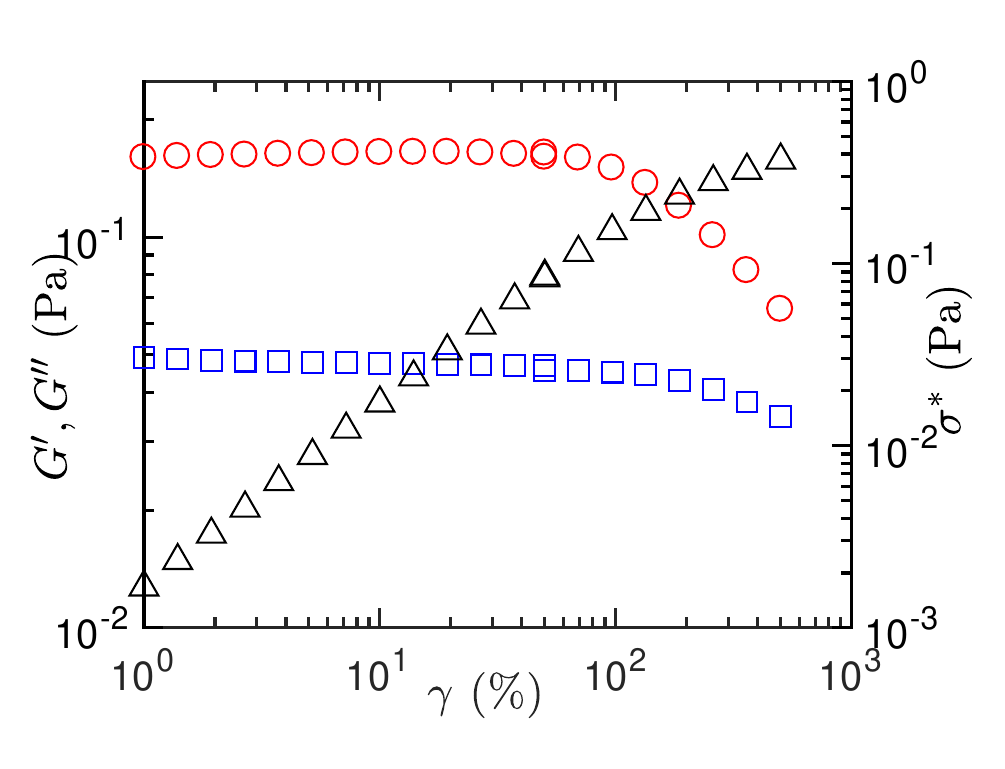}
	\end{center}
	\vspace*{-7mm}
	\caption{Strain controlled amplitude sweep for 1\% VES at \textcolor{black}{$\pi$ rad/s} frequency at \textcolor{black}{25C}. Red circles: $G'$, Blue squares: $G''$, Black triangle: $\sigma^*$}
	\label{laos2}
\end{figure} 
 Another detail is the increase in $G''$ with $\gamma\%$ in the nonlinear regime for the 8\% VES. Before we discuss this further, it is imperative to mention that data in the nonlinear regime is not drawn from steady state responses and as such the values of $G''$ at amplitudes in that regime might change if we wait for long enough time to achieve steadiness. However, we expect that the presence of an overshoot should not be affected. The overshoot in loss modulus is a feature that is observed in many different soft materials (See \cite{donley2020elucidating} and references contained therein) and thus the mechanism for it should be structure agnostic \cite{donley2020elucidating}. One such mechanism proposed that could apply to our micellar gels, is that as strain amplitude is increased, the nature of strain acquired by the network changes from being recoverable to unrecoverable. Note that the overshoot (usually called Type III) response is different from what is expected from conventional viscoelastic worm-like micellar solutions which show no overshoot (type I) response. We find that VES revert back to type-1 response at a concentrations of 1\% as seen in Fig.~\ref{laos2}. Thus, the type-III response shown by VES at high concentration is akin to the  yielding displayed by soft gels and viscoelastic solids, whereas at lower concentrations VES's behaviour in strain-amplitude tests is similar to viscoelastic liquids. 

\subsection{Shear Rheology}
\begin{figure}[!h]
	\begin{center}
		\includegraphics*[width=75mm]{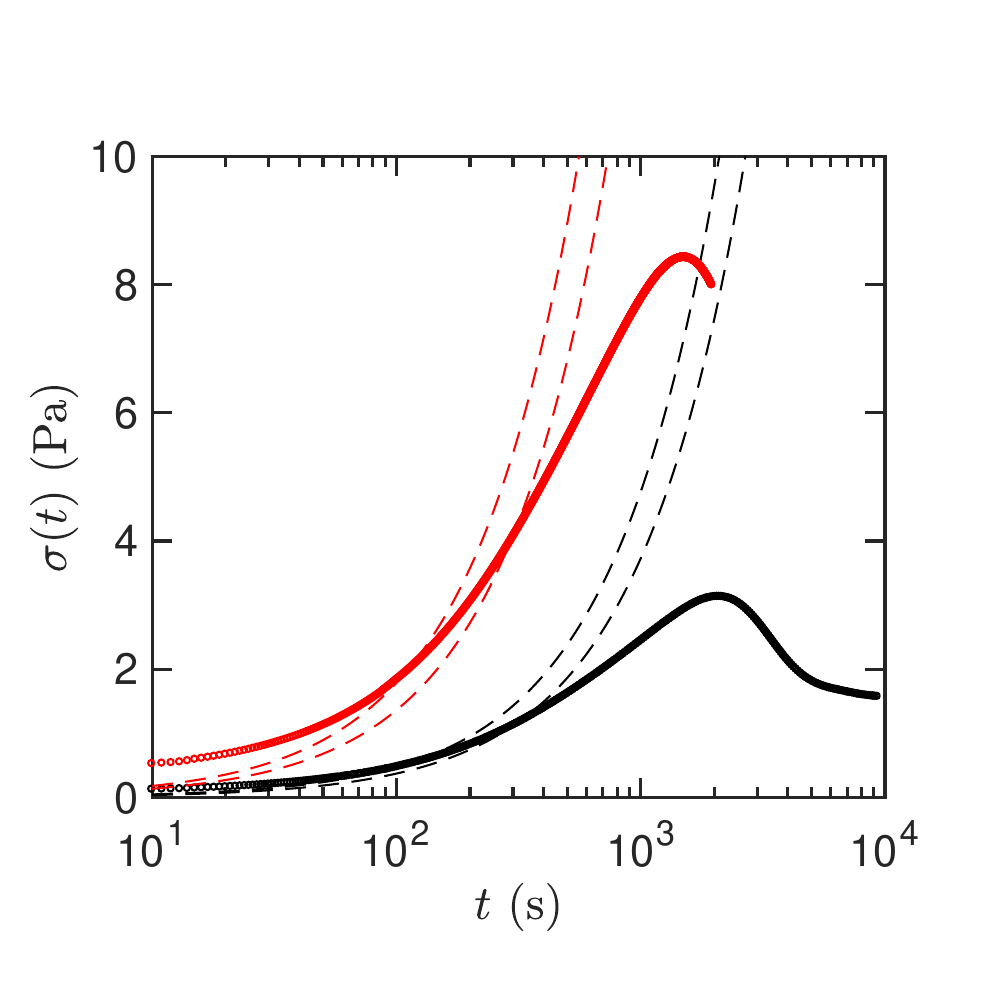}
		\end{center}
		\vspace*{-7mm}
	\caption{Shear-startup results for $\dot{\gamma} = 10^{-3} s^{-1}$ for red: 8\% VES, black: 4.5\% at \textcolor{black}{25C}. Dashed lines are for $\sigma = G'\dot{\gamma} t$. Further description in text.}
	\label{ss1}
\end{figure}

After probing the rheological behaviour of wormlike micellar gels in some detail using oscillation-based tests, we now move on to discussing the shear rheology of these materials. We first report behaviour in shear-startup tests. For this test we follow the same pre-shear+rest protocol outlined in the methods section except that during the rest period, we monitor the evolution of $G',G''$, by imposing oscillations at 1\% strain and \textcolor{black}{$\pi$ rad/s} frequency. After the end of this protocol (i.e at $t$=0), we impose a constant shear-rate, $\dot{\gamma}$ and record the evolution of the shear-stress signal, $\sigma(t)$. Because we use a stress-controlled rheometer to impose a constant $\dot{\gamma}$, there is a delay between when the rate imposed by the rheometer, $\dot{\gamma}_{imp}$ matches the desired shear-rate, $\dot{\gamma}_{des}$. Thus, we report $\sigma(t)$ from $t=10s$ onwards, after which $\dot{\gamma}_{imp}$ is much closer to $\dot{\gamma}_{des}$. The two solid curves seen in Fig.~\ref{ss1} show a characteristic \textit{stress overshoot} phenomenon - a growth of stress upto a maximum followed by a decrease of $\sigma$ towards an eventual steady state. Note that in the data presented in Fig.~\ref{ss1}, $\sigma$ has not reached steady state and it typically takes a very long time to attain it at such low values of $\dot{\gamma}$. The thin lines in Fig.~\ref{ss1} are stress curves for an elastic response i.e a linear growth $\sigma(t) = G'\dot{\gamma} t = G\gamma$. The two dashed lines for each VES concentration differ in the values of $G'$, the lower one extracted from initial time experimental data and the higher one has $G'=G'(t=0)$ taken from the end of the rest stage. \textcolor{black}{The mismatch between the experimental curve and dashed lines (especially for 8\% VES) in Fig.~\ref{ss1} possibly results from residual stresses post the rest stage that is imposed before shear}. An interesting feature of the solid curves in Fig.~\ref{ss1} is their deviation from an elastic driven linear growth after $t\gtrsim$ 500s i.e $\gamma \gtrsim$ 0.5 or 50\% ( (200s, 20\%) if we consider the higher dashed line), after which the value of $\sigma(t)$ is \textit{less} than that predicted by a purely elastic response.

In an elastic response, strain is accumulated in recoverable fashion. The deviation that preempts eventual "failure" in the material is thus characteristic of the material accumulating strain which is \textit{irrecoverable} in nature. This harkens back nicely to results discussed in the previous section where we showed that for 8\% VES (Fig.\ref{laos}), there is an overshoot in $G''$ with increasing strain amplitude and noted that a possible reason for this overshoot in the loss modulus is acquisition of irrecoverable strain \cite{donley2020elucidating}.  Note that in Fig.\ref{laos}, $G''$ starts increasing with strain amplitude at an approx. $\gamma$\% which correlates well with the $\gamma = \dot\gamma t$ at which the solid red line in Fig.~\ref{ss1} begins to lag behind the elastic response, signalling onset of irrecoverable strain acquisition. 
\begin{figure}[!h]
	\begin{center}
		\includegraphics*[width=75mm]{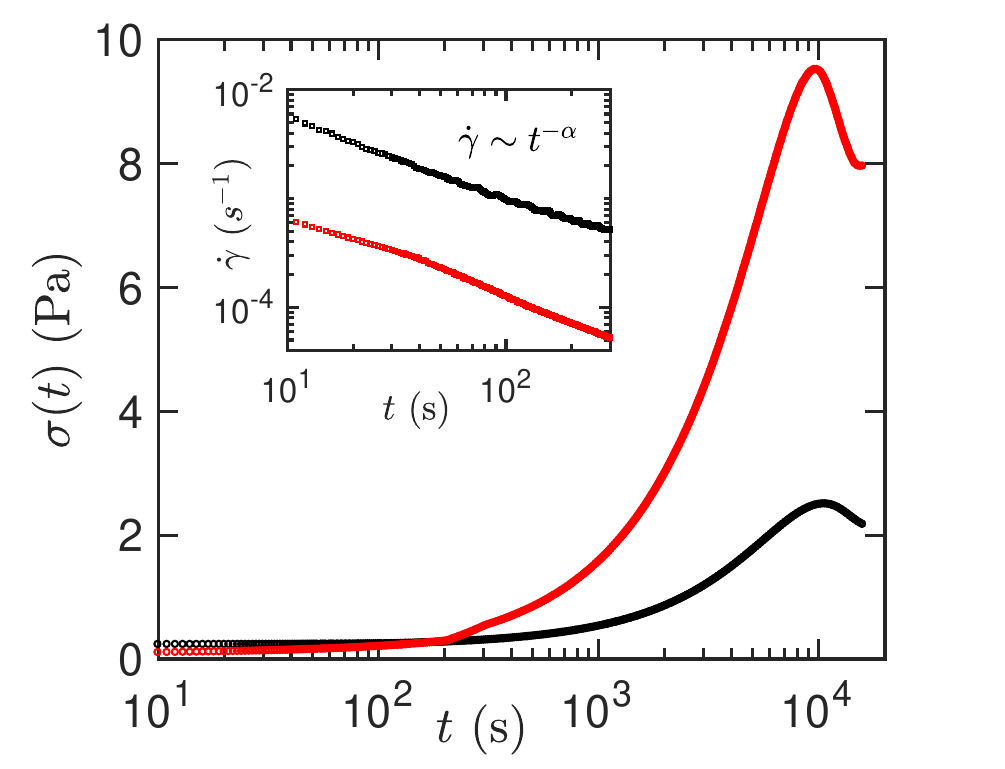}
         \end{center}
	\vspace*{-5mm}
	\caption{Shear-startup results for $\dot{\gamma} = 10^{-4} s^{-1}$ for red: 8\% VES, black: 4.5\% at \textcolor{black}{25C}. Inset shows a typical creep response to an imposed shear-stress of 2Pa by plotting the shear-rate recorded by the rheometer vs time for :  8\% (red) and 4.5\% (black) VES at \textcolor{black}{25C}. }
	\label{ss2}
\end{figure} 
The start-up shear response in Fig.~\ref{ss1} is similar to the response of carbopol microgel reported in \cite{divoux2011stress} where the authors interpreted deviation from an elastic response as the accumulation of \textit{plastic} strain that eventually leads to an elasto-plastic type failure. This conclusion was backed by velocimetry data. Additionally, there is evidence of their carbopol microgel undergoing a primary creep response marked by $\dot{\gamma}\sim t^{-\alpha}$ - a power-law scaling called Andrade creep \cite{andrade1910viscous} which often indicates plasticity in the system \cite{miguel2002dislocation,divoux2011stress}. Our wormlike micellar gels also show such a power-law creep scaling under the influence of an imposed shear-stress (Seen inset of Fig.~\ref{ss2}). We must note that, in general, an irrecoverable strain can be acquired by not only plastic events but also viscous flow within the material network. Indeed, a power-law creep was associated to the latter process in the case of Casein based protein-gels \cite{leocmach2014creep}. Details of power-law creep, its possible genesis and related flow will be discussed in a forthcoming publication. 

Next, we carry out start-up shear tests at higher shear-rates ($\dot{\gamma}>0.01s^{-1}$). From Fig.~\ref{ss3} below, we see that the response in this regime is markedly different from that discussed above for very low $\dot{\gamma}$. Here, a very short-time initial response (low deformation) is elastic and is followed by a the micellar network hardening under shear. it is unclear at this point why the effect of such strain-hardening was not captured in tests discussed in the previous section. Under the deformation imposed in this regime, $\sigma$ keeps increasing upto a peak value, $\sigma_c$ and then \textit{suddenly} drops at a value of strain $\gamma\% = \gamma_c\%$ as the network fractures.

\begin{figure}[!h]
	\begin{center}
		\includegraphics*[width=80mm]{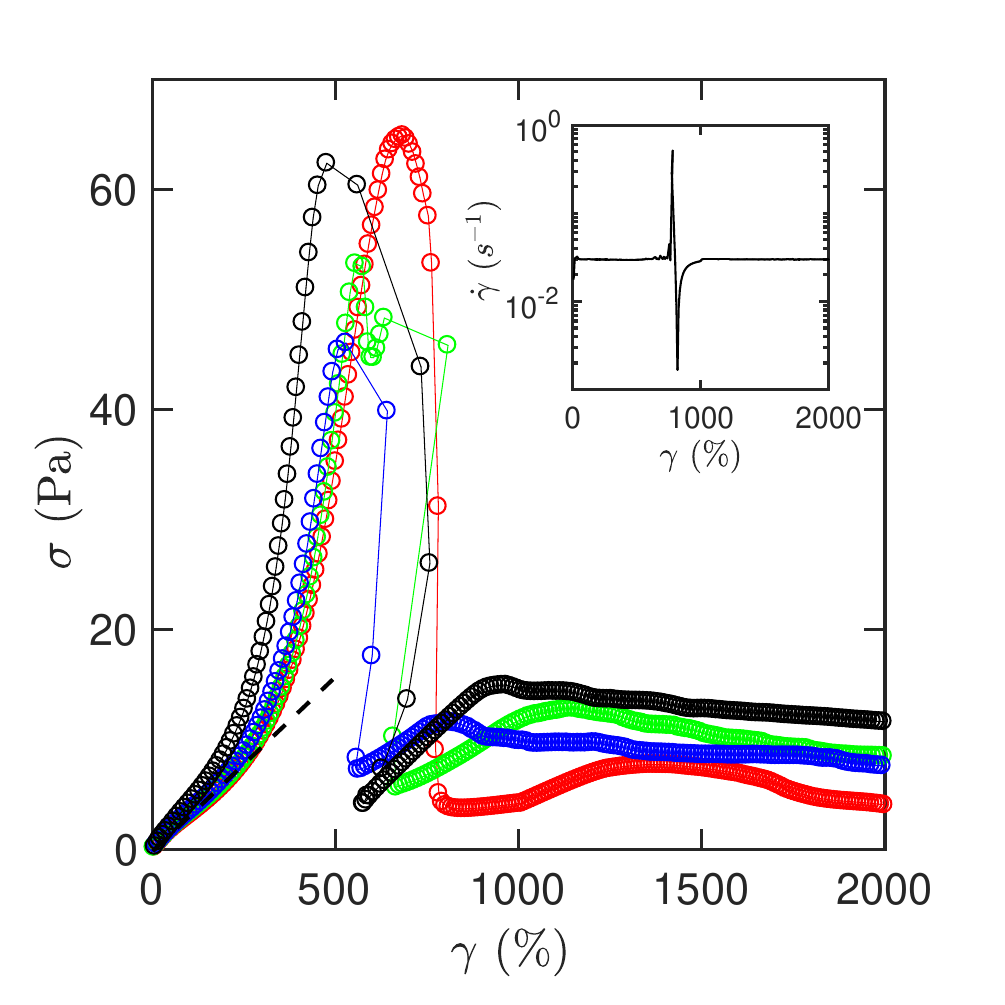}
	\end{center}
	\vspace*{-5mm}
	\caption{Shear start-up results plotted as $\sigma$ vs shear- strain, $\gamma$\% for 4.5\% VES solution at different values of $\dot{\gamma}$ at \textcolor{black}{25C}: Red - 0.03$s^{-1}$, Green - 0.05$s^{-1}$, Blue - 0.1$s^{-1}$, Black - 0.3$s^{-1}$. Dashed black line is obtained by extending the early time elastic response to guide the eye (No fitting). Inset : $\dot{\gamma}_{imp}$ vs. $\gamma$\% for $\dot{\gamma}_{des}$ = 0.03$s^{-1}$ }
	\label{ss3}
\end{figure} 
At, $\gamma\% = \gamma_c\%$, the rotor bob rapidly accelerates due the fracture and this is evident in the inset of Fig.~\ref{ss3}, in which for the specific case shown, $\dot{\gamma}_{imp}$ rapidly increases by an order of magnitude around $\gamma_c\%$. After this, the rheometer adjusts to the desired shear-rate, $\dot{\gamma}_{des}$ = 0.03. An interesting aspect ascertained from Fig.~\ref{ss3} is that both $\sigma_c$ and $\gamma_c\%$ do not seem to follow any trend with respect to the shear-rate employed to induce deformation. Rather, for one decade in $\dot{\gamma}$, we see that $\gamma_c\in(5-6.5)$ and $\sigma_c\in(45-65)$Pa. For experiments carried out with 8$\%$ VES, $\gamma_c\in(4-5)$ and $\sigma_c\in(100-135)$Pa. We note that this spread in $\gamma_c$ and $\sigma_c$ can likely result from inherent stochasticity that may characterise the fracture process. Indeed, we find a spread in times taken by the micellar gel to `fluidize' under a constant shear-stress as well, much like that observed in the fracture process governing polymer gels \cite{skrzeszewska2010fracture}. Fracture in micellar networks showing slow dynamics has also been shown to have a rate-independent $\gamma_c$ with some spread (See Fig.13 in \cite{olsson2010slow}). However, that \textit{both} $\gamma_c$ and $\sigma_c$ are insensitive to applied rate of deformation certainly merits discussion. To the best of our knowledge, shear-startup studies for system akin to our surfactant gels have not been systematically carried out. The study of fracture in complex fluids has however been an active field of research \cite{ligoure2013fractures}. One of the models that is routinely employed to model fracture phenomenon in transient networks is the Activated Bond Rupture (ABR) in which the dissociation of bonds in a network is enhanced due to an applied stress. The model was invoked for the case of polymer gels in \cite{skrzeszewska2010fracture} and it predicts the critical strain for fracture $\gamma_c\sim C_1$ln$(C_2\dot\gamma)$ and thus $\sigma_c\sim C_3$ln$(C_4\dot\gamma)$ ($C_{1-4}$ are material dependent constants) i.e - the rupture strain and stress have a logarithmic dependence on applied rate of deformation. As mentioned before however, no such dependence is found in our experiments. A shear-rate independent rupture strain was found in a lecithin-cyclohexane system that forms a wormlike micellar network \cite{olsson2010slow}, but the shear-stress at rupture strongly depends on $\dot{\gamma}$ in the fracture regime. The authors proposed that the fracture in this system is characterized by a \textit{uniformly} accelerating fracture driven by breakage of micellar strands. Their post-fracture state consists of a nematic phase populated by disentangled and highly aligned micelles. 

\begin{figure}[!h]
	\begin{center}
		\includegraphics*[width=85mm]{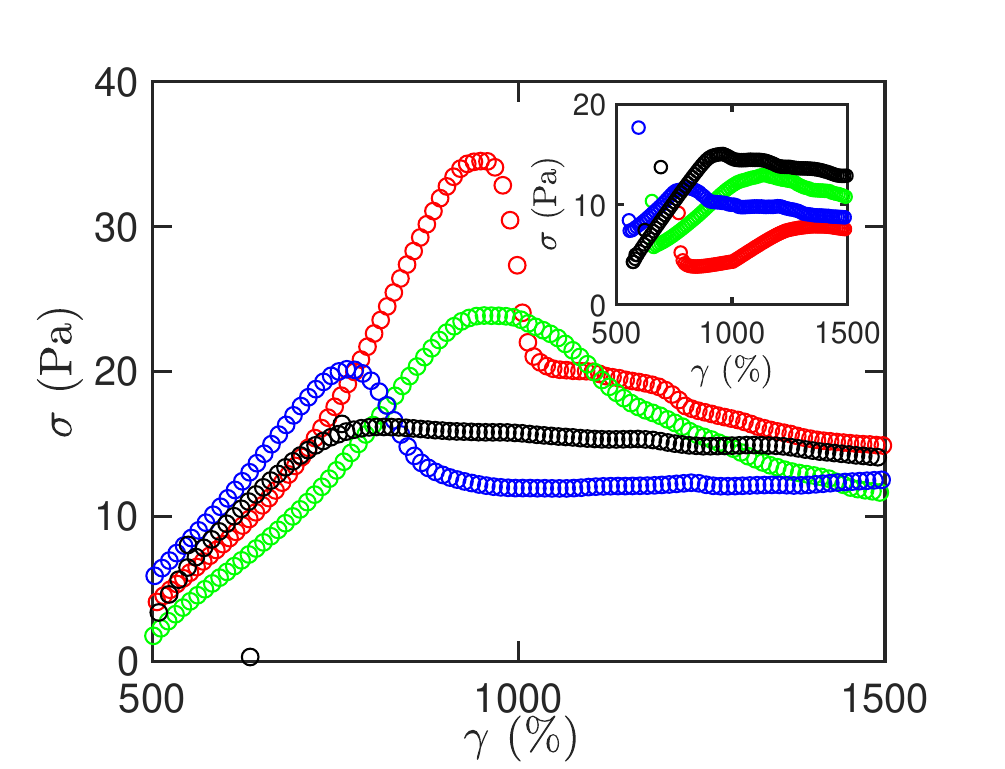}
	\end{center}
	\vspace*{-5mm}
	\caption{Shear start-up results post-fracture plotted as $\sigma$ vs shear- strain, $\gamma$\% for 8\% VES solution at different values of $\dot{\gamma}$ at \textcolor{black}{25C}: Red - 0.03$s^{-1}$, Green - 0.05$s^{-1}$, Blue - 0.1$s^{-1}$, Black - 0.3$s^{-1}$. Inset : Same as main figure but for 4.5\% VES. }
	\label{ss4}
\end{figure} 
This fracture mechanism does not apply to our system for two reasons - (1): it is unlikely that micellar breakage plays a role in the fracture as it was argued that for order unity deformations, micellar chains between entanglements store an elastic energy $\sim O(k_bT/l_e)$ \cite{cates1990nonlinear} which for \textcolor{black}{$l_e \sim O(10^2) nm$} is much smaller than the scission energies (even if the effective scission energy is reduced by an applied force). In any case the achieved $\sigma$ in our rheology experiments seem to be much smaller than micellar breaking stresses reported in previous studies \cite{rothstein2003transient,mandal2018stretch} (2): Our results clearly show that after the stress rapidly drops, the sheared material has a distinct \textit{elastic} growth of stress followed by a weak overshoot en-route to the final stress \textcolor{black}{as clearly demonstrated in Fig.~\ref{ss4} where we have plotted $\sigma$ vs $\gamma$ after the rapid drop of stress caused by a fracture.} This points towards an absence of a nematic state as a low viscosity nematic state cannot store an elastic stress \cite{olsson2010slow} and suggests that the post-fracture state is marked by structures that can contribute an elastic stress. In the absence of velocimetry data, we can only conjecture the mechanism of fracture in micellar gels. For $\dot{\gamma}$ in the range of the fracture regime, the micellar network is deformed quickly enough that there isn't enough time for re-arrangements (and re-entanglements) and its original elastic structure is forced to change. Since disentanglement has to occur before the formation of a nematic state, it is likely that the micellar disentanglement drives the eventual fracture of the network. We propose that the network fails by strain localization \cite{cheng2012shear,erk2012extreme}. The sudden drop after attaining a stress peak (Fig.\ref{ss3}) along with the absence of a nematic state of shear-aligned micelles suggests that this is not an homogeneous event as suggested in \cite{olsson2010slow} but a \textit{localized} crack or fracture \cite{berret2001evidence}. If we assume that this localized crack, presumably a region that is entanglement depleted, can then rupture and release stress according to the Griffith's criterion \cite{griffith1921vi}, it completes the picture of the shear-induced fracture dynamics discussed so far. This picture explains a couple of features of our data - (1): The critical rupture stress predicted by the Griffith's criterion, $\sigma_{rup}$ is independent of $\dot\gamma$ and this can explain the insensitivity to $\dot{\gamma}$ seen in our results, (2): $\sigma_{rup}\propto a$, where $a$ is an estimate of the size of the initial micro-crack that eventually leads to fracture. Since we do not have control over $a$ in our experiments and it could likely depend on a number of factors like thermal fluctuations or material history/preparation, this can explain why we see a spread in $\gamma_c,\sigma_c$ in our data. Clearly, there is much to uncover about the failure scenario in this regime. Preliminary experiments suggest that the presence of a fracture can be altered by the nature of rest before imposing a shear-rate. Indeed, such fracture dynamics have been shown to depend on boundary conditions and material history \cite{fielding2021yielding,benzi2021continuum}. Disentangling the role of these in the phenomenon displayed by our micellar gels is out of scope of the present paper. In order to better understand this interesting fracture regime and rule out other possible mechanisms like wall-slip, further detailed studies are required, starting with characterizing the dynamics of these wormlike micellar gels under an imposed shear stress. Such a study is currently underway \cite{gupta2020rheology}. We end this section on shear-rheology with a discussion on `flow-curves' exhibited by VES solutions. 
 
\begin{figure}[!h]
	\begin{center}
		\includegraphics*[width=85mm]{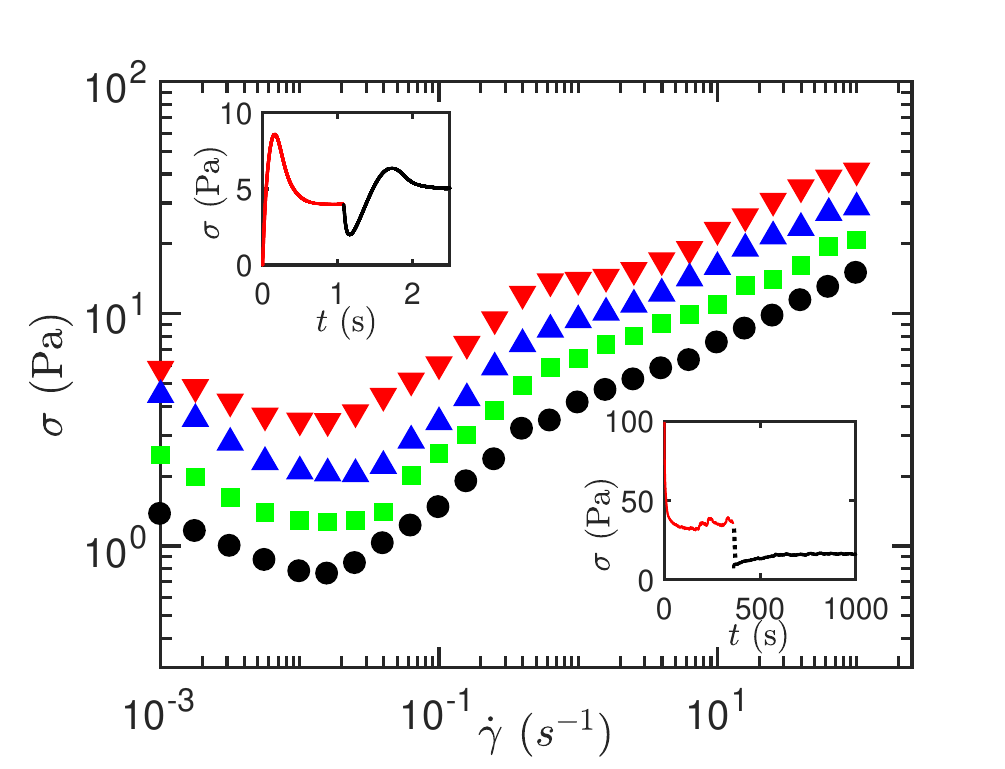}
	\end{center}
	\vspace*{-5mm}
	\caption{Flow curves at \textcolor{black}{25C} for different VES concentration, $c$ - Red - 10\%, Blue - 8\%, Green - 6\%, Black - 4.5\%. Inset (bottom) : A shear-rate step down test for 8\% VES - $\dot{\gamma}$ = 100 (red) to $\dot{\gamma}$ = 10 (black), Inset (top) : Same as bottom inset but for $\dot{\gamma} = 10^{-3}s^{-1}$ (red) to $\dot{\gamma} = 10^{-4}s^{-1}$ (black) and x axis is in units of $10^4 (s)$}
	\label{flowcurve}
\end{figure}

 For flow-curves, we sweep downwards from $\dot{\gamma} = 10^2s^{-1}-10^{-3}s^{-1}$ by adjusting the waiting time at each $\dot{\gamma}$. We report results upto $\dot\gamma = 10^{-3}s^{-1}$ because at lower rates, $\sigma$ takes very long time to reach steady-state. For $\dot\gamma>1s^{-1}$, VES show a marked shear-thinning behaviour, $\eta\sim\dot{\gamma}^{-\alpha}$ or $\sigma\sim\dot{\gamma}^{1-\alpha}$. The average value of the exponent extracted from the data in Fig.~\ref{flowcurve} is $\alpha \simeq 0.72$. To obtain a proper estimate of $\alpha$ would require conducting multiple flow-curve experiments and averaging over independent runs at each $c$, so our estimate of $\alpha$ should at best be looked as a rough estimate. But, we note that this scaling lies somewhere in between $\alpha = 0.77$ and 0.66, which are predictions made for flexible micelles and more rigid micelles respectively \cite{zhang2013mesoscale}. \textcolor{black}{A scaling of $\alpha \sim 0.66$ was also reported for the flow of EHAC \cite{padding2008dynamics}.} A typical step-down in shear experiment ($\dot{\gamma} = 10^2 -10 s^{-1}$) offers some insight into the state of VES when deformed by a $\dot\gamma$ in the shear-thinning regime. As seen in the bottom inset of Fig.~\ref{flowcurve}, when $\dot{\gamma}$ is stepped down, there is a rapid drop in stress, followed by a build-up to the steady-state stress at $\dot{\gamma} = 10s^{-1}$. This rapid drop indicates that for $\dot{\gamma}$ in this regime, the shear-thinning behaviour arises due to aligned micelles that form a viscous nematic phase which allows stress to relax almost instantaneously when $\dot{\gamma}$ is dropped. The stress build-up post drop-off indicates re-building of stress-bearing microstructure. Taken together, this behaviour is suggestive of a thixotropic response in this regime \cite{mewis2009thixotropy,larson2015constitutive,larson2019review}. We also note that for $\dot\gamma$ between $0.05-1s^{-1}$, the value of $\alpha$ is markedly smaller than that in the shear-thinning regime . This range of $\dot\gamma$ coincides with the range in which we have characterised a fracture behaviour earlier and the transition in the value of $\alpha$ can indicate a structural change in the micelles, the physical mechanism for which remains elusive at this point. 
 
 Another interesting features of the flow-curves in Fig.~\ref{flowcurve} is their non-monotonicity wherein for $\gamma \lesssim 0.01s^{-1}$, $\sigma$ \textit{increases} with decreasing $\dot{\gamma}$. A caveat is necessary before we proceed with discussing this feature. At these low values of $\dot{\gamma}$, the time taken to reach steady-state becomes very large and this makes steady-state flow-curve experiments difficult. Thus, the points \textcolor{black}{($\dot{\gamma}\lesssim 0.01s^{-1}$)} in the flow-curve likely represent a $transient$ scenario, rather than a true steady state. However, we can still gain some valuable insight from these experiments. 

In principle, an intrinsic non-monotonic flow-curve is often a signature of \textit{shear-banding} in the system, but the experimental manifestation of this phenomenon is a stress-plateau. The region of the flow-curve where non-monotonicity occurs is experimentally inaccessible because for shear-rates in this regime, the material separates into two or more shear-bands \cite{spenley1996nonmonotonic,britton1997two}. However, a measurable non-monotonicty in the flow-curve can be closely related to \textit{transient} shear-banding. In \cite{martin2012transient}, the authors showed that a competition in breakup and restoration (ageing) of stress-bearing microstructure can lead to regions where measured $\sigma$ increases with decreasing imposed $\dot{\gamma}$ and this directly correlates with banding observed in their Laponite system. Indeed, as suggested in \cite{martin2012transient}, direct analogies can be drawn between their colloidal gel system and our wormlike micellar gel, where disentanglement/re-entanglement can play the role breakup/ageing and the stress-bearing microstructure is essentially a network of long entangled micelles. Prominent stress-overshoot at very low $\dot\gamma$ along with the very long times to reach steady in this regime, both generally features of shear-banding often observed in wormlike micellar solutions \cite{lerouge1998shear,lerouge2000correlations,grand1997slow}, are also observed in our surfactant gels. This supports the occurence of shear-banding in our system. Note, that disentanglement of micellar strands is at the heart of a proposed mechanism for shear-banding in WLM solutions, wherein progressive disentanglement leads to gradual development of shear-bands \cite{hu2005kinetics}. The cup-bob geometry aids this mechanism due to its inherent stress gradient and has been reported to aid the formation of shear-bands in systems with flow-curves that are nearly non-monotonic \cite{martin2012transient,greco1997shear}, further lending credence to the presence of shear-banding in our system. The above invoked mechanism is likely distinct from the one suggested for causing fracture for higher values of $\dot\gamma$. It is possible that fractures do not form in the low $\dot\gamma$ regime because re-entanglement can 'heal' zones of depleted entanglements \cite{erk2012extreme,de2007melt}. \textcolor{black}{It is imperative to mention here that the `measured' non-monotonicity in the flow-curves for the VES are distinct from the flow-curves in granular media \cite{dijksman2011jamming}, microemulsions of oil-water \cite{michel2001unstable} and waxy crude oils \cite{dimitriou2014comprehensive}, where the resulting measurable non-monotonicity is usually indicative of an unstable flow. While we cannot confirm the nature of shear-bands with certainty in this paper owing to absence of local data and long timescales involved, it is likely that the `transient' bands that cause non-monotonicty in our flow curves, persist over long-times into a permanent band at steady state, finally manifesting rheologically as a stress-plateau.} Finally, we end this section by noting that there is precedent for non-monotonic flow-curves in a surfactant gel-system formed by long-chained zwitterionic surfactants.

In \cite{chu2010amidosulfobetaine}, authors employed EDAS - a zwitterionic amidosulfobetaine surfactant gel, and demonstrated that flow-curve experiments with this system showed non-monotonicty. The authors tied this to the presence of shear-banding, by visually confirming bands (although no comment was made on if these bands are transient or steady). There are some crucial differences between results reported in \cite{chu2010amidosulfobetaine} and our paper. 

\begin{figure}[!h]
	\begin{center}
		\includegraphics*[width=85mm]{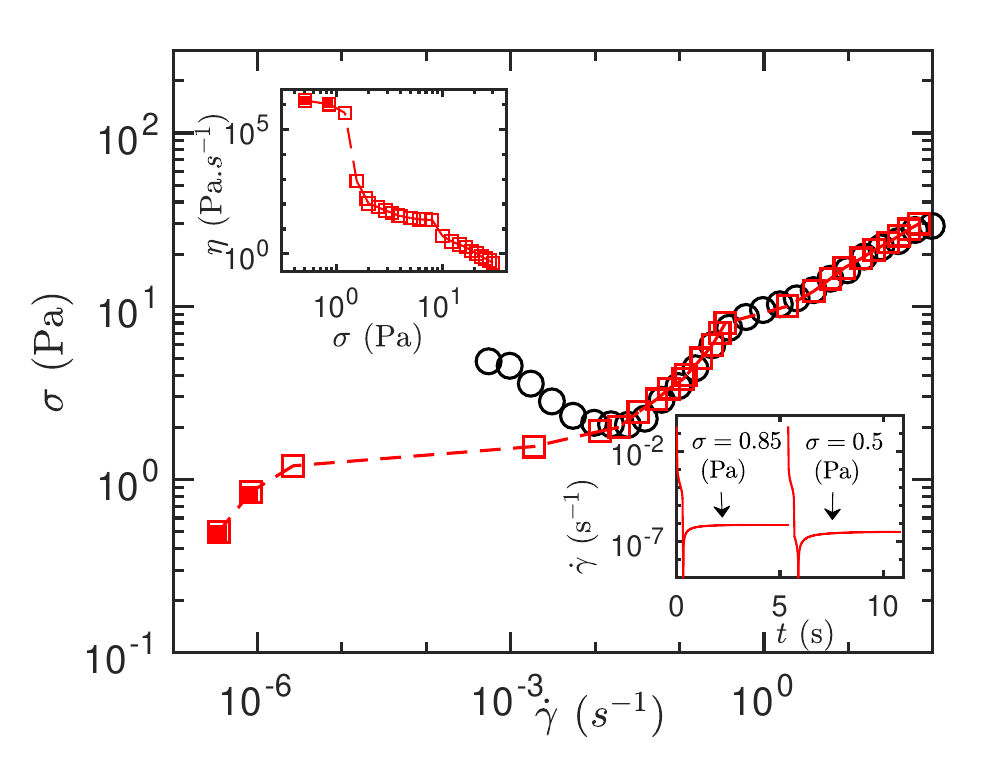}
	\end{center}
		\vspace*{-5mm}
	\caption{Flow curves at \textcolor{black}{25C} for 8\% VES in controlled rate (black circles) and controlled stress mode (red square). Top inset: Shear Viscosity vs shear stress for the controlled stress experiment in main figure. Bottom inset: imposed shear stress vs measured shear-rate for the two lowest values of $\sigma$ (two highest values of $\eta$) which correspond the colored data points in the respective plots }
	\label{flowcurve2}
\end{figure} 
First, they were able to experimentally access the `Newtonian' regime of the system in which $\sigma\propto\dot{\gamma}$ and no-overshoot appears in shear-startup tests. This guarantees that the surfactant `gel' isn't a true bulk gel and its yielding characteristics are crucially tied to the shear-banding transition and the peak stress extracted from the non-monotonic region of the flow-curve. We observe an overshoot for the lowest shear-rate tested $\dot\gamma = 10^{-4}(s^{-1})$ (As seen in Fig.~\ref{ss2}) and owing to long experimental times, could not access the Newtonian regime. However, as seen in Fig.~\ref{flowcurve2}, for the controlled-stress run, we are close to probing the Newtonian regime. The inset in Fig.~\ref{flowcurve2} confirms that steady state is reached for the lowest $\sigma$ values (Although we are close to the operating limit of the rheometer with such low values of $\sigma,\dot{\gamma}$). This suggests that our micellar gels too are not bulk gels. It is likely that the apparent yield-stress in our system too can be tied to a banding transition and gel-like rheological behaviour can be rationalized by noting that the cross-over frequencies for the system are vastly out of experimental range as explained in \cite{chu2010amidosulfobetaine}. Last, EDAS is a branched micellar system whereas our surfactant gel is an entangled WLM. It is possible that this difference in micellar morphology can lead to differences in observed dynamical behaviour. For instance \cite{chu2010amidosulfobetaine} proposes that their yielding transition is mediated by the creation of shear-bands with nematic order. However, our proposed mechanism avoids the creation of nematic bands. If there were a nematic band, the stress relaxation in step-down experiments should have a fast-relaxation component corresponding to the nematic band's relative inability to store elastic stress \cite{olsson2010slow}. As seen in the top inset of Fig.~\ref{flowcurve}, such a relaxation process is absent. (We do not see it even if the $\dot\gamma$ is stepped down to 0).

\section*{Conclusion}
In this paper we conducted a detailed linear and non-linear rheological study of surfactant gel formed by long-chained zwitterionic surfactants. We showed that this system has a microstructure made up of long entangled wormlike micelles and demonstrated gel-like behaviour in its linear-rheology at 25$C$ and on visual inspection, much like that displayed by similar systems \cite{raghavan2001highly,kumar2007wormlike,chu2010wormlike}. Our analysis of the viscoelastic state at $60C$ and higher temperatures allowed us to calculate time-scales associated with the system's rheological response as well as various activation energies that govern these quantities. We showed that these timescales do not obey scalings derived for more conventional surfactant systems. Our results call for further theoretical and experimental studies to resolve measured deviations from current theory for wormlike micellar systems and to gain better understanding of the self-assembly of long-chained surfactants. We also showed the absence of strain-stiffening in these surfactant gels, a behaviour contrary to that exhibited by similar systems \cite{tung2008self,tung2008strain} but in agreement with a system closest to ours - EDAB \cite{private}. Our shear rheology results revealed an interesting fracture regime reminiscent of brittle type fracture seen in amorphous materials \cite{barlow2020ductile,singh2020brittle}, the specifics of which have not been earlier reported for a micellar system to the best of our knowledge. We compared this behaviour to similar dynamics reported for other systems and outlined a possible mechanism for the same. A similar shear-induced fracture has been reported for certain telechelic protein based hydrogels \cite{olsen2010yielding}. Here, the related shear-banding behaviour is beneficial as it shields cells embedded in tissue matrices made from these hydrogels from damage during injection. As surfactant gels are being seen as alternatives to polymer based hydrogels for biomedical applications \cite{lee2001hydrogels}, insight into shear-induced failure in such materials will inform design of smart and efficient soft matter. 

While, in the fracture regime, it is likely that the velocity field is inhomogeneous (in addition to/because of shear-banding), our flow-curves provide indirect evidence of transient shear-banding in our wormlike micellar gels. Extremely slow transients mark the stress response in this regime and it is difficult to measure a steady-state response; instead we see a non-monotonic flow-curve. If indeed, micellar disentanglement drives eventual response in both the fracture and possible transient-banding regime, this surfactant gel is then an appropriate system to study two different mechanisms of strain localization \cite{erk2012extreme}. Finally, we show that it is likely that surfactant gels are not true `gels' but essentially `fluids' with ultra-long relaxation times (If we go by Fig.\ref{flowcurve2}, relaxation time $\sim O(10^{5}) s$) and this can explain the long transients in the flow-curve's non-monotonic region. This finding makes an advance in answering the conundrum of gelation induced \textit{only} by entanglements \cite{raghavan2012conundrum} and suggests that while a surfactant gel might show many gel-like features in its appearance and rheology, and even if it may sometimes functionally behave like a gel owing to its very high relaxation time, it is \textit{not} a permanent gel \cite{chu2010amidosulfobetaine}. Overall, this paper establishes wormlike micellar gels formed by long-chained zwitterionic surfactants as a fertile model system to explore a gamut of phenomena prevalent across soft matter. In the future, we plan to better elucidate the failure/fracture mechanisms of these gels - \textcolor{black}{in particular the possible roles of disentanglement in the bulk and wall-slip in this driving this phenomenon}. We also plan to check if our wormlike micellar gels display novel behaviour in extensional flow and other canonical flow scenarios. The results presented in this paper are likely to inform these studies and other such studies that deal with the formulation, characterization and flow dynamics of similar systems.

\begin{acknowledgments}
This research was made possible by research funding from Schlumberger and NSERC under the CRD program, project 505549-16. Experimental infrastructure was funded by the Canada Foundation for Innovation and the BC Knowledge Fund, grant number CFI JELF 36069. This funding is gratefully acknowledged. We thank Schlumberger Ltd. for providing us with the surfactant used for the experiments in the paper. R.G and R.M would also like to thank the University of British Columbia (UBC) for a 4YF Doctoral Fellowship, Claire Atkinson at UBC's Faculty of Medicine for technical expertise in preparing Cryo-TEM images, and John Casola at Netzsch for help in using the Kinexus rheometer. 

\end{acknowledgments}

\bibliography{EDAB_rheology}

\begin{thebibliography}{107}%
\makeatletter
\providecommand \@ifxundefined [1]{%
 \@ifx{#1\undefined}
}%
\providecommand \@ifnum [1]{%
 \ifnum #1\expandafter \@firstoftwo
 \else \expandafter \@secondoftwo
 \fi
}%
\providecommand \@ifx [1]{%
 \ifx #1\expandafter \@firstoftwo
 \else \expandafter \@secondoftwo
 \fi
}%
\providecommand \natexlab [1]{#1}%
\providecommand \enquote  [1]{``#1''}%
\providecommand \bibnamefont  [1]{#1}%
\providecommand \bibfnamefont [1]{#1}%
\providecommand \citenamefont [1]{#1}%
\providecommand \href@noop [0]{\@secondoftwo}%
\providecommand \href [0]{\begingroup \@sanitize@url \@href}%
\providecommand \@href[1]{\@@startlink{#1}\@@href}%
\providecommand \@@href[1]{\endgroup#1\@@endlink}%
\providecommand \@sanitize@url [0]{\catcode `\\12\catcode `\$12\catcode
  `\&12\catcode `\#12\catcode `\^12\catcode `\_12\catcode `\%12\relax}%
\providecommand \@@startlink[1]{}%
\providecommand \@@endlink[0]{}%
\providecommand \url  [0]{\begingroup\@sanitize@url \@url }%
\providecommand \@url [1]{\endgroup\@href {#1}{\urlprefix }}%
\providecommand \urlprefix  [0]{URL }%
\providecommand \Eprint [0]{\href }%
\providecommand \doibase [0]{http://dx.doi.org/}%
\providecommand \selectlanguage [0]{\@gobble}%
\providecommand \bibinfo  [0]{\@secondoftwo}%
\providecommand \bibfield  [0]{\@secondoftwo}%
\providecommand \translation [1]{[#1]}%
\providecommand \BibitemOpen [0]{}%
\providecommand \bibitemStop [0]{}%
\providecommand \bibitemNoStop [0]{.\EOS\space}%
\providecommand \EOS [0]{\spacefactor3000\relax}%
\providecommand \BibitemShut  [1]{\csname bibitem#1\endcsname}%
\let\auto@bib@innerbib\@empty
\bibitem [{\citenamefont {Fardin}\ and\ \citenamefont
  {Lerouge}(2014)}]{fardin2014flows}%
  \BibitemOpen
  \bibfield  {author} {\bibinfo {author} {\bibfnamefont {M.-A.}\ \bibnamefont
  {Fardin}}\ and\ \bibinfo {author} {\bibfnamefont {S.}~\bibnamefont
  {Lerouge}},\ }\bibfield  {title} {\enquote {\bibinfo {title} {Flows of living
  polymer fluids},}\ }\href@noop {} {\bibfield  {journal} {\bibinfo  {journal}
  {Soft Matter}\ }\textbf {\bibinfo {volume} {10}},\ \bibinfo {pages}
  {8789--8799} (\bibinfo {year} {2014})}\BibitemShut {NoStop}%
\bibitem [{\citenamefont {Cates}\ and\ \citenamefont
  {Fielding}(2006)}]{cates2006rheology}%
  \BibitemOpen
  \bibfield  {author} {\bibinfo {author} {\bibfnamefont {M.~E.}\ \bibnamefont
  {Cates}}\ and\ \bibinfo {author} {\bibfnamefont {S.~M.}\ \bibnamefont
  {Fielding}},\ }\bibfield  {title} {\enquote {\bibinfo {title} {Rheology of
  giant micelles},}\ }\href@noop {} {\bibfield  {journal} {\bibinfo  {journal}
  {Advances in Physics}\ }\textbf {\bibinfo {volume} {55}},\ \bibinfo {pages}
  {799--879} (\bibinfo {year} {2006})}\BibitemShut {NoStop}%
\bibitem [{\citenamefont {Rehage}\ and\ \citenamefont
  {Hoffmann}(1991)}]{rehage1991viscoelastic}%
  \BibitemOpen
  \bibfield  {author} {\bibinfo {author} {\bibfnamefont {H.}~\bibnamefont
  {Rehage}}\ and\ \bibinfo {author} {\bibfnamefont {H.}~\bibnamefont
  {Hoffmann}},\ }\bibfield  {title} {\enquote {\bibinfo {title} {Viscoelastic
  surfactant solutions: model systems for rheological research},}\ }\href@noop
  {} {\bibfield  {journal} {\bibinfo  {journal} {Molecular Physics}\ }\textbf
  {\bibinfo {volume} {74}},\ \bibinfo {pages} {933--973} (\bibinfo {year}
  {1991})}\BibitemShut {NoStop}%
\bibitem [{\citenamefont {Cates}\ and\ \citenamefont
  {Candau}(1990)}]{cates1990statics}%
  \BibitemOpen
  \bibfield  {author} {\bibinfo {author} {\bibfnamefont {M.}~\bibnamefont
  {Cates}}\ and\ \bibinfo {author} {\bibfnamefont {S.}~\bibnamefont {Candau}},\
  }\bibfield  {title} {\enquote {\bibinfo {title} {Statics and dynamics of
  worm-like surfactant micelles},}\ }\href@noop {} {\bibfield  {journal}
  {\bibinfo  {journal} {Journal of Physics: Condensed Matter}\ }\textbf
  {\bibinfo {volume} {2}},\ \bibinfo {pages} {6869} (\bibinfo {year}
  {1990})}\BibitemShut {NoStop}%
\bibitem [{\citenamefont {Cates}(1987)}]{cates1987reptation}%
  \BibitemOpen
  \bibfield  {author} {\bibinfo {author} {\bibfnamefont {M.}~\bibnamefont
  {Cates}},\ }\bibfield  {title} {\enquote {\bibinfo {title} {Reptation of
  living polymers: dynamics of entangled polymers in the presence of reversible
  chain-scission reactions},}\ }\href@noop {} {\bibfield  {journal} {\bibinfo
  {journal} {Macromolecules}\ }\textbf {\bibinfo {volume} {20}},\ \bibinfo
  {pages} {2289--2296} (\bibinfo {year} {1987})}\BibitemShut {NoStop}%
\bibitem [{\citenamefont {Sood}, \citenamefont {Bandyopadhyay},\ and\
  \citenamefont {Basappa}(1999)}]{sood1999linear}%
  \BibitemOpen
  \bibfield  {author} {\bibinfo {author} {\bibfnamefont {A.}~\bibnamefont
  {Sood}}, \bibinfo {author} {\bibfnamefont {R.}~\bibnamefont {Bandyopadhyay}},
  \ and\ \bibinfo {author} {\bibfnamefont {G.}~\bibnamefont {Basappa}},\
  }\bibfield  {title} {\enquote {\bibinfo {title} {Linear and nonlinear
  rheology of wormlike micelles},}\ }\href@noop {} {\bibfield  {journal}
  {\bibinfo  {journal} {Pramana}\ }\textbf {\bibinfo {volume} {53}},\ \bibinfo
  {pages} {223--235} (\bibinfo {year} {1999})}\BibitemShut {NoStop}%
\bibitem [{\citenamefont {Rothstein}(2008)}]{rothstein2008strong}%
  \BibitemOpen
  \bibfield  {author} {\bibinfo {author} {\bibfnamefont {J.~P.}\ \bibnamefont
  {Rothstein}},\ }\bibfield  {title} {\enquote {\bibinfo {title} {Strong flows
  of viscoelastic wormlike micelle solutions},}\ }\href@noop {} {\bibfield
  {journal} {\bibinfo  {journal} {Rheology Reviews}\ }\textbf {\bibinfo
  {volume} {2008}},\ \bibinfo {pages} {1--46} (\bibinfo {year}
  {2008})}\BibitemShut {NoStop}%
\bibitem [{\citenamefont {Fardin}\ and\ \citenamefont
  {Lerouge}(2012)}]{fardin2012instabilities}%
  \BibitemOpen
  \bibfield  {author} {\bibinfo {author} {\bibfnamefont {M.-A.}\ \bibnamefont
  {Fardin}}\ and\ \bibinfo {author} {\bibfnamefont {S.}~\bibnamefont
  {Lerouge}},\ }\bibfield  {title} {\enquote {\bibinfo {title} {Instabilities
  in wormlike micelle systems},}\ }\href@noop {} {\bibfield  {journal}
  {\bibinfo  {journal} {The European Physical Journal E}\ }\textbf {\bibinfo
  {volume} {35}},\ \bibinfo {pages} {1--29} (\bibinfo {year}
  {2012})}\BibitemShut {NoStop}%
\bibitem [{\citenamefont {Zhao}, \citenamefont {Cheung},\ and\ \citenamefont
  {Shen}(2014)}]{zhao2014microfluidic}%
  \BibitemOpen
  \bibfield  {author} {\bibinfo {author} {\bibfnamefont {Y.}~\bibnamefont
  {Zhao}}, \bibinfo {author} {\bibfnamefont {P.}~\bibnamefont {Cheung}}, \ and\
  \bibinfo {author} {\bibfnamefont {A.~Q.}\ \bibnamefont {Shen}},\ }\bibfield
  {title} {\enquote {\bibinfo {title} {Microfluidic flows of wormlike micellar
  solutions},}\ }\href@noop {} {\bibfield  {journal} {\bibinfo  {journal}
  {Advances in Colloid and Interface Science}\ }\textbf {\bibinfo {volume}
  {211}},\ \bibinfo {pages} {34--46} (\bibinfo {year} {2014})}\BibitemShut
  {NoStop}%
\bibitem [{\citenamefont {Berret}(2006)}]{berret2006rheology}%
  \BibitemOpen
  \bibfield  {author} {\bibinfo {author} {\bibfnamefont {J.-F.}\ \bibnamefont
  {Berret}},\ }\bibfield  {title} {\enquote {\bibinfo {title} {Rheology of
  wormlike micelles: Equilibrium properties and shear banding transitions},}\
  }in\ \href@noop {} {\emph {\bibinfo {booktitle} {Molecular gels}}}\ (\bibinfo
   {publisher} {Springer},\ \bibinfo {year} {2006})\ pp.\ \bibinfo {pages}
  {667--720}\BibitemShut {NoStop}%
\bibitem [{\citenamefont {Yang}(2002)}]{yang2002viscoelastic}%
  \BibitemOpen
  \bibfield  {author} {\bibinfo {author} {\bibfnamefont {J.}~\bibnamefont
  {Yang}},\ }\bibfield  {title} {\enquote {\bibinfo {title} {Viscoelastic
  wormlike micelles and their applications},}\ }\href@noop {} {\bibfield
  {journal} {\bibinfo  {journal} {Current Opinion in Colloid \& Interface
  Science}\ }\textbf {\bibinfo {volume} {7}},\ \bibinfo {pages} {276--281}
  (\bibinfo {year} {2002})}\BibitemShut {NoStop}%
\bibitem [{\citenamefont {Ezrahi}, \citenamefont {Tuval},\ and\ \citenamefont
  {Aserin}(2006)}]{ezrahi2006properties}%
  \BibitemOpen
  \bibfield  {author} {\bibinfo {author} {\bibfnamefont {S.}~\bibnamefont
  {Ezrahi}}, \bibinfo {author} {\bibfnamefont {E.}~\bibnamefont {Tuval}}, \
  and\ \bibinfo {author} {\bibfnamefont {A.}~\bibnamefont {Aserin}},\
  }\bibfield  {title} {\enquote {\bibinfo {title} {Properties, main
  applications and perspectives of worm micelles},}\ }\href@noop {} {\bibfield
  {journal} {\bibinfo  {journal} {Advances in Colloid and Interface Science}\
  }\textbf {\bibinfo {volume} {128}},\ \bibinfo {pages} {77--102} (\bibinfo
  {year} {2006})}\BibitemShut {NoStop}%
\bibitem [{\citenamefont {Virk}(1975)}]{virk1975drag}%
  \BibitemOpen
  \bibfield  {author} {\bibinfo {author} {\bibfnamefont {P.~S.}\ \bibnamefont
  {Virk}},\ }\bibfield  {title} {\enquote {\bibinfo {title} {Drag reduction
  fundamentals},}\ }\href@noop {} {\bibfield  {journal} {\bibinfo  {journal}
  {AIChE Journal}\ }\textbf {\bibinfo {volume} {21}},\ \bibinfo {pages}
  {625--656} (\bibinfo {year} {1975})}\BibitemShut {NoStop}%
\bibitem [{\citenamefont {White}\ and\ \citenamefont
  {Mungal}(2008)}]{white2008mechanics}%
  \BibitemOpen
  \bibfield  {author} {\bibinfo {author} {\bibfnamefont {C.~M.}\ \bibnamefont
  {White}}\ and\ \bibinfo {author} {\bibfnamefont {M.~G.}\ \bibnamefont
  {Mungal}},\ }\bibfield  {title} {\enquote {\bibinfo {title} {Mechanics and
  prediction of turbulent drag reduction with polymer additives},}\ }\href@noop
  {} {\bibfield  {journal} {\bibinfo  {journal} {Annu. Rev. Fluid Mech.}\
  }\textbf {\bibinfo {volume} {40}},\ \bibinfo {pages} {235--256} (\bibinfo
  {year} {2008})}\BibitemShut {NoStop}%
\bibitem [{\citenamefont {Li}\ \emph {et~al.}(2012)\citenamefont {Li},
  \citenamefont {Yu}, \citenamefont {Wei},\ and\ \citenamefont
  {Kawaguchi}}]{li2012turbulent}%
  \BibitemOpen
  \bibfield  {author} {\bibinfo {author} {\bibfnamefont {F.-C.}\ \bibnamefont
  {Li}}, \bibinfo {author} {\bibfnamefont {B.}~\bibnamefont {Yu}}, \bibinfo
  {author} {\bibfnamefont {J.-J.}\ \bibnamefont {Wei}}, \ and\ \bibinfo
  {author} {\bibfnamefont {Y.}~\bibnamefont {Kawaguchi}},\ }\href@noop {}
  {\emph {\bibinfo {title} {Turbulent drag reduction by surfactant
  additives}}}\ (\bibinfo  {publisher} {John Wiley \& Sons},\ \bibinfo {year}
  {2012})\BibitemShut {NoStop}%
\bibitem [{\citenamefont {Sullivan}\ \emph {et~al.}(2007)\citenamefont
  {Sullivan}, \citenamefont {Nelson}, \citenamefont {Anderson},\ and\
  \citenamefont {Hughes}}]{sullivan2007oilfield}%
  \BibitemOpen
  \bibfield  {author} {\bibinfo {author} {\bibfnamefont {P.}~\bibnamefont
  {Sullivan}}, \bibinfo {author} {\bibfnamefont {E.~B.}\ \bibnamefont
  {Nelson}}, \bibinfo {author} {\bibfnamefont {V.}~\bibnamefont {Anderson}}, \
  and\ \bibinfo {author} {\bibfnamefont {T.}~\bibnamefont {Hughes}},\
  }\bibfield  {title} {\enquote {\bibinfo {title} {Oilfield applications of
  giant micelles},}\ }in\ \href@noop {} {\emph {\bibinfo {booktitle} {Giant
  Micelles}}}\ (\bibinfo  {publisher} {CRC Press},\ \bibinfo {year} {2007})\
  pp.\ \bibinfo {pages} {453--472}\BibitemShut {NoStop}%
\bibitem [{\citenamefont {Kefi}\ \emph {et~al.}(2004)\citenamefont {Kefi},
  \citenamefont {Lee}, \citenamefont {Pope}, \citenamefont {Sullivan},
  \citenamefont {Nelson}, \citenamefont {Hernandez}, \citenamefont {Olsen},
  \citenamefont {Parlar}, \citenamefont {Powers}, \citenamefont {Roy} \emph
  {et~al.}}]{kefi2004expanding}%
  \BibitemOpen
  \bibfield  {author} {\bibinfo {author} {\bibfnamefont {S.}~\bibnamefont
  {Kefi}}, \bibinfo {author} {\bibfnamefont {J.}~\bibnamefont {Lee}}, \bibinfo
  {author} {\bibfnamefont {T.}~\bibnamefont {Pope}}, \bibinfo {author}
  {\bibfnamefont {P.}~\bibnamefont {Sullivan}}, \bibinfo {author}
  {\bibfnamefont {E.}~\bibnamefont {Nelson}}, \bibinfo {author} {\bibfnamefont
  {A.}~\bibnamefont {Hernandez}}, \bibinfo {author} {\bibfnamefont
  {T.}~\bibnamefont {Olsen}}, \bibinfo {author} {\bibfnamefont
  {M.}~\bibnamefont {Parlar}}, \bibinfo {author} {\bibfnamefont
  {B.}~\bibnamefont {Powers}}, \bibinfo {author} {\bibfnamefont
  {A.}~\bibnamefont {Roy}},  \emph {et~al.},\ }\bibfield  {title} {\enquote
  {\bibinfo {title} {Expanding applications for viscoelastic surfactants},}\
  }\href@noop {} {\bibfield  {journal} {\bibinfo  {journal} {Oilfield Rev.}\
  }\textbf {\bibinfo {volume} {16}},\ \bibinfo {pages} {10--23} (\bibinfo
  {year} {2004})}\BibitemShut {NoStop}%
\bibitem [{\citenamefont {Maitland}(2000)}]{maitland2000oil}%
  \BibitemOpen
  \bibfield  {author} {\bibinfo {author} {\bibfnamefont {G.}~\bibnamefont
  {Maitland}},\ }\bibfield  {title} {\enquote {\bibinfo {title} {Oil and gas
  production},}\ }\href@noop {} {\bibfield  {journal} {\bibinfo  {journal}
  {Current Opinion in Colloid \& Interface Science}\ }\textbf {\bibinfo
  {volume} {5}},\ \bibinfo {pages} {301--311} (\bibinfo {year}
  {2000})}\BibitemShut {NoStop}%
\bibitem [{\citenamefont {Boek}\ \emph {et~al.}(2002)\citenamefont {Boek},
  \citenamefont {Jusufi}, \citenamefont {L{\"o}wen},\ and\ \citenamefont
  {Maitland}}]{boek2002molecular}%
  \BibitemOpen
  \bibfield  {author} {\bibinfo {author} {\bibfnamefont {E.}~\bibnamefont
  {Boek}}, \bibinfo {author} {\bibfnamefont {A.}~\bibnamefont {Jusufi}},
  \bibinfo {author} {\bibfnamefont {H.}~\bibnamefont {L{\"o}wen}}, \ and\
  \bibinfo {author} {\bibfnamefont {G.}~\bibnamefont {Maitland}},\ }\bibfield
  {title} {\enquote {\bibinfo {title} {Molecular design of responsive fluids:
  molecular dynamics studies of viscoelastic surfactant solutions},}\
  }\href@noop {} {\bibfield  {journal} {\bibinfo  {journal} {Journal of
  Physics: Condensed Matter}\ }\textbf {\bibinfo {volume} {14}},\ \bibinfo
  {pages} {9413} (\bibinfo {year} {2002})}\BibitemShut {NoStop}%
\bibitem [{\citenamefont {Raghavan}\ and\ \citenamefont
  {Kaler}(2001)}]{raghavan2001highly}%
  \BibitemOpen
  \bibfield  {author} {\bibinfo {author} {\bibfnamefont {S.~R.}\ \bibnamefont
  {Raghavan}}\ and\ \bibinfo {author} {\bibfnamefont {E.~W.}\ \bibnamefont
  {Kaler}},\ }\bibfield  {title} {\enquote {\bibinfo {title} {Highly
  viscoelastic wormlike micellar solutions formed by cationic surfactants with
  long unsaturated tails},}\ }\href@noop {} {\bibfield  {journal} {\bibinfo
  {journal} {Langmuir}\ }\textbf {\bibinfo {volume} {17}},\ \bibinfo {pages}
  {300--306} (\bibinfo {year} {2001})}\BibitemShut {NoStop}%
\bibitem [{\citenamefont {Kumar}\ \emph {et~al.}(2007)\citenamefont {Kumar},
  \citenamefont {Kalur}, \citenamefont {Ziserman}, \citenamefont {Danino},\
  and\ \citenamefont {Raghavan}}]{kumar2007wormlike}%
  \BibitemOpen
  \bibfield  {author} {\bibinfo {author} {\bibfnamefont {R.}~\bibnamefont
  {Kumar}}, \bibinfo {author} {\bibfnamefont {G.~C.}\ \bibnamefont {Kalur}},
  \bibinfo {author} {\bibfnamefont {L.}~\bibnamefont {Ziserman}}, \bibinfo
  {author} {\bibfnamefont {D.}~\bibnamefont {Danino}}, \ and\ \bibinfo {author}
  {\bibfnamefont {S.~R.}\ \bibnamefont {Raghavan}},\ }\bibfield  {title}
  {\enquote {\bibinfo {title} {Wormlike micelles of a c22-tailed zwitterionic
  betaine surfactant: from viscoelastic solutions to elastic gels},}\
  }\href@noop {} {\bibfield  {journal} {\bibinfo  {journal} {Langmuir}\
  }\textbf {\bibinfo {volume} {23}},\ \bibinfo {pages} {12849--12856} (\bibinfo
  {year} {2007})}\BibitemShut {NoStop}%
\bibitem [{\citenamefont {Chu}\ \emph {et~al.}(2010)\citenamefont {Chu},
  \citenamefont {Feng}, \citenamefont {Su},\ and\ \citenamefont
  {Han}}]{chu2010wormlike}%
  \BibitemOpen
  \bibfield  {author} {\bibinfo {author} {\bibfnamefont {Z.}~\bibnamefont
  {Chu}}, \bibinfo {author} {\bibfnamefont {Y.}~\bibnamefont {Feng}}, \bibinfo
  {author} {\bibfnamefont {X.}~\bibnamefont {Su}}, \ and\ \bibinfo {author}
  {\bibfnamefont {Y.}~\bibnamefont {Han}},\ }\bibfield  {title} {\enquote
  {\bibinfo {title} {Wormlike micelles and solution properties of a c22-tailed
  amidosulfobetaine surfactant},}\ }\href@noop {} {\bibfield  {journal}
  {\bibinfo  {journal} {Langmuir}\ }\textbf {\bibinfo {volume} {26}},\ \bibinfo
  {pages} {7783--7791} (\bibinfo {year} {2010})}\BibitemShut {NoStop}%
\bibitem [{\citenamefont {Chu}\ and\ \citenamefont
  {Feng}(2011)}]{chu2011thermo}%
  \BibitemOpen
  \bibfield  {author} {\bibinfo {author} {\bibfnamefont {Z.}~\bibnamefont
  {Chu}}\ and\ \bibinfo {author} {\bibfnamefont {Y.}~\bibnamefont {Feng}},\
  }\bibfield  {title} {\enquote {\bibinfo {title} {Thermo-switchable surfactant
  gel},}\ }\href@noop {} {\bibfield  {journal} {\bibinfo  {journal} {Chemical
  Communications}\ }\textbf {\bibinfo {volume} {47}},\ \bibinfo {pages}
  {7191--7193} (\bibinfo {year} {2011})}\BibitemShut {NoStop}%
\bibitem [{\citenamefont {Han}\ \emph {et~al.}(2011)\citenamefont {Han},
  \citenamefont {Feng}, \citenamefont {Sun}, \citenamefont {Li}, \citenamefont
  {Han},\ and\ \citenamefont {Wang}}]{han2011wormlike}%
  \BibitemOpen
  \bibfield  {author} {\bibinfo {author} {\bibfnamefont {Y.}~\bibnamefont
  {Han}}, \bibinfo {author} {\bibfnamefont {Y.}~\bibnamefont {Feng}}, \bibinfo
  {author} {\bibfnamefont {H.}~\bibnamefont {Sun}}, \bibinfo {author}
  {\bibfnamefont {Z.}~\bibnamefont {Li}}, \bibinfo {author} {\bibfnamefont
  {Y.}~\bibnamefont {Han}}, \ and\ \bibinfo {author} {\bibfnamefont
  {H.}~\bibnamefont {Wang}},\ }\bibfield  {title} {\enquote {\bibinfo {title}
  {Wormlike micelles formed by sodium erucate in the presence of a
  tetraalkylammonium hydrotrope},}\ }\href@noop {} {\bibfield  {journal}
  {\bibinfo  {journal} {The Journal of Physical Chemistry B}\ }\textbf
  {\bibinfo {volume} {115}},\ \bibinfo {pages} {6893--6902} (\bibinfo {year}
  {2011})}\BibitemShut {NoStop}%
\bibitem [{\citenamefont {Raghavan}\ and\ \citenamefont
  {Douglas}(2012)}]{raghavan2012conundrum}%
  \BibitemOpen
  \bibfield  {author} {\bibinfo {author} {\bibfnamefont {S.~R.}\ \bibnamefont
  {Raghavan}}\ and\ \bibinfo {author} {\bibfnamefont {J.~F.}\ \bibnamefont
  {Douglas}},\ }\bibfield  {title} {\enquote {\bibinfo {title} {The conundrum
  of gel formation by molecular nanofibers, wormlike micelles, and filamentous
  proteins: gelation without cross-links?}}\ }\href@noop {} {\bibfield
  {journal} {\bibinfo  {journal} {Soft Matter}\ }\textbf {\bibinfo {volume}
  {8}},\ \bibinfo {pages} {8539--8546} (\bibinfo {year} {2012})}\BibitemShut
  {NoStop}%
\bibitem [{\citenamefont {Raghavan}\ and\ \citenamefont
  {Feng}(2017)}]{raghavan2017wormlike}%
  \BibitemOpen
  \bibfield  {author} {\bibinfo {author} {\bibfnamefont {S.~R.}\ \bibnamefont
  {Raghavan}}\ and\ \bibinfo {author} {\bibfnamefont {Y.}~\bibnamefont
  {Feng}},\ }\bibfield  {title} {\enquote {\bibinfo {title} {Wormlike micelles:
  Solutions, gels, or both?}}\ }in\ \href@noop {} {\emph {\bibinfo {booktitle}
  {Wormlike Micelles}}}\ (\bibinfo  {publisher} {Royal Society of Chemistry},\
  \bibinfo {year} {2017})\ pp.\ \bibinfo {pages} {9--30}\BibitemShut {NoStop}%
\bibitem [{\citenamefont {Chu}, \citenamefont {Dreiss},\ and\ \citenamefont
  {Feng}(2013)}]{chu2013smart}%
  \BibitemOpen
  \bibfield  {author} {\bibinfo {author} {\bibfnamefont {Z.}~\bibnamefont
  {Chu}}, \bibinfo {author} {\bibfnamefont {C.~A.}\ \bibnamefont {Dreiss}}, \
  and\ \bibinfo {author} {\bibfnamefont {Y.}~\bibnamefont {Feng}},\ }\bibfield
  {title} {\enquote {\bibinfo {title} {Smart wormlike micelles},}\ }\href@noop
  {} {\bibfield  {journal} {\bibinfo  {journal} {Chemical Society Reviews}\
  }\textbf {\bibinfo {volume} {42}},\ \bibinfo {pages} {7174--7203} (\bibinfo
  {year} {2013})}\BibitemShut {NoStop}%
\bibitem [{\citenamefont {Lee}\ and\ \citenamefont
  {Mooney}(2001)}]{lee2001hydrogels}%
  \BibitemOpen
  \bibfield  {author} {\bibinfo {author} {\bibfnamefont {K.~Y.}\ \bibnamefont
  {Lee}}\ and\ \bibinfo {author} {\bibfnamefont {D.~J.}\ \bibnamefont
  {Mooney}},\ }\bibfield  {title} {\enquote {\bibinfo {title} {Hydrogels for
  tissue engineering},}\ }\href@noop {} {\bibfield  {journal} {\bibinfo
  {journal} {Chemical Reviews}\ }\textbf {\bibinfo {volume} {101}},\ \bibinfo
  {pages} {1869--1880} (\bibinfo {year} {2001})}\BibitemShut {NoStop}%
\bibitem [{\citenamefont {Raghavan}(2009)}]{raghavan2009distinct}%
  \BibitemOpen
  \bibfield  {author} {\bibinfo {author} {\bibfnamefont {S.~R.}\ \bibnamefont
  {Raghavan}},\ }\bibfield  {title} {\enquote {\bibinfo {title} {Distinct
  character of surfactant gels: a smooth progression from micelles to fibrillar
  networks},}\ }\href@noop {} {\bibfield  {journal} {\bibinfo  {journal}
  {Langmuir}\ }\textbf {\bibinfo {volume} {25}},\ \bibinfo {pages} {8382--8385}
  (\bibinfo {year} {2009})}\BibitemShut {NoStop}%
\bibitem [{\citenamefont {MacKintosh}\ and\ \citenamefont
  {Janmey}(1997)}]{mackintosh1997actin}%
  \BibitemOpen
  \bibfield  {author} {\bibinfo {author} {\bibfnamefont {F.~C.}\ \bibnamefont
  {MacKintosh}}\ and\ \bibinfo {author} {\bibfnamefont {P.~A.}\ \bibnamefont
  {Janmey}},\ }\bibfield  {title} {\enquote {\bibinfo {title} {Actin gels},}\
  }\href@noop {} {\bibfield  {journal} {\bibinfo  {journal} {Current Opinion in
  Solid State and Materials Science}\ }\textbf {\bibinfo {volume} {2}},\
  \bibinfo {pages} {350--357} (\bibinfo {year} {1997})}\BibitemShut {NoStop}%
\bibitem [{\citenamefont {Tung}, \citenamefont {Huang},\ and\ \citenamefont
  {Raghavan}(2008)}]{tung2008self}%
  \BibitemOpen
  \bibfield  {author} {\bibinfo {author} {\bibfnamefont {S.-H.}\ \bibnamefont
  {Tung}}, \bibinfo {author} {\bibfnamefont {Y.-E.}\ \bibnamefont {Huang}}, \
  and\ \bibinfo {author} {\bibfnamefont {S.~R.}\ \bibnamefont {Raghavan}},\
  }\bibfield  {title} {\enquote {\bibinfo {title} {Self-assembled organogels
  obtained by adding minute concentrations of a bile salt to aot reverse
  micelles},}\ }\href@noop {} {\bibfield  {journal} {\bibinfo  {journal} {Soft
  Matter}\ }\textbf {\bibinfo {volume} {4}},\ \bibinfo {pages} {1086--1093}
  (\bibinfo {year} {2008})}\BibitemShut {NoStop}%
\bibitem [{\citenamefont {Bhattacharya}\ and\ \citenamefont
  {Samanta}(2011)}]{bhattacharya2011surfactants}%
  \BibitemOpen
  \bibfield  {author} {\bibinfo {author} {\bibfnamefont {S.}~\bibnamefont
  {Bhattacharya}}\ and\ \bibinfo {author} {\bibfnamefont {S.~K.}\ \bibnamefont
  {Samanta}},\ }\bibfield  {title} {\enquote {\bibinfo {title} {Surfactants
  possessing multiple polar heads. a perspective on their unique aggregation
  behavior and applications},}\ }\href@noop {} {\bibfield  {journal} {\bibinfo
  {journal} {The Journal of Physical Chemistry Letters}\ }\textbf {\bibinfo
  {volume} {2}},\ \bibinfo {pages} {914--920} (\bibinfo {year}
  {2011})}\BibitemShut {NoStop}%
\bibitem [{\citenamefont {Xie}\ \emph {et~al.}(2017)\citenamefont {Xie},
  \citenamefont {Ayoubi}, \citenamefont {Lu}, \citenamefont {Wang},
  \citenamefont {Huang},\ and\ \citenamefont {Wang}}]{xie2017unique}%
  \BibitemOpen
  \bibfield  {author} {\bibinfo {author} {\bibfnamefont {H.}~\bibnamefont
  {Xie}}, \bibinfo {author} {\bibfnamefont {M.~A.}\ \bibnamefont {Ayoubi}},
  \bibinfo {author} {\bibfnamefont {W.}~\bibnamefont {Lu}}, \bibinfo {author}
  {\bibfnamefont {J.}~\bibnamefont {Wang}}, \bibinfo {author} {\bibfnamefont
  {J.}~\bibnamefont {Huang}}, \ and\ \bibinfo {author} {\bibfnamefont
  {W.}~\bibnamefont {Wang}},\ }\bibfield  {title} {\enquote {\bibinfo {title}
  {A unique thermo-induced gel-to-gel transition in a ph-sensitive
  small-molecule hydrogel},}\ }\href@noop {} {\bibfield  {journal} {\bibinfo
  {journal} {Scientific Reports}\ }\textbf {\bibinfo {volume} {7}},\ \bibinfo
  {pages} {1--6} (\bibinfo {year} {2017})}\BibitemShut {NoStop}%
\bibitem [{\citenamefont {Christov}\ \emph {et~al.}(2004)\citenamefont
  {Christov}, \citenamefont {Denkov}, \citenamefont {Kralchevsky},
  \citenamefont {Ananthapadmanabhan},\ and\ \citenamefont
  {Lips}}]{christov2004synergistic}%
  \BibitemOpen
  \bibfield  {author} {\bibinfo {author} {\bibfnamefont {N.}~\bibnamefont
  {Christov}}, \bibinfo {author} {\bibfnamefont {N.}~\bibnamefont {Denkov}},
  \bibinfo {author} {\bibfnamefont {P.}~\bibnamefont {Kralchevsky}}, \bibinfo
  {author} {\bibfnamefont {K.}~\bibnamefont {Ananthapadmanabhan}}, \ and\
  \bibinfo {author} {\bibfnamefont {A.}~\bibnamefont {Lips}},\ }\bibfield
  {title} {\enquote {\bibinfo {title} {Synergistic sphere-to-rod micelle
  transition in mixed solutions of sodium dodecyl sulfate and cocoamidopropyl
  betaine},}\ }\href@noop {} {\bibfield  {journal} {\bibinfo  {journal}
  {Langmuir}\ }\textbf {\bibinfo {volume} {20}},\ \bibinfo {pages} {565--571}
  (\bibinfo {year} {2004})}\BibitemShut {NoStop}%
\bibitem [{\citenamefont {Zhou}\ \emph {et~al.}(2018)\citenamefont {Zhou},
  \citenamefont {Ma}, \citenamefont {Zhang}, \citenamefont {Wang},
  \citenamefont {Zhang}, \citenamefont {Luan}, \citenamefont {Zhu},\ and\
  \citenamefont {Zhang}}]{zhou2018surface}%
  \BibitemOpen
  \bibfield  {author} {\bibinfo {author} {\bibfnamefont {Z.-H.}\ \bibnamefont
  {Zhou}}, \bibinfo {author} {\bibfnamefont {D.-S.}\ \bibnamefont {Ma}},
  \bibinfo {author} {\bibfnamefont {Q.}~\bibnamefont {Zhang}}, \bibinfo
  {author} {\bibfnamefont {H.-Z.}\ \bibnamefont {Wang}}, \bibinfo {author}
  {\bibfnamefont {L.}~\bibnamefont {Zhang}}, \bibinfo {author} {\bibfnamefont
  {H.-x.}\ \bibnamefont {Luan}}, \bibinfo {author} {\bibfnamefont
  {Y.}~\bibnamefont {Zhu}}, \ and\ \bibinfo {author} {\bibfnamefont
  {L.}~\bibnamefont {Zhang}},\ }\bibfield  {title} {\enquote {\bibinfo {title}
  {Surface dilational rheology of betaine surfactants: Effect of molecular
  structures},}\ }\href@noop {} {\bibfield  {journal} {\bibinfo  {journal}
  {Colloids and Surfaces A: Physicochemical and Engineering Aspects}\ }\textbf
  {\bibinfo {volume} {538}},\ \bibinfo {pages} {739--747} (\bibinfo {year}
  {2018})}\BibitemShut {NoStop}%
\bibitem [{\citenamefont {Goyal}, \citenamefont {Elfring},\ and\ \citenamefont
  {Frigaard}(2017)}]{goyal2017}%
  \BibitemOpen
  \bibfield  {author} {\bibinfo {author} {\bibfnamefont {G.}~\bibnamefont
  {Goyal}}, \bibinfo {author} {\bibfnamefont {G.~J.}\ \bibnamefont {Elfring}},
  \ and\ \bibinfo {author} {\bibfnamefont {I.~A.}\ \bibnamefont {Frigaard}},\
  }\bibfield  {title} {\enquote {\bibinfo {title} {{Rheology and flow studies
  of drag-reducing gravel packing fluids}},}\ }\href {\doibase
  10.1007/s00397-017-1041-0} {\bibfield  {journal} {\bibinfo  {journal}
  {Rheologica Acta}\ }\textbf {\bibinfo {volume} {56}},\ \bibinfo {pages}
  {905--914} (\bibinfo {year} {2017})}\BibitemShut {NoStop}%
\bibitem [{\citenamefont {McCoy}\ \emph {et~al.}(2016)\citenamefont {McCoy},
  \citenamefont {Valiakhmetova}, \citenamefont {Pottage}, \citenamefont
  {Garvey}, \citenamefont {Campo}, \citenamefont {Rehm}, \citenamefont
  {Kuryashov},\ and\ \citenamefont {Tabor}}]{mccoy2016structural}%
  \BibitemOpen
  \bibfield  {author} {\bibinfo {author} {\bibfnamefont {T.~M.}\ \bibnamefont
  {McCoy}}, \bibinfo {author} {\bibfnamefont {A.}~\bibnamefont
  {Valiakhmetova}}, \bibinfo {author} {\bibfnamefont {M.~J.}\ \bibnamefont
  {Pottage}}, \bibinfo {author} {\bibfnamefont {C.~J.}\ \bibnamefont {Garvey}},
  \bibinfo {author} {\bibfnamefont {L.~d.}\ \bibnamefont {Campo}}, \bibinfo
  {author} {\bibfnamefont {C.}~\bibnamefont {Rehm}}, \bibinfo {author}
  {\bibfnamefont {D.~A.}\ \bibnamefont {Kuryashov}}, \ and\ \bibinfo {author}
  {\bibfnamefont {R.~F.}\ \bibnamefont {Tabor}},\ }\bibfield  {title} {\enquote
  {\bibinfo {title} {Structural evolution of wormlike micellar fluids formed by
  erucyl amidopropyl betaine with oil, salts, and surfactants},}\ }\href@noop
  {} {\bibfield  {journal} {\bibinfo  {journal} {Langmuir}\ }\textbf {\bibinfo
  {volume} {32}},\ \bibinfo {pages} {12423--12433} (\bibinfo {year}
  {2016})}\BibitemShut {NoStop}%
\bibitem [{\citenamefont {Beaumont}\ \emph {et~al.}(2013)\citenamefont
  {Beaumont}, \citenamefont {Louvet}, \citenamefont {Divoux}, \citenamefont
  {Fardin}, \citenamefont {Bodiguel}, \citenamefont {Lerouge}, \citenamefont
  {Manneville},\ and\ \citenamefont {Colin}}]{beaumont2013}%
  \BibitemOpen
  \bibfield  {author} {\bibinfo {author} {\bibfnamefont {J.}~\bibnamefont
  {Beaumont}}, \bibinfo {author} {\bibfnamefont {N.}~\bibnamefont {Louvet}},
  \bibinfo {author} {\bibfnamefont {T.}~\bibnamefont {Divoux}}, \bibinfo
  {author} {\bibfnamefont {M.~A.}\ \bibnamefont {Fardin}}, \bibinfo {author}
  {\bibfnamefont {H.}~\bibnamefont {Bodiguel}}, \bibinfo {author}
  {\bibfnamefont {S.}~\bibnamefont {Lerouge}}, \bibinfo {author} {\bibfnamefont
  {S.}~\bibnamefont {Manneville}}, \ and\ \bibinfo {author} {\bibfnamefont
  {A.}~\bibnamefont {Colin}},\ }\bibfield  {title} {\enquote {\bibinfo {title}
  {{Turbulent flows in highly elastic wormlike micelles}},}\ }\href {\doibase
  10.1039/c2sm26760h} {\bibfield  {journal} {\bibinfo  {journal} {Soft Matter}\
  }\textbf {\bibinfo {volume} {9}},\ \bibinfo {pages} {735--749} (\bibinfo
  {year} {2013})}\BibitemShut {NoStop}%
\bibitem [{\citenamefont {Gonz{\'a}lez}\ and\ \citenamefont
  {Kaler}(2005)}]{gonzalez2005cryo}%
  \BibitemOpen
  \bibfield  {author} {\bibinfo {author} {\bibfnamefont {Y.~I.}\ \bibnamefont
  {Gonz{\'a}lez}}\ and\ \bibinfo {author} {\bibfnamefont {E.~W.}\ \bibnamefont
  {Kaler}},\ }\bibfield  {title} {\enquote {\bibinfo {title} {Cryo-tem studies
  of worm-like micellar solutions},}\ }\href@noop {} {\bibfield  {journal}
  {\bibinfo  {journal} {Current Opinion in Colloid \& Interface Science}\
  }\textbf {\bibinfo {volume} {10}},\ \bibinfo {pages} {256--260} (\bibinfo
  {year} {2005})}\BibitemShut {NoStop}%
\bibitem [{\citenamefont {Kesselman}\ and\ \citenamefont
  {Danino}(2017)}]{kesselman2017direct}%
  \BibitemOpen
  \bibfield  {author} {\bibinfo {author} {\bibfnamefont {E.}~\bibnamefont
  {Kesselman}}\ and\ \bibinfo {author} {\bibfnamefont {D.}~\bibnamefont
  {Danino}},\ }\bibfield  {title} {\enquote {\bibinfo {title} {Direct-imaging
  cryo-transmission electron microscopy of wormlike micelles},}\ }in\
  \href@noop {} {\emph {\bibinfo {booktitle} {Wormlike Micelles}}}\ (\bibinfo
  {year} {2017})\ pp.\ \bibinfo {pages} {171--192}\BibitemShut {NoStop}%
\bibitem [{\citenamefont {Clausen}\ \emph {et~al.}(1992)\citenamefont
  {Clausen}, \citenamefont {Vinson}, \citenamefont {Minter}, \citenamefont
  {Davis}, \citenamefont {Talmon},\ and\ \citenamefont
  {Miller}}]{clausen1992viscoelastic}%
  \BibitemOpen
  \bibfield  {author} {\bibinfo {author} {\bibfnamefont {T.}~\bibnamefont
  {Clausen}}, \bibinfo {author} {\bibfnamefont {P.}~\bibnamefont {Vinson}},
  \bibinfo {author} {\bibfnamefont {J.}~\bibnamefont {Minter}}, \bibinfo
  {author} {\bibfnamefont {H.}~\bibnamefont {Davis}}, \bibinfo {author}
  {\bibfnamefont {Y.}~\bibnamefont {Talmon}}, \ and\ \bibinfo {author}
  {\bibfnamefont {W.}~\bibnamefont {Miller}},\ }\bibfield  {title} {\enquote
  {\bibinfo {title} {Viscoelastic micellar solutions: microscopy and
  rheology},}\ }\href@noop {} {\bibfield  {journal} {\bibinfo  {journal} {The
  Journal of Physical Chemistry}\ }\textbf {\bibinfo {volume} {96}},\ \bibinfo
  {pages} {474--484} (\bibinfo {year} {1992})}\BibitemShut {NoStop}%
\bibitem [{\citenamefont {Liu}\ \emph {et~al.}(1996)\citenamefont {Liu},
  \citenamefont {Ramaswamy}, \citenamefont {Mason}, \citenamefont {Gang},\ and\
  \citenamefont {Weitz}}]{liu1996anomalous}%
  \BibitemOpen
  \bibfield  {author} {\bibinfo {author} {\bibfnamefont {A.~J.}\ \bibnamefont
  {Liu}}, \bibinfo {author} {\bibfnamefont {S.}~\bibnamefont {Ramaswamy}},
  \bibinfo {author} {\bibfnamefont {T.}~\bibnamefont {Mason}}, \bibinfo
  {author} {\bibfnamefont {H.}~\bibnamefont {Gang}}, \ and\ \bibinfo {author}
  {\bibfnamefont {D.}~\bibnamefont {Weitz}},\ }\bibfield  {title} {\enquote
  {\bibinfo {title} {Anomalous viscous loss in emulsions},}\ }\href@noop {}
  {\bibfield  {journal} {\bibinfo  {journal} {Physical Review Letters}\
  }\textbf {\bibinfo {volume} {76}},\ \bibinfo {pages} {3017} (\bibinfo {year}
  {1996})}\BibitemShut {NoStop}%
\bibitem [{\citenamefont {Divoux}, \citenamefont {Barentin},\ and\
  \citenamefont {Manneville}(2011)}]{divoux2011stress}%
  \BibitemOpen
  \bibfield  {author} {\bibinfo {author} {\bibfnamefont {T.}~\bibnamefont
  {Divoux}}, \bibinfo {author} {\bibfnamefont {C.}~\bibnamefont {Barentin}}, \
  and\ \bibinfo {author} {\bibfnamefont {S.}~\bibnamefont {Manneville}},\
  }\bibfield  {title} {\enquote {\bibinfo {title} {From stress-induced
  fluidization processes to herschel-bulkley behaviour in simple yield stress
  fluids},}\ }\href@noop {} {\bibfield  {journal} {\bibinfo  {journal} {Soft
  Matter}\ }\textbf {\bibinfo {volume} {7}},\ \bibinfo {pages} {8409--8418}
  (\bibinfo {year} {2011})}\BibitemShut {NoStop}%
\bibitem [{\citenamefont {Conley}\ \emph {et~al.}(2019)\citenamefont {Conley},
  \citenamefont {Zhang}, \citenamefont {Aebischer}, \citenamefont {Harden},\
  and\ \citenamefont {Scheffold}}]{conley2019relationship}%
  \BibitemOpen
  \bibfield  {author} {\bibinfo {author} {\bibfnamefont {G.~M.}\ \bibnamefont
  {Conley}}, \bibinfo {author} {\bibfnamefont {C.}~\bibnamefont {Zhang}},
  \bibinfo {author} {\bibfnamefont {P.}~\bibnamefont {Aebischer}}, \bibinfo
  {author} {\bibfnamefont {J.~L.}\ \bibnamefont {Harden}}, \ and\ \bibinfo
  {author} {\bibfnamefont {F.}~\bibnamefont {Scheffold}},\ }\bibfield  {title}
  {\enquote {\bibinfo {title} {Relationship between rheology and structure of
  interpenetrating, deforming and compressing microgels},}\ }\href@noop {}
  {\bibfield  {journal} {\bibinfo  {journal} {Nature Communications}\ }\textbf
  {\bibinfo {volume} {10}},\ \bibinfo {pages} {1--8} (\bibinfo {year}
  {2019})}\BibitemShut {NoStop}%
\bibitem [{\citenamefont {Migliozzi}\ \emph {et~al.}(2020)\citenamefont
  {Migliozzi}, \citenamefont {Meridiano}, \citenamefont {Angeli},\ and\
  \citenamefont {Mazzei}}]{migliozzi2020investigation}%
  \BibitemOpen
  \bibfield  {author} {\bibinfo {author} {\bibfnamefont {S.}~\bibnamefont
  {Migliozzi}}, \bibinfo {author} {\bibfnamefont {G.}~\bibnamefont
  {Meridiano}}, \bibinfo {author} {\bibfnamefont {P.}~\bibnamefont {Angeli}}, \
  and\ \bibinfo {author} {\bibfnamefont {L.}~\bibnamefont {Mazzei}},\
  }\bibfield  {title} {\enquote {\bibinfo {title} {Investigation of the swollen
  state of carbopol molecules in non-aqueous solvents through rheological
  characterization},}\ }\href@noop {} {\bibfield  {journal} {\bibinfo
  {journal} {Soft Matter}\ } (\bibinfo {year} {2020})}\BibitemShut {NoStop}%
\bibitem [{\citenamefont {Lee}\ \emph {et~al.}(2019)\citenamefont {Lee},
  \citenamefont {Song}, \citenamefont {Subbiah}, \citenamefont {Chung},
  \citenamefont {Choi}, \citenamefont {Park}, \citenamefont {Kim},\ and\
  \citenamefont {Oh}}]{lee2019effect}%
  \BibitemOpen
  \bibfield  {author} {\bibinfo {author} {\bibfnamefont {J.}~\bibnamefont
  {Lee}}, \bibinfo {author} {\bibfnamefont {B.}~\bibnamefont {Song}}, \bibinfo
  {author} {\bibfnamefont {R.}~\bibnamefont {Subbiah}}, \bibinfo {author}
  {\bibfnamefont {J.~J.}\ \bibnamefont {Chung}}, \bibinfo {author}
  {\bibfnamefont {U.~H.}\ \bibnamefont {Choi}}, \bibinfo {author}
  {\bibfnamefont {K.}~\bibnamefont {Park}}, \bibinfo {author} {\bibfnamefont
  {S.-H.}\ \bibnamefont {Kim}}, \ and\ \bibinfo {author} {\bibfnamefont
  {S.~J.}\ \bibnamefont {Oh}},\ }\bibfield  {title} {\enquote {\bibinfo {title}
  {Effect of chain flexibility on cell adhesion: Semi-flexible model-based
  analysis of cell adhesion to hydrogels},}\ }\href@noop {} {\bibfield
  {journal} {\bibinfo  {journal} {Scientific Reports}\ }\textbf {\bibinfo
  {volume} {9}},\ \bibinfo {pages} {1--8} (\bibinfo {year} {2019})}\BibitemShut
  {NoStop}%
\bibitem [{\citenamefont {MacKintosh}, \citenamefont {K{\"a}s},\ and\
  \citenamefont {Janmey}(1995)}]{mackintosh1995}%
  \BibitemOpen
  \bibfield  {author} {\bibinfo {author} {\bibfnamefont {F.}~\bibnamefont
  {MacKintosh}}, \bibinfo {author} {\bibfnamefont {J.}~\bibnamefont {K{\"a}s}},
  \ and\ \bibinfo {author} {\bibfnamefont {P.}~\bibnamefont {Janmey}},\
  }\bibfield  {title} {\enquote {\bibinfo {title} {Elasticity of semiflexible
  biopolymer networks},}\ }\href@noop {} {\bibfield  {journal} {\bibinfo
  {journal} {Physical Review Letters}\ }\textbf {\bibinfo {volume} {75}},\
  \bibinfo {pages} {4425} (\bibinfo {year} {1995})}\BibitemShut {NoStop}%
\bibitem [{\citenamefont {Poling-Skutvik}\ \emph {et~al.}(2020)\citenamefont
  {Poling-Skutvik}, \citenamefont {McEvoy}, \citenamefont {Shenoy},\ and\
  \citenamefont {Osuji}}]{poling2020yielding}%
  \BibitemOpen
  \bibfield  {author} {\bibinfo {author} {\bibfnamefont {R.}~\bibnamefont
  {Poling-Skutvik}}, \bibinfo {author} {\bibfnamefont {E.}~\bibnamefont
  {McEvoy}}, \bibinfo {author} {\bibfnamefont {V.}~\bibnamefont {Shenoy}}, \
  and\ \bibinfo {author} {\bibfnamefont {C.~O.}\ \bibnamefont {Osuji}},\
  }\bibfield  {title} {\enquote {\bibinfo {title} {Yielding and bifurcated
  aging in nanofibrillar networks},}\ }\href@noop {} {\bibfield  {journal}
  {\bibinfo  {journal} {Physical Review Materials}\ }\textbf {\bibinfo {volume}
  {4}},\ \bibinfo {pages} {102601} (\bibinfo {year} {2020})}\BibitemShut
  {NoStop}%
\bibitem [{\citenamefont {Doi}, \citenamefont {Edwards},\ and\ \citenamefont
  {Edwards}(1988)}]{doi1988theory}%
  \BibitemOpen
  \bibfield  {author} {\bibinfo {author} {\bibfnamefont {M.}~\bibnamefont
  {Doi}}, \bibinfo {author} {\bibfnamefont {S.~F.}\ \bibnamefont {Edwards}}, \
  and\ \bibinfo {author} {\bibfnamefont {S.~F.}\ \bibnamefont {Edwards}},\
  }\href@noop {} {\emph {\bibinfo {title} {The Theory of Polymer Dynamics}}},\
  Vol.~\bibinfo {volume} {73}\ (\bibinfo  {publisher} {Oxford University
  Press},\ \bibinfo {year} {1988})\BibitemShut {NoStop}%
\bibitem [{\citenamefont {Buchanan}\ \emph {et~al.}(2005)\citenamefont
  {Buchanan}, \citenamefont {Atakhorrami}, \citenamefont {Palierne},
  \citenamefont {MacKintosh},\ and\ \citenamefont
  {Schmidt}}]{buchanan2005high}%
  \BibitemOpen
  \bibfield  {author} {\bibinfo {author} {\bibfnamefont {M.}~\bibnamefont
  {Buchanan}}, \bibinfo {author} {\bibfnamefont {M.}~\bibnamefont
  {Atakhorrami}}, \bibinfo {author} {\bibfnamefont {J.}~\bibnamefont
  {Palierne}}, \bibinfo {author} {\bibfnamefont {F.}~\bibnamefont
  {MacKintosh}}, \ and\ \bibinfo {author} {\bibfnamefont {C.}~\bibnamefont
  {Schmidt}},\ }\bibfield  {title} {\enquote {\bibinfo {title} {High-frequency
  microrheology of wormlike micelles},}\ }\href@noop {} {\bibfield  {journal}
  {\bibinfo  {journal} {Physical Review E}\ }\textbf {\bibinfo {volume} {72}},\
  \bibinfo {pages} {011504} (\bibinfo {year} {2005})}\BibitemShut {NoStop}%
\bibitem [{\citenamefont {Granek}\ and\ \citenamefont
  {Cates}(1992)}]{granek1992stress}%
  \BibitemOpen
  \bibfield  {author} {\bibinfo {author} {\bibfnamefont {R.}~\bibnamefont
  {Granek}}\ and\ \bibinfo {author} {\bibfnamefont {M.}~\bibnamefont {Cates}},\
  }\bibfield  {title} {\enquote {\bibinfo {title} {Stress relaxation in living
  polymers: Results from a poisson renewal model},}\ }\href@noop {} {\bibfield
  {journal} {\bibinfo  {journal} {The Journal of Chemical Physics}\ }\textbf
  {\bibinfo {volume} {96}},\ \bibinfo {pages} {4758--4767} (\bibinfo {year}
  {1992})}\BibitemShut {NoStop}%
\bibitem [{\citenamefont {Cates}(1988)}]{cates1988dynamics}%
  \BibitemOpen
  \bibfield  {author} {\bibinfo {author} {\bibfnamefont {M.}~\bibnamefont
  {Cates}},\ }\bibfield  {title} {\enquote {\bibinfo {title} {Dynamics of
  living polymers and flexible surfactant micelles: scaling laws for
  dilution},}\ }\href@noop {} {\bibfield  {journal} {\bibinfo  {journal}
  {Journal de Physique}\ }\textbf {\bibinfo {volume} {49}},\ \bibinfo {pages}
  {1593--1600} (\bibinfo {year} {1988})}\BibitemShut {NoStop}%
\bibitem [{\citenamefont {Kern}\ \emph {et~al.}(1994)\citenamefont {Kern},
  \citenamefont {Lequeux}, \citenamefont {Zana},\ and\ \citenamefont
  {Candau}}]{kern1994dynamic}%
  \BibitemOpen
  \bibfield  {author} {\bibinfo {author} {\bibfnamefont {F.}~\bibnamefont
  {Kern}}, \bibinfo {author} {\bibfnamefont {F.}~\bibnamefont {Lequeux}},
  \bibinfo {author} {\bibfnamefont {R.}~\bibnamefont {Zana}}, \ and\ \bibinfo
  {author} {\bibfnamefont {S.}~\bibnamefont {Candau}},\ }\bibfield  {title}
  {\enquote {\bibinfo {title} {Dynamic properties of salt-free viscoelastic
  micellar solutions},}\ }\href@noop {} {\bibfield  {journal} {\bibinfo
  {journal} {Langmuir}\ }\textbf {\bibinfo {volume} {10}},\ \bibinfo {pages}
  {1714--1723} (\bibinfo {year} {1994})}\BibitemShut {NoStop}%
\bibitem [{\citenamefont {Soltero}, \citenamefont {Puig},\ and\ \citenamefont
  {Manero}(1996)}]{soltero1996rheology}%
  \BibitemOpen
  \bibfield  {author} {\bibinfo {author} {\bibfnamefont {J.}~\bibnamefont
  {Soltero}}, \bibinfo {author} {\bibfnamefont {J.}~\bibnamefont {Puig}}, \
  and\ \bibinfo {author} {\bibfnamefont {O.}~\bibnamefont {Manero}},\
  }\bibfield  {title} {\enquote {\bibinfo {title} {Rheology of the
  cetyltrimethylammonium tosilate- water system. 2. linear viscoelastic
  regime},}\ }\href@noop {} {\bibfield  {journal} {\bibinfo  {journal}
  {Langmuir}\ }\textbf {\bibinfo {volume} {12}},\ \bibinfo {pages} {2654--2662}
  (\bibinfo {year} {1996})}\BibitemShut {NoStop}%
\bibitem [{\citenamefont {Candau}\ and\ \citenamefont
  {Oda}(2001)}]{candau2001linear}%
  \BibitemOpen
  \bibfield  {author} {\bibinfo {author} {\bibfnamefont {S.}~\bibnamefont
  {Candau}}\ and\ \bibinfo {author} {\bibfnamefont {R.}~\bibnamefont {Oda}},\
  }\bibfield  {title} {\enquote {\bibinfo {title} {Linear viscoelasticity of
  salt-free wormlike micellar solutions},}\ }\href@noop {} {\bibfield
  {journal} {\bibinfo  {journal} {Colloids and Surfaces A: Physicochemical and
  Engineering Aspects}\ }\textbf {\bibinfo {volume} {183}},\ \bibinfo {pages}
  {5--14} (\bibinfo {year} {2001})}\BibitemShut {NoStop}%
\bibitem [{\citenamefont {Zhang}\ \emph {et~al.}(2013)\citenamefont {Zhang},
  \citenamefont {Luo}, \citenamefont {Wang}, \citenamefont {Zhang},\ and\
  \citenamefont {Feng}}]{zhang2013single}%
  \BibitemOpen
  \bibfield  {author} {\bibinfo {author} {\bibfnamefont {Y.}~\bibnamefont
  {Zhang}}, \bibinfo {author} {\bibfnamefont {Y.}~\bibnamefont {Luo}}, \bibinfo
  {author} {\bibfnamefont {Y.}~\bibnamefont {Wang}}, \bibinfo {author}
  {\bibfnamefont {J.}~\bibnamefont {Zhang}}, \ and\ \bibinfo {author}
  {\bibfnamefont {Y.}~\bibnamefont {Feng}},\ }\bibfield  {title} {\enquote
  {\bibinfo {title} {Single-component wormlike micellar system formed by a
  carboxylbetaine surfactant with c22 saturated tail},}\ }\href@noop {}
  {\bibfield  {journal} {\bibinfo  {journal} {Colloids and Surfaces A:
  Physicochemical and Engineering Aspects}\ }\textbf {\bibinfo {volume}
  {436}},\ \bibinfo {pages} {71--79} (\bibinfo {year} {2013})}\BibitemShut
  {NoStop}%
\bibitem [{\citenamefont {Wang}\ \emph {et~al.}(2017)\citenamefont {Wang},
  \citenamefont {Feng}, \citenamefont {Agrawal},\ and\ \citenamefont
  {Raghavan}}]{wang2017wormlike}%
  \BibitemOpen
  \bibfield  {author} {\bibinfo {author} {\bibfnamefont {J.}~\bibnamefont
  {Wang}}, \bibinfo {author} {\bibfnamefont {Y.}~\bibnamefont {Feng}}, \bibinfo
  {author} {\bibfnamefont {N.~R.}\ \bibnamefont {Agrawal}}, \ and\ \bibinfo
  {author} {\bibfnamefont {S.~R.}\ \bibnamefont {Raghavan}},\ }\bibfield
  {title} {\enquote {\bibinfo {title} {Wormlike micelles versus water-soluble
  polymers as rheology-modifiers: similarities and differences},}\ }\href@noop
  {} {\bibfield  {journal} {\bibinfo  {journal} {Physical Chemistry Chemical
  Physics}\ }\textbf {\bibinfo {volume} {19}},\ \bibinfo {pages} {24458--24466}
  (\bibinfo {year} {2017})}\BibitemShut {NoStop}%
\bibitem [{\citenamefont {Sarmiento-Gomez}, \citenamefont {Lopez-Diaz},\ and\
  \citenamefont {Castillo}(2010)}]{sarmiento2010microrheology}%
  \BibitemOpen
  \bibfield  {author} {\bibinfo {author} {\bibfnamefont {E.}~\bibnamefont
  {Sarmiento-Gomez}}, \bibinfo {author} {\bibfnamefont {D.}~\bibnamefont
  {Lopez-Diaz}}, \ and\ \bibinfo {author} {\bibfnamefont {R.}~\bibnamefont
  {Castillo}},\ }\bibfield  {title} {\enquote {\bibinfo {title} {Microrheology
  and characteristic lengths in wormlike micelles made of a zwitterionic
  surfactant and sds in brine},}\ }\href@noop {} {\bibfield  {journal}
  {\bibinfo  {journal} {The Journal of Physical Chemistry B}\ }\textbf
  {\bibinfo {volume} {114}},\ \bibinfo {pages} {12193--12202} (\bibinfo {year}
  {2010})}\BibitemShut {NoStop}%
\bibitem [{\citenamefont {Khatory}\ \emph {et~al.}(1993)\citenamefont
  {Khatory}, \citenamefont {Lequeux}, \citenamefont {Kern},\ and\ \citenamefont
  {Candau}}]{khatory1993linear}%
  \BibitemOpen
  \bibfield  {author} {\bibinfo {author} {\bibfnamefont {A.}~\bibnamefont
  {Khatory}}, \bibinfo {author} {\bibfnamefont {F.}~\bibnamefont {Lequeux}},
  \bibinfo {author} {\bibfnamefont {F.}~\bibnamefont {Kern}}, \ and\ \bibinfo
  {author} {\bibfnamefont {S.}~\bibnamefont {Candau}},\ }\bibfield  {title}
  {\enquote {\bibinfo {title} {Linear and nonlinear viscoelasticity of
  semidilute solutions of wormlike micelles at high salt content},}\
  }\href@noop {} {\bibfield  {journal} {\bibinfo  {journal} {Langmuir}\
  }\textbf {\bibinfo {volume} {9}},\ \bibinfo {pages} {1456--1464} (\bibinfo
  {year} {1993})}\BibitemShut {NoStop}%
\bibitem [{\citenamefont {Kern}\ \emph {et~al.}(1992)\citenamefont {Kern},
  \citenamefont {Lemarechal}, \citenamefont {Candau},\ and\ \citenamefont
  {Cates}}]{kern1992rheological}%
  \BibitemOpen
  \bibfield  {author} {\bibinfo {author} {\bibfnamefont {F.}~\bibnamefont
  {Kern}}, \bibinfo {author} {\bibfnamefont {P.}~\bibnamefont {Lemarechal}},
  \bibinfo {author} {\bibfnamefont {S.}~\bibnamefont {Candau}}, \ and\ \bibinfo
  {author} {\bibfnamefont {M.}~\bibnamefont {Cates}},\ }\bibfield  {title}
  {\enquote {\bibinfo {title} {Rheological properties of semidilute and
  concentrated aqueous solutions of cetyltrimethylammonium bromide in the
  presence of potassium bromide},}\ }\href@noop {} {\bibfield  {journal}
  {\bibinfo  {journal} {Langmuir}\ }\textbf {\bibinfo {volume} {8}},\ \bibinfo
  {pages} {437--440} (\bibinfo {year} {1992})}\BibitemShut {NoStop}%
\bibitem [{\citenamefont {Berret}, \citenamefont {Appell},\ and\ \citenamefont
  {Porte}(1993)}]{berret1993linear}%
  \BibitemOpen
  \bibfield  {author} {\bibinfo {author} {\bibfnamefont {J.~F.}\ \bibnamefont
  {Berret}}, \bibinfo {author} {\bibfnamefont {J.}~\bibnamefont {Appell}}, \
  and\ \bibinfo {author} {\bibfnamefont {G.}~\bibnamefont {Porte}},\ }\bibfield
   {title} {\enquote {\bibinfo {title} {Linear rheology of entangled wormlike
  micelles},}\ }\href@noop {} {\bibfield  {journal} {\bibinfo  {journal}
  {Langmuir}\ }\textbf {\bibinfo {volume} {9}},\ \bibinfo {pages} {2851--2854}
  (\bibinfo {year} {1993})}\BibitemShut {NoStop}%
\bibitem [{\citenamefont {Tan}\ \emph {et~al.}(2021)\citenamefont {Tan},
  \citenamefont {Zou}, \citenamefont {Weaver},\ and\ \citenamefont
  {Larson}}]{tan2021determining}%
  \BibitemOpen
  \bibfield  {author} {\bibinfo {author} {\bibfnamefont {G.}~\bibnamefont
  {Tan}}, \bibinfo {author} {\bibfnamefont {W.}~\bibnamefont {Zou}}, \bibinfo
  {author} {\bibfnamefont {M.}~\bibnamefont {Weaver}}, \ and\ \bibinfo {author}
  {\bibfnamefont {R.~G.}\ \bibnamefont {Larson}},\ }\bibfield  {title}
  {\enquote {\bibinfo {title} {Determining threadlike micelle lengths from
  rheometry},}\ }\href@noop {} {\bibfield  {journal} {\bibinfo  {journal}
  {Journal of Rheology}\ }\textbf {\bibinfo {volume} {65}},\ \bibinfo {pages}
  {59--71} (\bibinfo {year} {2021})}\BibitemShut {NoStop}%
\bibitem [{\citenamefont {Zou}\ and\ \citenamefont
  {Larson}(2014)}]{zou2014mesoscopic}%
  \BibitemOpen
  \bibfield  {author} {\bibinfo {author} {\bibfnamefont {W.}~\bibnamefont
  {Zou}}\ and\ \bibinfo {author} {\bibfnamefont {R.~G.}\ \bibnamefont
  {Larson}},\ }\bibfield  {title} {\enquote {\bibinfo {title} {A mesoscopic
  simulation method for predicting the rheology of semi-dilute wormlike
  micellar solutions},}\ }\href@noop {} {\bibfield  {journal} {\bibinfo
  {journal} {Journal of Rheology}\ }\textbf {\bibinfo {volume} {58}},\ \bibinfo
  {pages} {681--721} (\bibinfo {year} {2014})}\BibitemShut {NoStop}%
\bibitem [{\citenamefont {Willenbacher}\ \emph {et~al.}(2007)\citenamefont
  {Willenbacher}, \citenamefont {Oelschlaeger}, \citenamefont {Schopferer},
  \citenamefont {Fischer}, \citenamefont {Cardinaux},\ and\ \citenamefont
  {Scheffold}}]{willenbacher2007broad}%
  \BibitemOpen
  \bibfield  {author} {\bibinfo {author} {\bibfnamefont {N.}~\bibnamefont
  {Willenbacher}}, \bibinfo {author} {\bibfnamefont {C.}~\bibnamefont
  {Oelschlaeger}}, \bibinfo {author} {\bibfnamefont {M.}~\bibnamefont
  {Schopferer}}, \bibinfo {author} {\bibfnamefont {P.}~\bibnamefont {Fischer}},
  \bibinfo {author} {\bibfnamefont {F.}~\bibnamefont {Cardinaux}}, \ and\
  \bibinfo {author} {\bibfnamefont {F.}~\bibnamefont {Scheffold}},\ }\bibfield
  {title} {\enquote {\bibinfo {title} {Broad bandwidth optical and mechanical
  rheometry of wormlike micelle solutions},}\ }\href@noop {} {\bibfield
  {journal} {\bibinfo  {journal} {Physical Review Letters}\ }\textbf {\bibinfo
  {volume} {99}},\ \bibinfo {pages} {068302} (\bibinfo {year}
  {2007})}\BibitemShut {NoStop}%
\bibitem [{\citenamefont {Couillet}\ \emph {et~al.}(2004)\citenamefont
  {Couillet}, \citenamefont {Hughes}, \citenamefont {Maitland}, \citenamefont
  {Candau},\ and\ \citenamefont {Candau}}]{couillet2004growth}%
  \BibitemOpen
  \bibfield  {author} {\bibinfo {author} {\bibfnamefont {I.}~\bibnamefont
  {Couillet}}, \bibinfo {author} {\bibfnamefont {T.}~\bibnamefont {Hughes}},
  \bibinfo {author} {\bibfnamefont {G.}~\bibnamefont {Maitland}}, \bibinfo
  {author} {\bibfnamefont {F.}~\bibnamefont {Candau}}, \ and\ \bibinfo {author}
  {\bibfnamefont {S.~J.}\ \bibnamefont {Candau}},\ }\bibfield  {title}
  {\enquote {\bibinfo {title} {Growth and scission energy of wormlike micelles
  formed by a cationic surfactant with long unsaturated tails},}\ }\href@noop
  {} {\bibfield  {journal} {\bibinfo  {journal} {Langmuir}\ }\textbf {\bibinfo
  {volume} {20}},\ \bibinfo {pages} {9541--9550} (\bibinfo {year}
  {2004})}\BibitemShut {NoStop}%
\bibitem [{\citenamefont {Jiang}\ \emph {et~al.}(2018)\citenamefont {Jiang},
  \citenamefont {Vogtt}, \citenamefont {Thomas}, \citenamefont {Beaucage},\
  and\ \citenamefont {Mulderig}}]{jiang2018enthalpy}%
  \BibitemOpen
  \bibfield  {author} {\bibinfo {author} {\bibfnamefont {H.}~\bibnamefont
  {Jiang}}, \bibinfo {author} {\bibfnamefont {K.}~\bibnamefont {Vogtt}},
  \bibinfo {author} {\bibfnamefont {J.~B.}\ \bibnamefont {Thomas}}, \bibinfo
  {author} {\bibfnamefont {G.}~\bibnamefont {Beaucage}}, \ and\ \bibinfo
  {author} {\bibfnamefont {A.}~\bibnamefont {Mulderig}},\ }\bibfield  {title}
  {\enquote {\bibinfo {title} {Enthalpy and entropy of scission in wormlike
  micelles},}\ }\href@noop {} {\bibfield  {journal} {\bibinfo  {journal}
  {Langmuir}\ }\textbf {\bibinfo {volume} {34}},\ \bibinfo {pages}
  {13956--13964} (\bibinfo {year} {2018})}\BibitemShut {NoStop}%
\bibitem [{\citenamefont {Storm}\ \emph {et~al.}(2005)\citenamefont {Storm},
  \citenamefont {Pastore}, \citenamefont {MacKintosh}, \citenamefont
  {Lubensky},\ and\ \citenamefont {Janmey}}]{storm2005nonlinear}%
  \BibitemOpen
  \bibfield  {author} {\bibinfo {author} {\bibfnamefont {C.}~\bibnamefont
  {Storm}}, \bibinfo {author} {\bibfnamefont {J.~J.}\ \bibnamefont {Pastore}},
  \bibinfo {author} {\bibfnamefont {F.~C.}\ \bibnamefont {MacKintosh}},
  \bibinfo {author} {\bibfnamefont {T.~C.}\ \bibnamefont {Lubensky}}, \ and\
  \bibinfo {author} {\bibfnamefont {P.~A.}\ \bibnamefont {Janmey}},\ }\bibfield
   {title} {\enquote {\bibinfo {title} {Nonlinear elasticity in biological
  gels},}\ }\href@noop {} {\bibfield  {journal} {\bibinfo  {journal} {Nature}\
  }\textbf {\bibinfo {volume} {435}},\ \bibinfo {pages} {191--194} (\bibinfo
  {year} {2005})}\BibitemShut {NoStop}%
\bibitem [{\citenamefont {Tung}\ and\ \citenamefont
  {Raghavan}(2008)}]{tung2008strain}%
  \BibitemOpen
  \bibfield  {author} {\bibinfo {author} {\bibfnamefont {S.-H.}\ \bibnamefont
  {Tung}}\ and\ \bibinfo {author} {\bibfnamefont {S.~R.}\ \bibnamefont
  {Raghavan}},\ }\bibfield  {title} {\enquote {\bibinfo {title}
  {Strain-stiffening response in transient networks formed by reverse wormlike
  micelles},}\ }\href@noop {} {\bibfield  {journal} {\bibinfo  {journal}
  {Langmuir}\ }\textbf {\bibinfo {volume} {24}},\ \bibinfo {pages} {8405--8408}
  (\bibinfo {year} {2008})}\BibitemShut {NoStop}%
\bibitem [{\citenamefont {Donley}\ \emph {et~al.}(2020)\citenamefont {Donley},
  \citenamefont {Singh}, \citenamefont {Shetty},\ and\ \citenamefont
  {Rogers}}]{donley2020elucidating}%
  \BibitemOpen
  \bibfield  {author} {\bibinfo {author} {\bibfnamefont {G.~J.}\ \bibnamefont
  {Donley}}, \bibinfo {author} {\bibfnamefont {P.~K.}\ \bibnamefont {Singh}},
  \bibinfo {author} {\bibfnamefont {A.}~\bibnamefont {Shetty}}, \ and\ \bibinfo
  {author} {\bibfnamefont {S.~A.}\ \bibnamefont {Rogers}},\ }\bibfield  {title}
  {\enquote {\bibinfo {title} {Elucidating the g ?overshoot in soft materials
  with a yield transition via a time-resolved experimental strain
  decomposition},}\ }\href@noop {} {\bibfield  {journal} {\bibinfo  {journal}
  {Proceedings of the National Academy of Sciences}\ }\textbf {\bibinfo
  {volume} {117}},\ \bibinfo {pages} {21945--21952} (\bibinfo {year}
  {2020})}\BibitemShut {NoStop}%
\bibitem [{\citenamefont {Andrade}(1910)}]{andrade1910viscous}%
  \BibitemOpen
  \bibfield  {author} {\bibinfo {author} {\bibfnamefont {E.~N. D.~C.}\
  \bibnamefont {Andrade}},\ }\bibfield  {title} {\enquote {\bibinfo {title} {On
  the viscous flow in metals, and allied phenomena},}\ }\href@noop {}
  {\bibfield  {journal} {\bibinfo  {journal} {Proceedings of the Royal Society
  of London. Series A, Containing Papers of a Mathematical and Physical
  Character}\ }\textbf {\bibinfo {volume} {84}},\ \bibinfo {pages} {1--12}
  (\bibinfo {year} {1910})}\BibitemShut {NoStop}%
\bibitem [{\citenamefont {Miguel}\ \emph {et~al.}(2002)\citenamefont {Miguel},
  \citenamefont {Vespignani}, \citenamefont {Zaiser},\ and\ \citenamefont
  {Zapperi}}]{miguel2002dislocation}%
  \BibitemOpen
  \bibfield  {author} {\bibinfo {author} {\bibfnamefont {M.-C.}\ \bibnamefont
  {Miguel}}, \bibinfo {author} {\bibfnamefont {A.}~\bibnamefont {Vespignani}},
  \bibinfo {author} {\bibfnamefont {M.}~\bibnamefont {Zaiser}}, \ and\ \bibinfo
  {author} {\bibfnamefont {S.}~\bibnamefont {Zapperi}},\ }\bibfield  {title}
  {\enquote {\bibinfo {title} {Dislocation jamming and andrade creep},}\
  }\href@noop {} {\bibfield  {journal} {\bibinfo  {journal} {Physical Review
  Letters}\ }\textbf {\bibinfo {volume} {89}},\ \bibinfo {pages} {165501}
  (\bibinfo {year} {2002})}\BibitemShut {NoStop}%
\bibitem [{\citenamefont {Leocmach}\ \emph {et~al.}(2014)\citenamefont
  {Leocmach}, \citenamefont {Perge}, \citenamefont {Divoux},\ and\
  \citenamefont {Manneville}}]{leocmach2014creep}%
  \BibitemOpen
  \bibfield  {author} {\bibinfo {author} {\bibfnamefont {M.}~\bibnamefont
  {Leocmach}}, \bibinfo {author} {\bibfnamefont {C.}~\bibnamefont {Perge}},
  \bibinfo {author} {\bibfnamefont {T.}~\bibnamefont {Divoux}}, \ and\ \bibinfo
  {author} {\bibfnamefont {S.}~\bibnamefont {Manneville}},\ }\bibfield  {title}
  {\enquote {\bibinfo {title} {Creep and fracture of a protein gel under
  stress},}\ }\href@noop {} {\bibfield  {journal} {\bibinfo  {journal}
  {Physical Review Letters}\ }\textbf {\bibinfo {volume} {113}},\ \bibinfo
  {pages} {038303} (\bibinfo {year} {2014})}\BibitemShut {NoStop}%
\bibitem [{\citenamefont {Skrzeszewska}\ \emph {et~al.}(2010)\citenamefont
  {Skrzeszewska}, \citenamefont {Sprakel}, \citenamefont {de~Wolf},
  \citenamefont {Fokkink}, \citenamefont {Cohen~Stuart},\ and\ \citenamefont
  {van~der Gucht}}]{skrzeszewska2010fracture}%
  \BibitemOpen
  \bibfield  {author} {\bibinfo {author} {\bibfnamefont {P.~J.}\ \bibnamefont
  {Skrzeszewska}}, \bibinfo {author} {\bibfnamefont {J.}~\bibnamefont
  {Sprakel}}, \bibinfo {author} {\bibfnamefont {F.~A.}\ \bibnamefont
  {de~Wolf}}, \bibinfo {author} {\bibfnamefont {R.}~\bibnamefont {Fokkink}},
  \bibinfo {author} {\bibfnamefont {M.~A.}\ \bibnamefont {Cohen~Stuart}}, \
  and\ \bibinfo {author} {\bibfnamefont {J.}~\bibnamefont {van~der Gucht}},\
  }\bibfield  {title} {\enquote {\bibinfo {title} {Fracture and self-healing in
  a well-defined self-assembled polymer network},}\ }\href@noop {} {\bibfield
  {journal} {\bibinfo  {journal} {Macromolecules}\ }\textbf {\bibinfo {volume}
  {43}},\ \bibinfo {pages} {3542--3548} (\bibinfo {year} {2010})}\BibitemShut
  {NoStop}%
\bibitem [{\citenamefont {Olsson}\ \emph {et~al.}(2010)\citenamefont {Olsson},
  \citenamefont {B{\"o}rjesson}, \citenamefont {Angelico}, \citenamefont
  {Ceglie},\ and\ \citenamefont {Palazzo}}]{olsson2010slow}%
  \BibitemOpen
  \bibfield  {author} {\bibinfo {author} {\bibfnamefont {U.}~\bibnamefont
  {Olsson}}, \bibinfo {author} {\bibfnamefont {J.}~\bibnamefont
  {B{\"o}rjesson}}, \bibinfo {author} {\bibfnamefont {R.}~\bibnamefont
  {Angelico}}, \bibinfo {author} {\bibfnamefont {A.}~\bibnamefont {Ceglie}}, \
  and\ \bibinfo {author} {\bibfnamefont {G.}~\bibnamefont {Palazzo}},\
  }\bibfield  {title} {\enquote {\bibinfo {title} {Slow dynamics of wormlike
  micelles},}\ }\href@noop {} {\bibfield  {journal} {\bibinfo  {journal} {Soft
  Matter}\ }\textbf {\bibinfo {volume} {6}},\ \bibinfo {pages} {1769--1777}
  (\bibinfo {year} {2010})}\BibitemShut {NoStop}%
\bibitem [{\citenamefont {Ligoure}\ and\ \citenamefont
  {Mora}(2013)}]{ligoure2013fractures}%
  \BibitemOpen
  \bibfield  {author} {\bibinfo {author} {\bibfnamefont {C.}~\bibnamefont
  {Ligoure}}\ and\ \bibinfo {author} {\bibfnamefont {S.}~\bibnamefont {Mora}},\
  }\bibfield  {title} {\enquote {\bibinfo {title} {Fractures in complex fluids:
  The case of transient networks},}\ }\href@noop {} {\bibfield  {journal}
  {\bibinfo  {journal} {Rheologica Acta}\ }\textbf {\bibinfo {volume} {52}},\
  \bibinfo {pages} {91--114} (\bibinfo {year} {2013})}\BibitemShut {NoStop}%
\bibitem [{\citenamefont {Cates}(1990)}]{cates1990nonlinear}%
  \BibitemOpen
  \bibfield  {author} {\bibinfo {author} {\bibfnamefont {M.}~\bibnamefont
  {Cates}},\ }\bibfield  {title} {\enquote {\bibinfo {title} {Nonlinear
  viscoelasticity of wormlike micelles (and other reversibly breakable
  polymers)},}\ }\href@noop {} {\bibfield  {journal} {\bibinfo  {journal}
  {Journal of Physical Chemistry}\ }\textbf {\bibinfo {volume} {94}},\ \bibinfo
  {pages} {371--375} (\bibinfo {year} {1990})}\BibitemShut {NoStop}%
\bibitem [{\citenamefont {Rothstein}(2003)}]{rothstein2003transient}%
  \BibitemOpen
  \bibfield  {author} {\bibinfo {author} {\bibfnamefont {J.~P.}\ \bibnamefont
  {Rothstein}},\ }\bibfield  {title} {\enquote {\bibinfo {title} {Transient
  extensional rheology of wormlike micelle solutions},}\ }\href@noop {}
  {\bibfield  {journal} {\bibinfo  {journal} {Journal of Rheology}\ }\textbf
  {\bibinfo {volume} {47}},\ \bibinfo {pages} {1227--1247} (\bibinfo {year}
  {2003})}\BibitemShut {NoStop}%
\bibitem [{\citenamefont {Mandal}\ and\ \citenamefont
  {Larson}(2018)}]{mandal2018stretch}%
  \BibitemOpen
  \bibfield  {author} {\bibinfo {author} {\bibfnamefont {T.}~\bibnamefont
  {Mandal}}\ and\ \bibinfo {author} {\bibfnamefont {R.~G.}\ \bibnamefont
  {Larson}},\ }\bibfield  {title} {\enquote {\bibinfo {title} {Stretch and
  breakage of wormlike micelles under uniaxial strain: A simulation study and
  comparison with experimental results},}\ }\href@noop {} {\bibfield  {journal}
  {\bibinfo  {journal} {Langmuir}\ }\textbf {\bibinfo {volume} {34}},\ \bibinfo
  {pages} {12600--12608} (\bibinfo {year} {2018})}\BibitemShut {NoStop}%
\bibitem [{\citenamefont {Cheng}\ and\ \citenamefont
  {Wang}(2012)}]{cheng2012shear}%
  \BibitemOpen
  \bibfield  {author} {\bibinfo {author} {\bibfnamefont {S.}~\bibnamefont
  {Cheng}}\ and\ \bibinfo {author} {\bibfnamefont {S.-Q.}\ \bibnamefont
  {Wang}},\ }\bibfield  {title} {\enquote {\bibinfo {title} {Is shear banding a
  metastable property of well-entangled polymer solutions?}}\ }\href@noop {}
  {\bibfield  {journal} {\bibinfo  {journal} {Journal of Rheology}\ }\textbf
  {\bibinfo {volume} {56}},\ \bibinfo {pages} {1413--1428} (\bibinfo {year}
  {2012})}\BibitemShut {NoStop}%
\bibitem [{\citenamefont {Erk}\ \emph {et~al.}(2012)\citenamefont {Erk},
  \citenamefont {Martin}, \citenamefont {Hu},\ and\ \citenamefont
  {Shull}}]{erk2012extreme}%
  \BibitemOpen
  \bibfield  {author} {\bibinfo {author} {\bibfnamefont {K.~A.}\ \bibnamefont
  {Erk}}, \bibinfo {author} {\bibfnamefont {J.~D.}\ \bibnamefont {Martin}},
  \bibinfo {author} {\bibfnamefont {Y.~T.}\ \bibnamefont {Hu}}, \ and\ \bibinfo
  {author} {\bibfnamefont {K.~R.}\ \bibnamefont {Shull}},\ }\bibfield  {title}
  {\enquote {\bibinfo {title} {Extreme strain localization and sliding friction
  in physically associating polymer gels},}\ }\href@noop {} {\bibfield
  {journal} {\bibinfo  {journal} {Langmuir}\ }\textbf {\bibinfo {volume}
  {28}},\ \bibinfo {pages} {4472--4478} (\bibinfo {year} {2012})}\BibitemShut
  {NoStop}%
\bibitem [{\citenamefont {Berret}\ and\ \citenamefont
  {S{\'e}r{\'e}ro}(2001)}]{berret2001evidence}%
  \BibitemOpen
  \bibfield  {author} {\bibinfo {author} {\bibfnamefont {J.-F.}\ \bibnamefont
  {Berret}}\ and\ \bibinfo {author} {\bibfnamefont {Y.}~\bibnamefont
  {S{\'e}r{\'e}ro}},\ }\bibfield  {title} {\enquote {\bibinfo {title} {Evidence
  of shear-induced fluid fracture in telechelic polymer networks},}\
  }\href@noop {} {\bibfield  {journal} {\bibinfo  {journal} {Physical Review
  Letters}\ }\textbf {\bibinfo {volume} {87}},\ \bibinfo {pages} {048303}
  (\bibinfo {year} {2001})}\BibitemShut {NoStop}%
\bibitem [{\citenamefont {Griffith}(1921)}]{griffith1921vi}%
  \BibitemOpen
  \bibfield  {author} {\bibinfo {author} {\bibfnamefont {A.~A.}\ \bibnamefont
  {Griffith}},\ }\bibfield  {title} {\enquote {\bibinfo {title} {Vi. the
  phenomena of rupture and flow in solids},}\ }\href@noop {} {\bibfield
  {journal} {\bibinfo  {journal} {Philosophical Transactions Of The Royal
  Society Of London. Series A}\ }\textbf {\bibinfo {volume} {221}},\ \bibinfo
  {pages} {163--198} (\bibinfo {year} {1921})}\BibitemShut {NoStop}%
\bibitem [{\citenamefont {Fielding}(2021)}]{fielding2021yielding}%
  \BibitemOpen
  \bibfield  {author} {\bibinfo {author} {\bibfnamefont {S.~M.}\ \bibnamefont
  {Fielding}},\ }\href@noop {} {\enquote {\bibinfo {title} {Yielding, shear
  banding and brittle failure of amorphous materials},}\ } (\bibinfo {year}
  {2021}),\ \Eprint {http://arxiv.org/abs/2103.06782} {arXiv:2103.06782
  [cond-mat.stat-mech]} \BibitemShut {NoStop}%
\bibitem [{\citenamefont {Benzi}\ \emph {et~al.}(2021)\citenamefont {Benzi},
  \citenamefont {Divoux}, \citenamefont {Barentin}, \citenamefont {Manneville},
  \citenamefont {Sbragaglia},\ and\ \citenamefont
  {Toschi}}]{benzi2021continuum}%
  \BibitemOpen
  \bibfield  {author} {\bibinfo {author} {\bibfnamefont {R.}~\bibnamefont
  {Benzi}}, \bibinfo {author} {\bibfnamefont {T.}~\bibnamefont {Divoux}},
  \bibinfo {author} {\bibfnamefont {C.}~\bibnamefont {Barentin}}, \bibinfo
  {author} {\bibfnamefont {S.}~\bibnamefont {Manneville}}, \bibinfo {author}
  {\bibfnamefont {M.}~\bibnamefont {Sbragaglia}}, \ and\ \bibinfo {author}
  {\bibfnamefont {F.}~\bibnamefont {Toschi}},\ }\href@noop {} {\enquote
  {\bibinfo {title} {Continuum modelling of shear start-up in soft glassy
  materials},}\ } (\bibinfo {year} {2021}),\ \Eprint
  {http://arxiv.org/abs/2103.17071} {arXiv:2103.17071 [cond-mat.soft]}
  \BibitemShut {NoStop}%
\bibitem [{\citenamefont {Gupta}, \citenamefont {Elfring},\ and\ \citenamefont
  {Frigaard}(2020)}]{gupta2020rheology}%
  \BibitemOpen
  \bibfield  {author} {\bibinfo {author} {\bibfnamefont {R.}~\bibnamefont
  {Gupta}}, \bibinfo {author} {\bibfnamefont {G.}~\bibnamefont {Elfring}}, \
  and\ \bibinfo {author} {\bibfnamefont {I.}~\bibnamefont {Frigaard}},\
  }\bibfield  {title} {\enquote {\bibinfo {title} {Rheology and creep dynamics
  of wormlike micellar gels},}\ }\href@noop {} {\bibfield  {journal} {\bibinfo
  {journal} {Bulletin of the American Physical Society}\ } (\bibinfo {year}
  {2020})}\BibitemShut {NoStop}%
\bibitem [{\citenamefont {Zhang}\ and\ \citenamefont
  {Wei}(2013)}]{zhang2013mesoscale}%
  \BibitemOpen
  \bibfield  {author} {\bibinfo {author} {\bibfnamefont {C.}~\bibnamefont
  {Zhang}}\ and\ \bibinfo {author} {\bibfnamefont {J.}~\bibnamefont {Wei}},\
  }\bibfield  {title} {\enquote {\bibinfo {title} {Mesoscale simulation study
  of the structure and rheology of dilute solutions of flexible micelles},}\
  }\href@noop {} {\bibfield  {journal} {\bibinfo  {journal} {Chemical
  Engineering Science}\ }\textbf {\bibinfo {volume} {102}},\ \bibinfo {pages}
  {544--550} (\bibinfo {year} {2013})}\BibitemShut {NoStop}%
\bibitem [{\citenamefont {Padding}, \citenamefont {Boek},\ and\ \citenamefont
  {Briels}(2008)}]{padding2008dynamics}%
  \BibitemOpen
  \bibfield  {author} {\bibinfo {author} {\bibfnamefont {J.~T.}\ \bibnamefont
  {Padding}}, \bibinfo {author} {\bibfnamefont {E.~S.}\ \bibnamefont {Boek}}, \
  and\ \bibinfo {author} {\bibfnamefont {W.~J.}\ \bibnamefont {Briels}},\
  }\bibfield  {title} {\enquote {\bibinfo {title} {Dynamics and rheology of
  wormlike micelles emerging from particulate computer simulations},}\
  }\href@noop {} {\bibfield  {journal} {\bibinfo  {journal} {The Journal of
  chemical physics}\ }\textbf {\bibinfo {volume} {129}},\ \bibinfo {pages}
  {074903} (\bibinfo {year} {2008})}\BibitemShut {NoStop}%
\bibitem [{\citenamefont {Mewis}\ and\ \citenamefont
  {Wagner}(2009)}]{mewis2009thixotropy}%
  \BibitemOpen
  \bibfield  {author} {\bibinfo {author} {\bibfnamefont {J.}~\bibnamefont
  {Mewis}}\ and\ \bibinfo {author} {\bibfnamefont {N.~J.}\ \bibnamefont
  {Wagner}},\ }\bibfield  {title} {\enquote {\bibinfo {title} {Thixotropy},}\
  }\href@noop {} {\bibfield  {journal} {\bibinfo  {journal} {Advances in
  Colloid and Interface Science}\ }\textbf {\bibinfo {volume} {147}},\ \bibinfo
  {pages} {214--227} (\bibinfo {year} {2009})}\BibitemShut {NoStop}%
\bibitem [{\citenamefont {Larson}(2015)}]{larson2015constitutive}%
  \BibitemOpen
  \bibfield  {author} {\bibinfo {author} {\bibfnamefont {R.}~\bibnamefont
  {Larson}},\ }\bibfield  {title} {\enquote {\bibinfo {title} {Constitutive
  equations for thixotropic fluids},}\ }\href@noop {} {\bibfield  {journal}
  {\bibinfo  {journal} {Journal of Rheology}\ }\textbf {\bibinfo {volume}
  {59}},\ \bibinfo {pages} {595--611} (\bibinfo {year} {2015})}\BibitemShut
  {NoStop}%
\bibitem [{\citenamefont {Larson}\ and\ \citenamefont
  {Wei}(2019)}]{larson2019review}%
  \BibitemOpen
  \bibfield  {author} {\bibinfo {author} {\bibfnamefont {R.~G.}\ \bibnamefont
  {Larson}}\ and\ \bibinfo {author} {\bibfnamefont {Y.}~\bibnamefont {Wei}},\
  }\bibfield  {title} {\enquote {\bibinfo {title} {A review of thixotropy and
  its rheological modeling},}\ }\href@noop {} {\bibfield  {journal} {\bibinfo
  {journal} {Journal of Rheology}\ }\textbf {\bibinfo {volume} {63}},\ \bibinfo
  {pages} {477--501} (\bibinfo {year} {2019})}\BibitemShut {NoStop}%
\bibitem [{\citenamefont {Spenley}, \citenamefont {Yuan},\ and\ \citenamefont
  {Cates}(1996)}]{spenley1996nonmonotonic}%
  \BibitemOpen
  \bibfield  {author} {\bibinfo {author} {\bibfnamefont {N.}~\bibnamefont
  {Spenley}}, \bibinfo {author} {\bibfnamefont {X.}~\bibnamefont {Yuan}}, \
  and\ \bibinfo {author} {\bibfnamefont {M.}~\bibnamefont {Cates}},\ }\bibfield
   {title} {\enquote {\bibinfo {title} {Nonmonotonic constitutive laws and the
  formation of shear-banded flows},}\ }\href@noop {} {\bibfield  {journal}
  {\bibinfo  {journal} {Journal de Physique II}\ }\textbf {\bibinfo {volume}
  {6}},\ \bibinfo {pages} {551--571} (\bibinfo {year} {1996})}\BibitemShut
  {NoStop}%
\bibitem [{\citenamefont {Britton}\ and\ \citenamefont
  {Callaghan}(1997)}]{britton1997two}%
  \BibitemOpen
  \bibfield  {author} {\bibinfo {author} {\bibfnamefont {M.~M.}\ \bibnamefont
  {Britton}}\ and\ \bibinfo {author} {\bibfnamefont {P.~T.}\ \bibnamefont
  {Callaghan}},\ }\bibfield  {title} {\enquote {\bibinfo {title} {Two-phase
  shear band structures at uniform stress},}\ }\href@noop {} {\bibfield
  {journal} {\bibinfo  {journal} {Physical Review Letters}\ }\textbf {\bibinfo
  {volume} {78}},\ \bibinfo {pages} {4930} (\bibinfo {year}
  {1997})}\BibitemShut {NoStop}%
\bibitem [{\citenamefont {Martin}\ and\ \citenamefont
  {Hu}(2012)}]{martin2012transient}%
  \BibitemOpen
  \bibfield  {author} {\bibinfo {author} {\bibfnamefont {J.~D.}\ \bibnamefont
  {Martin}}\ and\ \bibinfo {author} {\bibfnamefont {Y.~T.}\ \bibnamefont
  {Hu}},\ }\bibfield  {title} {\enquote {\bibinfo {title} {Transient and
  steady-state shear banding in aging soft glassy materials},}\ }\href@noop {}
  {\bibfield  {journal} {\bibinfo  {journal} {Soft Matter}\ }\textbf {\bibinfo
  {volume} {8}},\ \bibinfo {pages} {6940--6949} (\bibinfo {year}
  {2012})}\BibitemShut {NoStop}%
\bibitem [{\citenamefont {Lerouge}, \citenamefont {Decruppe},\ and\
  \citenamefont {Humbert}(1998)}]{lerouge1998shear}%
  \BibitemOpen
  \bibfield  {author} {\bibinfo {author} {\bibfnamefont {S.}~\bibnamefont
  {Lerouge}}, \bibinfo {author} {\bibfnamefont {J.}~\bibnamefont {Decruppe}}, \
  and\ \bibinfo {author} {\bibfnamefont {C.}~\bibnamefont {Humbert}},\
  }\bibfield  {title} {\enquote {\bibinfo {title} {Shear banding in a micellar
  solution under transient flow},}\ }\href@noop {} {\bibfield  {journal}
  {\bibinfo  {journal} {Physical Review Letters}\ }\textbf {\bibinfo {volume}
  {81}},\ \bibinfo {pages} {5457} (\bibinfo {year} {1998})}\BibitemShut
  {NoStop}%
\bibitem [{\citenamefont {Lerouge}, \citenamefont {Decruppe},\ and\
  \citenamefont {Berret}(2000)}]{lerouge2000correlations}%
  \BibitemOpen
  \bibfield  {author} {\bibinfo {author} {\bibfnamefont {S.}~\bibnamefont
  {Lerouge}}, \bibinfo {author} {\bibfnamefont {J.-P.}\ \bibnamefont
  {Decruppe}}, \ and\ \bibinfo {author} {\bibfnamefont {J.-F.}\ \bibnamefont
  {Berret}},\ }\bibfield  {title} {\enquote {\bibinfo {title} {Correlations
  between rheological and optical properties of a micellar solution under shear
  banding flow},}\ }\href@noop {} {\bibfield  {journal} {\bibinfo  {journal}
  {Langmuir}\ }\textbf {\bibinfo {volume} {16}},\ \bibinfo {pages} {6464--6474}
  (\bibinfo {year} {2000})}\BibitemShut {NoStop}%
\bibitem [{\citenamefont {Grand}, \citenamefont {Arrault},\ and\ \citenamefont
  {Cates}(1997)}]{grand1997slow}%
  \BibitemOpen
  \bibfield  {author} {\bibinfo {author} {\bibfnamefont {C.}~\bibnamefont
  {Grand}}, \bibinfo {author} {\bibfnamefont {J.}~\bibnamefont {Arrault}}, \
  and\ \bibinfo {author} {\bibfnamefont {M.}~\bibnamefont {Cates}},\ }\bibfield
   {title} {\enquote {\bibinfo {title} {Slow transients and metastability in
  wormlike micelle rheology},}\ }\href@noop {} {\bibfield  {journal} {\bibinfo
  {journal} {Journal de Physique II}\ }\textbf {\bibinfo {volume} {7}},\
  \bibinfo {pages} {1071--1086} (\bibinfo {year} {1997})}\BibitemShut {NoStop}%
\bibitem [{\citenamefont {Hu}\ and\ \citenamefont
  {Lips}(2005)}]{hu2005kinetics}%
  \BibitemOpen
  \bibfield  {author} {\bibinfo {author} {\bibfnamefont {Y.~T.}\ \bibnamefont
  {Hu}}\ and\ \bibinfo {author} {\bibfnamefont {A.}~\bibnamefont {Lips}},\
  }\bibfield  {title} {\enquote {\bibinfo {title} {Kinetics and mechanism of
  shear banding in an entangled micellar solution},}\ }\href@noop {} {\bibfield
   {journal} {\bibinfo  {journal} {Journal of Rheology}\ }\textbf {\bibinfo
  {volume} {49}},\ \bibinfo {pages} {1001--1027} (\bibinfo {year}
  {2005})}\BibitemShut {NoStop}%
\bibitem [{\citenamefont {Greco}\ and\ \citenamefont
  {Ball}(1997)}]{greco1997shear}%
  \BibitemOpen
  \bibfield  {author} {\bibinfo {author} {\bibfnamefont {F.}~\bibnamefont
  {Greco}}\ and\ \bibinfo {author} {\bibfnamefont {R.}~\bibnamefont {Ball}},\
  }\bibfield  {title} {\enquote {\bibinfo {title} {Shear-band formation in a
  non-newtonian fluid model with a constitutive instability},}\ }\href@noop {}
  {\bibfield  {journal} {\bibinfo  {journal} {Journal of Non-Newtonian Fluid
  Mechanics}\ }\textbf {\bibinfo {volume} {69}},\ \bibinfo {pages} {195--206}
  (\bibinfo {year} {1997})}\BibitemShut {NoStop}%
\bibitem [{\citenamefont {De~Gennes}(2007)}]{de2007melt}%
  \BibitemOpen
  \bibfield  {author} {\bibinfo {author} {\bibfnamefont {P.-G.}\ \bibnamefont
  {De~Gennes}},\ }\bibfield  {title} {\enquote {\bibinfo {title} {Melt fracture
  of entangled polymers},}\ }\href@noop {} {\bibfield  {journal} {\bibinfo
  {journal} {The European Physical Journal E}\ }\textbf {\bibinfo {volume}
  {23}},\ \bibinfo {pages} {3--5} (\bibinfo {year} {2007})}\BibitemShut
  {NoStop}%
\bibitem [{\citenamefont {Dijksman}\ \emph {et~al.}(2011)\citenamefont
  {Dijksman}, \citenamefont {Wortel}, \citenamefont {van Dellen}, \citenamefont
  {Dauchot},\ and\ \citenamefont {van Hecke}}]{dijksman2011jamming}%
  \BibitemOpen
  \bibfield  {author} {\bibinfo {author} {\bibfnamefont {J.~A.}\ \bibnamefont
  {Dijksman}}, \bibinfo {author} {\bibfnamefont {G.~H.}\ \bibnamefont
  {Wortel}}, \bibinfo {author} {\bibfnamefont {L.~T.}\ \bibnamefont {van
  Dellen}}, \bibinfo {author} {\bibfnamefont {O.}~\bibnamefont {Dauchot}}, \
  and\ \bibinfo {author} {\bibfnamefont {M.}~\bibnamefont {van Hecke}},\
  }\bibfield  {title} {\enquote {\bibinfo {title} {Jamming, yielding, and
  rheology of weakly vibrated granular media},}\ }\href@noop {} {\bibfield
  {journal} {\bibinfo  {journal} {Physical review letters}\ }\textbf {\bibinfo
  {volume} {107}},\ \bibinfo {pages} {108303} (\bibinfo {year}
  {2011})}\BibitemShut {NoStop}%
\bibitem [{\citenamefont {Michel}\ \emph {et~al.}(2001)\citenamefont {Michel},
  \citenamefont {Appell}, \citenamefont {Molino}, \citenamefont {Kieffer},\
  and\ \citenamefont {Porte}}]{michel2001unstable}%
  \BibitemOpen
  \bibfield  {author} {\bibinfo {author} {\bibfnamefont {E.}~\bibnamefont
  {Michel}}, \bibinfo {author} {\bibfnamefont {J.}~\bibnamefont {Appell}},
  \bibinfo {author} {\bibfnamefont {F.}~\bibnamefont {Molino}}, \bibinfo
  {author} {\bibfnamefont {J.}~\bibnamefont {Kieffer}}, \ and\ \bibinfo
  {author} {\bibfnamefont {G.}~\bibnamefont {Porte}},\ }\bibfield  {title}
  {\enquote {\bibinfo {title} {Unstable flow and nonmonotonic flow curves of
  transient networks},}\ }\href@noop {} {\bibfield  {journal} {\bibinfo
  {journal} {Journal of Rheology}\ }\textbf {\bibinfo {volume} {45}},\ \bibinfo
  {pages} {1465--1477} (\bibinfo {year} {2001})}\BibitemShut {NoStop}%
\bibitem [{\citenamefont {Dimitriou}\ and\ \citenamefont
  {McKinley}(2014)}]{dimitriou2014comprehensive}%
  \BibitemOpen
  \bibfield  {author} {\bibinfo {author} {\bibfnamefont {C.~J.}\ \bibnamefont
  {Dimitriou}}\ and\ \bibinfo {author} {\bibfnamefont {G.~H.}\ \bibnamefont
  {McKinley}},\ }\bibfield  {title} {\enquote {\bibinfo {title} {A
  comprehensive constitutive law for waxy crude oil: a thixotropic yield stress
  fluid},}\ }\href@noop {} {\bibfield  {journal} {\bibinfo  {journal} {Soft
  Matter}\ }\textbf {\bibinfo {volume} {10}},\ \bibinfo {pages} {6619--6644}
  (\bibinfo {year} {2014})}\BibitemShut {NoStop}%
\bibitem [{\citenamefont {Chu}\ and\ \citenamefont
  {Feng}(2010)}]{chu2010amidosulfobetaine}%
  \BibitemOpen
  \bibfield  {author} {\bibinfo {author} {\bibfnamefont {Z.}~\bibnamefont
  {Chu}}\ and\ \bibinfo {author} {\bibfnamefont {Y.}~\bibnamefont {Feng}},\
  }\bibfield  {title} {\enquote {\bibinfo {title} {Amidosulfobetaine surfactant
  gels with shear banding transitions},}\ }\href@noop {} {\bibfield  {journal}
  {\bibinfo  {journal} {Soft Matter}\ }\textbf {\bibinfo {volume} {6}},\
  \bibinfo {pages} {6065--6067} (\bibinfo {year} {2010})}\BibitemShut {NoStop}%
\bibitem [{\citenamefont {Raghavan}()}]{private}%
  \BibitemOpen
  \bibfield  {author} {\bibinfo {author} {\bibfnamefont {S.~R.}\ \bibnamefont
  {Raghavan}},\ }\href@noop {} {}\bibinfo {howpublished} {Private
  Communication}\BibitemShut {NoStop}%
\bibitem [{\citenamefont {Barlow}, \citenamefont {Cochran},\ and\ \citenamefont
  {Fielding}(2020)}]{barlow2020ductile}%
  \BibitemOpen
  \bibfield  {author} {\bibinfo {author} {\bibfnamefont {H.~J.}\ \bibnamefont
  {Barlow}}, \bibinfo {author} {\bibfnamefont {J.~O.}\ \bibnamefont {Cochran}},
  \ and\ \bibinfo {author} {\bibfnamefont {S.~M.}\ \bibnamefont {Fielding}},\
  }\bibfield  {title} {\enquote {\bibinfo {title} {Ductile and brittle yielding
  in thermal and athermal amorphous materials},}\ }\href@noop {} {\bibfield
  {journal} {\bibinfo  {journal} {Physical Review Letters}\ }\textbf {\bibinfo
  {volume} {125}},\ \bibinfo {pages} {168003} (\bibinfo {year}
  {2020})}\BibitemShut {NoStop}%
\bibitem [{\citenamefont {Singh}, \citenamefont {Ozawa},\ and\ \citenamefont
  {Berthier}(2020)}]{singh2020brittle}%
  \BibitemOpen
  \bibfield  {author} {\bibinfo {author} {\bibfnamefont {M.}~\bibnamefont
  {Singh}}, \bibinfo {author} {\bibfnamefont {M.}~\bibnamefont {Ozawa}}, \ and\
  \bibinfo {author} {\bibfnamefont {L.}~\bibnamefont {Berthier}},\ }\bibfield
  {title} {\enquote {\bibinfo {title} {Brittle yielding of amorphous solids at
  finite shear rates},}\ }\href@noop {} {\bibfield  {journal} {\bibinfo
  {journal} {Physical Review Materials}\ }\textbf {\bibinfo {volume} {4}},\
  \bibinfo {pages} {025603} (\bibinfo {year} {2020})}\BibitemShut {NoStop}%
\bibitem [{\citenamefont {Olsen}, \citenamefont {Kornfield},\ and\
  \citenamefont {Tirrell}(2010)}]{olsen2010yielding}%
  \BibitemOpen
  \bibfield  {author} {\bibinfo {author} {\bibfnamefont {B.~D.}\ \bibnamefont
  {Olsen}}, \bibinfo {author} {\bibfnamefont {J.~A.}\ \bibnamefont
  {Kornfield}}, \ and\ \bibinfo {author} {\bibfnamefont {D.~A.}\ \bibnamefont
  {Tirrell}},\ }\bibfield  {title} {\enquote {\bibinfo {title} {Yielding
  behavior in injectable hydrogels from telechelic proteins},}\ }\href@noop {}
  {\bibfield  {journal} {\bibinfo  {journal} {Macromolecules}\ }\textbf
  {\bibinfo {volume} {43}},\ \bibinfo {pages} {9094--9099} (\bibinfo {year}
  {2010})}\BibitemShut {NoStop}%
\end{thebibliography}%

\end{document}